\documentclass{emulateapj}
\usepackage{apjfonts}
\newcommand{\scaleup}{\epsscale{1.1}}
\newcommand{\scaledown}{\epsscale{0.9}}
\newcommand{\plotter}{\plotone}
\newcommand{\plotterr}{\plotone}
\newcommand{\breaker}{}

\newcommand{\tableset}{deluxetable}

\newcommand{\Mdot}{\dot{M}}

\newcommand{\etal}{et al.}
\newcommand{\mbh}{M_{\rm BH}}

\newcommand{\msun}{M_{\sun}}

\newcommand{\qeos}{q_{\rm eos}}
\newcommand{\fgas}{f_{\rm gas}}

\newcommand{\re}{R_{\rm e}}

\newcommand{\mdisk}{M_{\rm disk}}
\newcommand{\scalelen}{R_{\rm d}}
\newcommand{\barangle}{\phi_{b}}
\newcommand{\orbitfreq}{\Omega_{\rm o}}
\newcommand{\diskfreq}{\Omega_{\rm d}}
\newcommand{\degree}{^{\circ}}

\shorttitle{How Do Disks Survive Mergers?}
\shortauthors{Hopkins \etal}
\slugcomment{Submitted to ApJ, June 5, 2008}
\begin{document}

\title{How Do Disks Survive Mergers?} 
\author{Philip F. Hopkins\altaffilmark{1}, 
Thomas J. Cox\altaffilmark{1,2}, 
Joshua D. Younger\altaffilmark{1}, 
\&\ Lars Hernquist\altaffilmark{1}
}
\altaffiltext{1}{Harvard-Smithsonian Center for Astrophysics, 
60 Garden Street, Cambridge, MA 02138}
\altaffiltext{2}{W.~M.\ Keck Postdoctoral Fellow at the 
Harvard-Smithsonian Center for Astrophysics}

\begin{abstract}

We develop a general physical model for how 
galactic disks survive and/or are destroyed in mergers and interactions. 
Based on simple dynamical arguments, we show that gas primarily loses 
angular momentum to internal torques in a merger, induced by 
the gravity of the secondary. Gas within 
some characteristic radius, determined by the efficiency of 
this angular momentum loss (itself a function of the orbital parameters, 
mass ratio, and gas fraction of the merging galaxies), will 
quickly lose angular momentum to the stars sharing the perturbed host disk, fall to the 
center and be consumed in a starburst. We 
use a similar analysis to determine where violent relaxation of the pre-merger 
stellar disks is efficient on final coalescence. 
Our model describes both the dissipational and dissipationless 
components of the merger, and allows us to predict, for a 
given arbitrary encounter, the stellar and gas content of the material
that will survive (without significant angular momentum loss or violent relaxation) 
to re-form a disk in the merger remnant, versus being dissipationlessly 
violently relaxed or dissipationally losing angular momentum and 
forming a compact central starburst. 
We test these predictions with a large library of hydrodynamic merger 
simulations, and show that they agree well (with small scatter) 
with the properties of simulated merger remnants as a function of 
merger mass ratio, orbital parameters, and gas distributions, 
in simulations which span a wide range of parameter space 
in these properties as well as prescriptions for gas physics, 
stellar and AGN feedback, halo and initial disk structural 
properties, redshift, and galaxy masses. 
We show that, in an immediate (short-term) sense, the amount of stellar 
or gaseous disk that survives or re-forms 
following a given interaction can be understood purely 
in terms of simple, well-understood gravitational physics, 
independent of the details of the ISM gas physics or 
stellar and AGN feedback. 
This allows us to demonstrate and quantify how these physics are 
in fact important, in an indirect sense, to enable disks to survive 
mergers, by lowering star formation efficiencies in low mass systems 
(allowing them to retain large gas fractions) and distributing the 
gas to large radii. The efficiency of disk destruction in mergers is a 
strong function of gas content -- our model allows us to explicitly 
predict and demonstrate how, in sufficiently gas rich mergers (with quite 
general orbital parameters), even 1:1 mass-ratio mergers can yield 
disk-dominated remnants, and more realistic 1:3-1:4 mass-ratio 
major mergers can yield systems with $<20\%$ of their mass in bulges. 
We discuss a number of implications of this modeling for 
the abundance and morphology of bulges as a function of mass and 
redshift, and provide simple prescriptions for the implementation of our 
results in analytic or semi-analytic models of galaxy formation. 

\end{abstract}

\keywords{galaxies: formation --- galaxies: evolution --- galaxies: active --- 
galaxies: spiral --- cosmology: theory}

\section{Introduction}
\label{sec:intro}

In the now established ``concordance'' $\Lambda$CDM cosmology, 
structure grows hierarchically \citep[e.g.][]{whiterees78}, making mergers and interactions 
between galaxies an essential and inescapable process in 
galaxy formation. Indeed, mergers are widely believed to be 
responsible for the morphologies of spheroids \citep[bulges in disks 
and elliptical galaxies;][]{toomre77}, and 
observations find recent merger remnants in 
considerable abundance in the local universe 
\citep{schweizer82,LakeDressler86,Doyon94,ShierFischer98,James99,
Genzel01,tacconi:ulirgs.sb.profiles,dasyra:mass.ratio.conditions,dasyra:pg.qso.dynamics,
rj:profiles,rothberg.joseph:kinematics} as well as e.g.\ faint 
shells and tidal features common around apparently ``normal'' 
galaxies \citep{malin80,malin83,schweizer80,
schweizerseitzer92,schweizer96}, which are thought to be
signatures of galaxy collisions \citep[e.g.][]{hernquistquinn88,hernquist.spergel.92}. 

From both theoretical grounds 
\citep[][and references therein]{ostrikertremaine75,maller:sph.merger.rates,
fakhouri:halo.merger.rates,stewart:mw.minor.accretion} and 
observations \citep[e.g.][]{lin:merger.fraction,barton:triggered.sf,
woods:tidal.triggering,
woods:minor.mergers} it appears that ``minor'' mergers 
of mass ratios $\lesssim 1:10$ are ubiquitous (there are 
almost no galaxies without mergers of at least this mass ratio in the 
last few Gyr), and moreover a large fraction ($\sim1/2$) of the $\sim L_{\ast}$ 
galaxy population is observed and expected 
to have experienced a ``major'' merger 
(mass ratio $\lesssim1:3$) since $z\sim2-3$ 
\citep{lotz:merger.fraction,bell:merger.fraction,
bridge:merger.fractions,lin:mergers.by.type,kartaltepe:pair.fractions}. 
With increasing redshift, kinematic and morphological 
indications of recent, violent disturbance in disk-dominated galaxies 
appear more frequent \citep{hammer:obs.disks.w.mergers,
flores:tf.evolution,puech:highz.vsigma.disks,puech:tf.evol}.

Far from there not being enough mergers to explain the abundance 
of bulges and ellipticals, this has led to the 
concern that there may be far too {\em many} mergers to 
explain the survival and abundance of galactic disks in the context 
of our present understanding of galaxy formation. 
\citet{toomre72} were among the first to point out that mergers 
are capable of dramatically altering the morphologies of disks, 
transforming them into elliptical galaxies. 
Although their neglect of the importance of dissipational star formation 
and gas dynamics in the mergers led to some 
controversy \citep[e.g.][]{ostriker80,carlberg:phase.space,gunn87,kormendy:dissipation}, 
it is now increasingly well-established that 
major mergers between spiral galaxies (similar to those observed 
locally and at $z\lesssim2-3$) with gas fractions comparable to those 
observed yields remnants in good agreement with 
essentially all observed properties of low and intermediate-mass 
local elliptical galaxies \citep[e.g.\ morphologies, shapes, sizes, kinematics, densities, colors, 
black hole properties, fundamental scaling relations, 
stellar populations, and halo gas;][]{hernquist.89,
barnes.hernquist.91,barneshernquist96,
hernquist:phasespace,mihos:gradients,mihos:starbursts.96,
dimatteo:msigma,naab:gas,jesseit:kinematics,
cox:xray.gas,cox:kinematics,robertson:fp,springel:red.galaxies,burkert:anisotropy,
hopkins:clustering,
hopkins:cusps.ell,hopkins:cores,hopkins:cusps.fp,hopkins:groups.ell,hopkins:cusps.mergers}. 

Many
intermediate and low-luminosity ``cusp'' 
ellipticals (encompassing $\sim80-90\%$ of the 
mass density in ellipticals) contain significant embedded disks
\citep[perhaps all such ellipticals, given projection effects; see][]{ferrarese:type12,lauer:centers}, 
and they form a continuous sequence with most S0 galaxies, known to 
have prominent stellar (and even gaseous) disks \citep{kormendy:spheroidal1,
bender:ell.kinematics,ferrarese:type12,kormendy94:review,lauer:95,
faber:ell.centers,kormendy99,ferrarese:profiles,emsellem:sauron.rotation}. 
Indeed, the existence of embedded disks in simulated merger remnants is critical 
to matching the properties described above. 

A wide variety of observations including stellar populations and star formation histories 
\citep[e.g.][]{bender89,trager:ages,
mcdermid:sauron.profiles} and kinematic and structural analysis of recent 
merger remnants \citep{schweizer83,schweizer83:review,schweizerseitzer92,
schweizer:ngc34.disk,hibbard.yun:excess.light,rj:profiles} demonstrate that 
most of these disks are not accreted in the standard cosmological 
fashion after the spheroid 
forms -- they must somehow survive the merger or form very quickly thereafter from 
gas already in and around the galaxies. Therefore, despite the destruction of a 
large portion of a stellar disk in major mergers, {\em some} disk must survive mergers, 
and the amount that does so is a critical component determining 
many of the photometric and kinematic properties of even bulge-dominated 
and elliptical galaxies. 

Moreover, ``minor'' mergers -- at least those with mass ratios $\lesssim10:1$ (below 
which the difference between ``merger'' and accretion becomes increasingly blurred) -- 
are not generally believed to entirely destroy disks, but they 
are almost an order of magnitude more frequent than major 
mergers and as such may pose a 
more severe a problem for disk survival. 
In the $\Lambda$CDM cosmology, 
and from observed satellite fractions, 
it is unlikely than any disk (let alone a large fraction of disk galaxies) 
with a significant stellar age 
has survived $\sim5-10$\,Gyr without 
experiencing a merger of mass ratio $10:1$ or larger. 
Simulations \citep{quinn.84,quinn86:dynfric.on.sats,quinn93.minor.mergers,
hernquist.mihos:minor.mergers,
walker:disk.fragility.minor.merger,velazquezwhite:disk.heating,
naab:minor.mergers,
bournaud:minor.mergers,younger:antitruncated.disks,younger:minor.mergers} 
and analytic 
arguments \citep{ostrikertremaine75,tothostriker:disk.heating,
sellwood:resonant.disk.thickening} 
suggest that gas-poor minor mergers can convert a considerable 
fraction of a stellar disk into bulge and cause significant 
perturbation (``puffing up'' via dynamical heating) to the disk. 
The observed coldness of galactic disks suggests that 
this may be a severe problem: 
\citet{tothostriker:disk.heating} argued that large disks such as 
that in the Milky Way could not have undergone a merger of 
mass ratio $\lesssim10:1$ in the last $\sim10\,$Gyr. More 
recently e.g.\ \citet{stewart:mw.minor.accretion} and 
\citet{hammer:mw.no.mergers} emphasized that the tension
between these constraints and the expectation in CDM models that a
number of such mergers should occur implies either a deficit in our
understanding of hierarchical disk formation or a challenge to the
concordance cosmological model.

Given the successes of the $\Lambda$CDM model on large scales, 
and the increasing observational confirmation that disks do undergo 
(and therefore must somehow survive) 
a large number of mergers, it is likely that the problem lies in 
our (still relatively poor) understanding of disk galaxy formation. 
This has led to a great deal of focus on the problem of forming realistic 
disks in a cosmological context, with many different attempts and 
debate on the missing elements necessary to produce disks in 
simulations. Various groups have argued 
that self-consistent treatment of gas physics and star formation 
along with implementation of 
feedback of different kinds is necessary, along with greatly improved 
numerical resolution \citep{weil98:cooling.suppression.key.to.disks,
sommerlarsen99:disk.sne.fb,sommerlarsen03:disk.sne.fb,
thackercouchman00,thackercouchman01,abadi03:disk.structure,
governato04:resolution.fx,governato:disk.formation,
robertson:cosmological.disk.formation,
okamoto:feedback.vs.disk.morphology,scannapieco:fb.disk.sims}, 
in order to enable disks to survive 
their expected violent merger histories without completely 
losing angular momentum and transforming into systems that 
are too compact and have too much bulge mass (relative 
to real observed disks) by $z=0$. 

It has been known for some time \citep[see e.g.][]{hernquist:kinematic.subsystems,
barneshernquist96} that (even 
without any feedback) some
fraction of the gas in even a major merger of two disks 
can survive and form new, embedded disks in the remnant -- 
i.e.\ despite the problems outlined above, disks are not necessarily 
completely destroyed in mergers. However, 
early studies of this were restricted to cases with low gas 
content ($f_{\rm gas}\lesssim10\%$ in the progenitor disks), most of which was 
rapidly consumed in star formation, yielding small remnant disks in 
strongly bulge-dominated remnants. 
In seminal work, \citet{springel:spiral.in.merger} and 
\citet{robertson:disk.formation} showed that, in idealized merger simulations 
with significant stellar feedback to allow 
the stable evolution of extremely gas rich disks ($f_{\rm gas}\sim1$), 
even a major merger can produce a 
disk-dominated remnant. 
This has since been confirmed in fully cosmological simulations 
\citep{governato:disk.formation}. 
Together with other recent investigations (see references above), these works have 
led to the growing consensus that a combination of strong stellar feedback and 
large gas content is essential to the survival of disk galaxies. 

A large number of open questions remain, 
however. How, exactly, does feedback 
allow disks to survive mergers? What are the most important physics? 
Does it require fine-tuning of feedback prescriptions? How might 
things vary as a function of galaxy mass, redshift, gas content, 
merger orbits, and environment? Fundamentally, should this be expected 
for typical cosmological circumstances, or are these cases 
pathological? 

The ambiguity largely owes to the fact that there is no deep physical 
understanding of how disks survive or re-form after mergers and 
interactions. It has only just become possible to conduct simulations 
with the requisite large gas fractions, and thus far theoretical explanations 
have largely been restricted to phenomenological analysis, with continued 
efforts to improve resolution and sub-resolution prescriptions. 
Moreover, without a 
full model for how disks behave in interactions, these simulations 
cannot be placed into the broader context of the emergence 
of the entire Hubble sequence (for example asking the question, are 
the disks in lenticulars and embedded disks in ellipticals 
survivors of their pre-merger disks? Are they re-accreted? What determines 
how large they are? What is the key physics that gives rise to 
realistic embedded disks, leading to bulge-dominated galaxies 
with kinematic and photometric properties similar to those in the 
real universe?) 
or within a fully cosmological context.

The resolution requirements 
for full models of disk formation are severe -- limiting any 
attempt to properly simulate a cosmological box and still achieve 
the resolution necessary to reliably model a disk population -- 
and so models of the population of disks, largely semi-analytic, 
are forced to adopt simplified and un-tested prescriptions 
for the behavior of disks in mergers. This, in turn, has led to 
other well-known problems in modeling disk populations 
(even where prescriptions can ensure no artificial angular 
momentum losses); even when the 
cumulative (morphology-independent) galaxy mass function is 
correctly predicted at the low-mass end, semi-analytic models widely overproduce 
the relative abundance of low-mass spheroids and underproduce 
disks \citep[even when satellites, which have other associated 
model uncertainties, are removed from consideration; see e.g.][]{somerville:sam,
somerville:new.sam,croton:sam,
bower:sam,delucia:sam}. Lacking a proper, physically motivated understanding of 
how low-mass or gas-rich disks may or may not survive mergers, 
attempts to address this problem in the models have been 
purely phenomenological and involve arbitrary prescriptions 
\citep[see e.g.][]{koda:disk.survival.prescriptions}.

Motivated by these concerns, in this paper
we develop a physical, dynamical model for how 
disks survive and are destroyed in mergers and interactions. 
We show that, in an immediate (short-term) sense, the amount of stellar 
or gaseous disk that survives or re-forms 
following a given interaction can be understood purely 
in terms of simple, well-understood gravitational physics. 
Knowing these physics, we develop an analytic model that allows 
us to accurately predict how much of a given pre-merger 
stellar and cold gas disk will survive a merger, as a function of 
the merger mass ratio, orbital parameters, pre-merger cold gas 
fraction, and mass distribution of the gas and stars. 
We compare these predictions to the 
results of a large library of hundreds of hydrodynamic simulations of 
galaxy mergers and interactions, spanning a wide parameter space 
in these properties as well as prescriptions for gas physics, 
stellar and AGN feedback, halo and initial disk structural 
properties, redshift, and absolute galaxy masses. Our numerical experiments 
confirm that the analytic scalings accurately describe 
the behavior and bulge formation/disk destruction in mergers 
over the entire dynamic range surveyed, and 
confirm that the parameters not explicitly included in our 
model do not systematically affect either the mean predictions or 
the scatter of simulations about those predictions. This allows us to 
understand the mean behavior of systems with different orbits and 
mass ratios, as well as why systems with large gas fractions can 
form little bulge in even major mergers. 

This is 
possible because gas, in mergers, primarily loses angular momentum to 
internal gravitational torques (from the stars in the same disk) 
owing to asymmetries in 
the galaxy induced by the merger. Hydrodynamic torques and 
the direct torquing of the secondary are second-order effects, 
and very inefficient. Once gas is drained of angular momentum, 
there is little alternative but for it to fall to the center of the galaxy and 
form stars, regardless of the details of the prescriptions for star 
formation and feedback (these may change things at the 
$\sim10-20\%$ level by blowing out some of the gas, 
but they cannot fundamentally alter the 
fact that cold gas with no angular momentum will be largely 
unable to form any sort of disk, or the fact that a galaxy's worth of 
gas compressed to high densities and small radii 
will inevitably form a large mass in stars). 
But if the systems are sufficiently gas-rich, then there is little 
stellar material sharing the disk to torque on the gas in the interaction, and 
little or no angular momentum is lost. 

Feedback can dramatically alter the ability of a disk to survive in a 
cosmological sense: by allowing galaxies to retain large 
gas fractions (as opposed to no-feedback scenarios, in which cold gas 
in a disk is usually quickly converted into stars), they are 
more gas-rich when they undergo interactions, allowing them 
to avoid angular momentum loss for the reason above. Moreover, 
we show that in detail (owing to the resonant structure of 
interactions), it is really gas within a certain radius of the stellar disk 
that is drained of angular momentum. The commonly-invoked stellar wind 
feedback then enables cosmological disk survival in a second fashion: 
by redistributing gas out to large radii, it prevents angular momentum loss 
and allows rapid re-formation of disks after a merger. 
Independent of any tuning, our model allows us to quantify the 
disks expected as a function of interactions of arbitrary properties, 
and to physically, explicitly quantify what the requirements are for feedback, 
in a cosmological scenario, to enable disk survival. 

In \S~\ref{sec:sims} we describe our library of gas-rich merger simulations, 
which we use to test our physical model for disk destruction and 
survival. In \S~\ref{sec:id} we demonstrate the existence of genuine disks 
in remnants of even major mergers and 
briefly consider their properties, and compare methods to 
separate the disks and bulges in merger remnants. 
In \S~\ref{sec:form.major} we consider the question of how these disks 
form in and survive mergers: we identify the key components of any 
merger remnant in \S~\ref{sec:form.major:components}, highlighting 
that these disks originate from a combination of undestroyed pre-merger 
stellar disks and gas which avoids angular momentum loss 
in the merger. In \S~\ref{sec:form.major:angloss} we discuss how, in detail, 
that angular momentum loss proceeds. We use this, 
in \S~\ref{sec:model.overview}, to build a physical model for how 
angular momentum loss proceeds in mergers and predict the 
surviving disk content of merger remnants: we model and test how this 
depends on the gas content of the pre-merger disks (\S~\ref{sec:model.gas}), 
and the orbital parameters (\S~\ref{sec:model.orbit}) and 
mass ratio (\S~\ref{sec:model.massratio}) of the encounter. We generalize to 
first passage and fly-by encounters (\S~\ref{sec:model.flyby}) 
and demonstrate that (for otherwise fixed conditions at the time of an encounter) 
our conclusions are purely dynamical, independent of 
feedback physics or details in our 
treatment of e.g.\ star formation and the ISM gas physics (\S~\ref{sec:model.feedback}), 
although we use our model to determine exactly how these choices 
can have dramatic {\em indirect} consequences for disk survival (by 
altering the state of systems leading into a merger). 
We discuss some exceptions and pathological cases in \S~\ref{sec:model.exceptions}, 
and relate our results to the long-term secular evolution of 
barred systems in \S~\ref{sec:model.secular}. 
In \S~\ref{sec:prescriptions}, we outline how these results can and should 
be applied in analytic and semi-analytic models of galaxy formation, and 
give appropriate prescriptions derived from our numerical experiments. 
Finally, we summarize our results and discuss some of their cosmological implications 
and applications to other models and observations in \S~\ref{sec:discussion}. 

Throughout, we assume a $\Omega_{\rm M}=0.3$, $\Omega_{\Lambda}=0.7$,
$H_{0}=70\,{\rm km\,s^{-1}\,Mpc^{-1}}$ cosmology, but this has
little effect on our conclusions.

\breaker
\section{The Simulations}
\label{sec:sims}

Our simulations were performed with the parallel TreeSPH code {\small
GADGET-2} \citep{springel:gadget}, employing the fully conservative
formulation \citep{springel:entropy} of smoothed particle
hydrodynamics (SPH), which conserves energy and entropy simultaneously
even when smoothing lengths evolve adaptively \citep[see
e.g.,][]{hernquist:sph.cautions,oshea:sph.tests}.  Our simulations
account for radiative cooling and incorporate a sub-resolution
model of a multiphase interstellar medium (ISM) to describe star
formation and supernova feedback \citep{springel:multiphase}.
Feedback from supernovae is captured in this sub-resolution model
through an effective equation of state for star-forming gas, enabling
us to stably evolve disks with arbitrary gas fractions \citep[see,
e.g.][]{springel:models,
springel:spiral.in.merger,robertson:disk.formation,robertson:msigma.evolution}.
This is described by the parameter $\qeos$, which ranges from
$\qeos=0$ for an isothermal gas with effective temperature of $10^4$
K, to $\qeos=1$ for our full multiphase model with an effective
temperature $\sim10^5$ K. We have also compared with a subset of simulations
which adopt the star formation feedback prescription from
\citet{mihos:cusps,mihos:gradients,
mihos:starbursts.94,mihos:starbursts.96}, in which
the ISM is treated as a single-phase isothermal medium and feedback
energy is deposited as a kinetic impulse. We examine the 
effects of these choices in \S~\ref{sec:model.feedback}, and find they are minimal. 

Likewise, although they make little difference to the analysis here, 
supermassive black holes are usually included at the centers of both
progenitor galaxies.  These black holes are represented by ``sink''
particles that accrete gas at a rate $\Mdot$ estimated from the local
gas density and sound speed using an Eddington-limited prescription
based on Bondi-Hoyle-Lyttleton accretion theory.  The bolometric
luminosity of the black hole is taken to be $L_{\rm
bol}=\epsilon_{r}\dot{M}\,c^{2}$, where $\epsilon_r=0.1$ is the
radiative efficiency. We assume that a small fraction (typically
$\approx 5\%$) of $L_{\rm bol}$ couples dynamically to the surrounding
gas, and that this feedback is injected into the gas as thermal
energy, weighted by the SPH smoothing kernel.  This fraction is a free
parameter, which we determine as in \citet{dimatteo:msigma} by
matching the observed $M_{\rm BH}-\sigma$ relation.  For now, we do
not resolve the small-scale dynamics of the gas in the immediate
vicinity of the black hole, but assume that the time-averaged
accretion rate can be estimated from the gas properties on the scale
of our spatial resolution (roughly $\approx 20$\,pc, in the best
cases). While the black holes can be indirectly important, owing to 
their feedback ejecting gas into the halo and thus preserving it from 
star formation until the final merger, we find that, for a given gas content 
at the time of the actual merger, our results are unchanged 
in a parallel suite of simulations without black holes. 

The progenitor galaxy models are described in
\citet{springel:models}, and we review their properties here.  For each
simulation, we generate two stable, isolated disk galaxies, each with
an extended dark matter halo with a \citet{hernquist:profile} profile,
motivated by cosmological simulations \citep{nfw:profile,busha:halomass}, 
an exponential disk of gas and stars, and (optionally) a
bulge.  The galaxies have total masses $M_{\rm vir}=V_{\rm
vir}^{3}/(10GH[z])$, with the baryonic disk having a mass
fraction $m_{\rm d}=0.041$, the bulge (when present) having $m_{\rm
b}=0.0136$, and the rest of the mass in dark matter.  The dark matter
halos are assigned a
concentration parameter scaled as in \citet{robertson:msigma.evolution} appropriately for the 
galaxy mass and redshift following \citet{bullock:concentrations}. We have also 
varied the concentration in a subset of simulations, and find it has little 
effect on our conclusions (because the central regions of the 
galaxy are, in any case, baryon-dominated), insofar as they pertain to 
disk survival in mergers (it has been demonstrated that halo concentrations 
are important for e.g.\ the exact sizes and velocity scalings of disks, and our 
predicted disk sizes scale accordingly). 
The initial disk scale-length is computed
based on an assumed spin parameter $\lambda=0.033$, chosen to be near
the mode in the $\lambda$ distribution measured in simulations \citep{vitvitska:spin},
and the scale-length of an initial bulge (when present) is set to $0.2$ times this.

Typically, each galaxy initially consists of 168000 dark matter halo
particles, 8000 bulge particles (when present), 40000 gas and 40000
stellar disk particles, and one BH particle.  We vary the numerical
resolution, with many simulations using twice, and a subset up to 128
times, as many particles. We choose the initial seed
mass of the black hole either in accord with the observed $M_{\rm
BH}$-$\sigma$ relation or to be sufficiently small that its presence
will not have an immediate dynamical effect, but we have varied the seed
mass to identify any systematic dependences.  Given the particle
numbers employed, the dark matter, gas, and star particles are all of
roughly equal mass, and central cusps in the dark matter and bulge
are reasonably well resolved.

We consider a series of several hundred simulations of colliding
galaxies, described in detail in 
\citet{robertson:fp,robertson:msigma.evolution} and
\citet{cox:xray.gas,cox:kinematics}.  We vary the numerical resolution, the orbit of the
encounter (disk inclinations, pericenter separation), the masses and
structural properties of the merging galaxies, presence or absence 
of bulges in the progenitor galaxies, initial gas fractions,
halo concentrations, the parameters describing star formation and
feedback from supernovae and black hole growth, and initial black hole
masses. 

The progenitor galaxies have virial velocities $V_{\rm vir}=55, 80, 113, 160,
226, 320,$ and $500\,{\rm km\,s^{-1}}$, and redshifts $z=0, 2, 3, {\rm
and}\ 6$, and the simulations span a range in final spheroid mass
$\mbh\sim10^{8}-10^{13}\,M_{\sun}$, covering essentially the
entire range of the observations we consider at all redshifts, and
allowing us to identify any systematic dependences in our models.  We
consider initial disk gas fractions (by mass) of $\fgas = 0.05,\ 0.1,\ 0.2,\ 0.4,\ 0.6,\ 
0.8,\ {\rm and}\ 1.0$ for several choices of virial velocities,
redshifts, and ISM equations of state. 
The results described in this
paper are based primarily on simulations of equal-mass mergers;
however, we examine in \S~\ref{sec:model.massratio} how our results scale with 
mass ratio in mixed encounters, down to mass ratios $\sim1:10$ or so, 
below which (as we show) the encounters have little noticeable 
effect. In detail, the simulations studied there are 
described in \citet{younger:minor.mergers} and constitute a complete 
subset of permutations of our standard galaxy models with mass ratios 
uniformly sampling the range $1:1$ to $1:8$. As in 
our larger set of $1:1$ mergers, at each mass ratio we 
systematically survey the effects of different absolute galaxy mass, 
orbital parameters, and disk gas fraction (resulting 
in a typical $\sim30-40$ simulations spanning the full range of 
orbital parameters and gas fractions of interest, around 
each mass ratio $\sim 1:1,\ 1:2,\ 1:4,$ and $1:8$). 
We have considered 
more limited studies of minor mergers where we vary 
e.g.\ the ISM equation of state, redshift, initial disk 
structural properties; as we find in our studies of these 
parameters in the larger suite of equal-mass mergers, 
they make no significant difference to our conclusions here.

\begin{\tableset}{lccccl}
\tabletypesize{\scriptsize}
\tablecaption{Disk Orientations\label{tbl:orbits}}
\tablewidth{0pt}
\tablehead{
\colhead{Name} &
\colhead{$\theta_{1}$} &
\colhead{$\phi_{1}$} &
\colhead{$\theta_{2}$} &
\colhead{$\phi_{2}$} &
\colhead{Comments} 
}
\startdata
{\bf b} & 180 & 0 & 0 & 0 & prograde-retrograde \\
{\bf c} & 180 & 0 & 180 & 0 & both retrograde \\
{\bf d} & 90 & 0 & 0 & 0 & polar-prograde \\
{\bf e} & 30 & 60 & -30 & 45 & ``random'' (prograde) \\
{\bf f} & 60 & 60 & 150 & 0 & tilted polar-retrograde \\
{\bf g} & 150 & 0 & -30 & 45 & retrograde-``random'' \\
{\bf h} & 0 & 0 & 0 & 0 & both prograde \\
\\
\hline \\
{\bf i} & 0 & 0 & 71 & 30 & Barnes orientations \\
{\bf j} & -109 & 90 & 71 & 90 &   \\
{\bf k} & -109 & -30 & 71 & -30 &   \\
{\bf l} & -109 & 30 & 180 & 0 &   \\
{\bf m} & 0 & 0 & 71 & 90 &   \\
{\bf n} & -109 & -30 & 71 & 30 &   \\
{\bf o} & -109 & 30 & 71 & -30 &   \\
{\bf p} & -109 & 90 & 180 & 0 &   \\
\\
\hline \\
{\bf m000} & 0 & 0 & -30 & 45 & Minor merger orientations \\
{\bf m030} & 30 & 0 & -30 & 45 &  \\
{\bf m090} & 90 & 0 & -30 & 45 &  \\
{\bf m150} & 150 & 0 & -30 & 45 &  \\
{\bf m180} & 180 & 0 & -30 & 45 &  \\
\enddata
\tablenotetext{\,}{List of disk galaxy orientations for major merger 
simulations. Columns show: (1) the orbit identification (used to 
refer to each orbit throughout); (2-3) the initial orientation 
of disk 1 (in standard spherical coordinates); (4-5) the initial 
orientation of disk 2; and (6) a brief description of some of the 
orientations.}
\end{\tableset}

Once built, pairs of galaxies are placed on parabolic orbits 
\citep[motivated by cosmological simulations; see e.g.][]{benson:cosmo.orbits,
khochfar:cosmo.orbits} with 
the spin axis of each disk specified by the angles $\theta$ and $\phi$ 
in standard spherical coordinates. Table~\ref{tbl:orbits} lists the 
orientations in different representative orbits we have sampled. 
The particular choice of orbits follows \citet{cox:kinematics}; 
there are seven idealized mergers (cases {\bf b}-{\bf h}) 
that represent orientations often seen in the literature 
(for example, case {\bf h}, where all the angular momentum 
vectors of the disks and orbit are initially aligned), the 
rest ({\bf i}-{\bf p}) follow \citet{barnes:disk.halo.mergers} 
by selecting unbiased initial disk orientations according 
to the coordinates of two oppositely directed tetrahedrons. 
These orbits are identical to those considered in various other studies, 
such as \citet{naab:minor.mergers}. 
For our series of orbits of various mass ratios from \citet{younger:minor.mergers}
(where, in minor 
mergers, the inclination of the secondary is less important than 
that of the primary) we survey the inclination of the primary 
in a systematic sense, considering all our mergers 
with $\theta_{1}=0,\,30,\,90,\,150,\,180\,\degree$ (cases {\bf m000-m180}). 
We examine the effect of 
orbits in detail in \S~\ref{sec:model.orbit}, and find that for random orbits, 
the differences are quantifiable but not strong -- pathological orbits 
(such as the aligned case {\bf h} above) are discussed in 
\S~\ref{sec:model.exceptions} (these pathological 
cases often, in fact, are the most efficient at destroying disks). 
We have also tested our predicted scalings with limited subsets of 
simulations that vary the pericentric passage distance and the 
energy of the orbit, described in \citet{robertson:disk.formation}
and \citet{cox:kinematics}, and find that our estimates are 
robust to these variations. 

Each simulation is evolved until the merger is complete and the remnants are 
fully relaxed, typically $\sim1-2$\,Gyr after the final merger 
and coalescence of the BHs. We then analyze the 
remnants following \citet{cox:kinematics}, in a manner designed to mirror 
the methods typically used by observers. For each remnant we project the 
stars onto a plane as if observed from a particular direction (we consider 
100 viewing angles to each remnant, which uniformly sample the unit sphere). 
When we plot projected quantities such as $\re$, $\sigma$, and $V_{c}$, we 
typically show just 
the median value for each simulation across all $\sim100$ viewing directions.
The sightline-to-sightline 
variation in these quantities is typically smaller than the 
simulation-to-simulation scatter, but we explicitly note where it is large.

\breaker
\section{The Existence of Disks in Major Merger Remnants}
\label{sec:id}

\citet{robertson:disk.formation} and \citet{springel:spiral.in.merger}, and 
subsequently \citet{governato:disk.formation} and \citet{naab:gas} 
have demonstrated that even major mergers can leave remnants with 
non-negligible disk components. Nevertheless, we wish to highlight 
several properties of these disks first, to establish their existence and 
nature. Moreover, 
we wish to ensure that we can 
robustly identify disks in our merger remnants, 
before going on to analyze the conditions for their survival. 
In order to do this, we have considered several methods, 
include e.g.\ fitting the surface brightness profiles to 
traditional bulge-disk decompositions, 
\citep[see e.g.][]{robertson:disk.formation,hopkins:cusps.mergers}, 
kinematic decompositions based on 
one and two-dimensional velocity maps 
\citep{cox:kinematics,hoffman:prep}, and 
three dimensional component fitting. 
These ultimately give similar results, although 
e.g.\ surface brightness profile fits and velocity decompositions 
can be considerably dependent on the viewing angle 
and are not especially robust at separating a small 
disk (in a bulge-dominated system) 
from e.g.\ other kinematic subcomponents 
or rotating bulges \citep[a well-known observational difficulty, 
see e.g.][and references therein]{balcells:bulge.xl,jk:profiles,
marinova:bar.frac.vs.freq,barazza:bar.colors}. 

We therefore choose to take advantage of our full three-dimensional 
information in the simulations to easily decompose bulges and 
disks in a simple, automated fashion. 
For convenience, let us consider the remnant in 
cylindrical coordinates ${\bf x}=(R,\,\phi)$ where the 
axis of symmetry ($\hat{z}$) is defined by the net angular momentum 
vector of the baryonic mass in the relaxed remnant. 
The effective rotational support of 
any given stellar or gas particle in the simulation is then 
\begin{equation}
\tilde{v}_{\rm rot} = \frac{v_{\phi}}{v_{c}(r)}
\end{equation}
where $v_{c}$ is the circular velocity 
\begin{equation}
v_{c} = \sqrt{\frac{G\,M_{\rm enc}(r)}{r}}
\end{equation}
(here $r$ is the {\em three dimensional} radius from the galaxy center). 
If we consider the distribution of 
baryonic mass in $\tilde{v}_{\rm rot}$, we find a clear 
segregation between bulge and disk components. 

\begin{figure*}
    \centering
    \scaleup
    \plotone{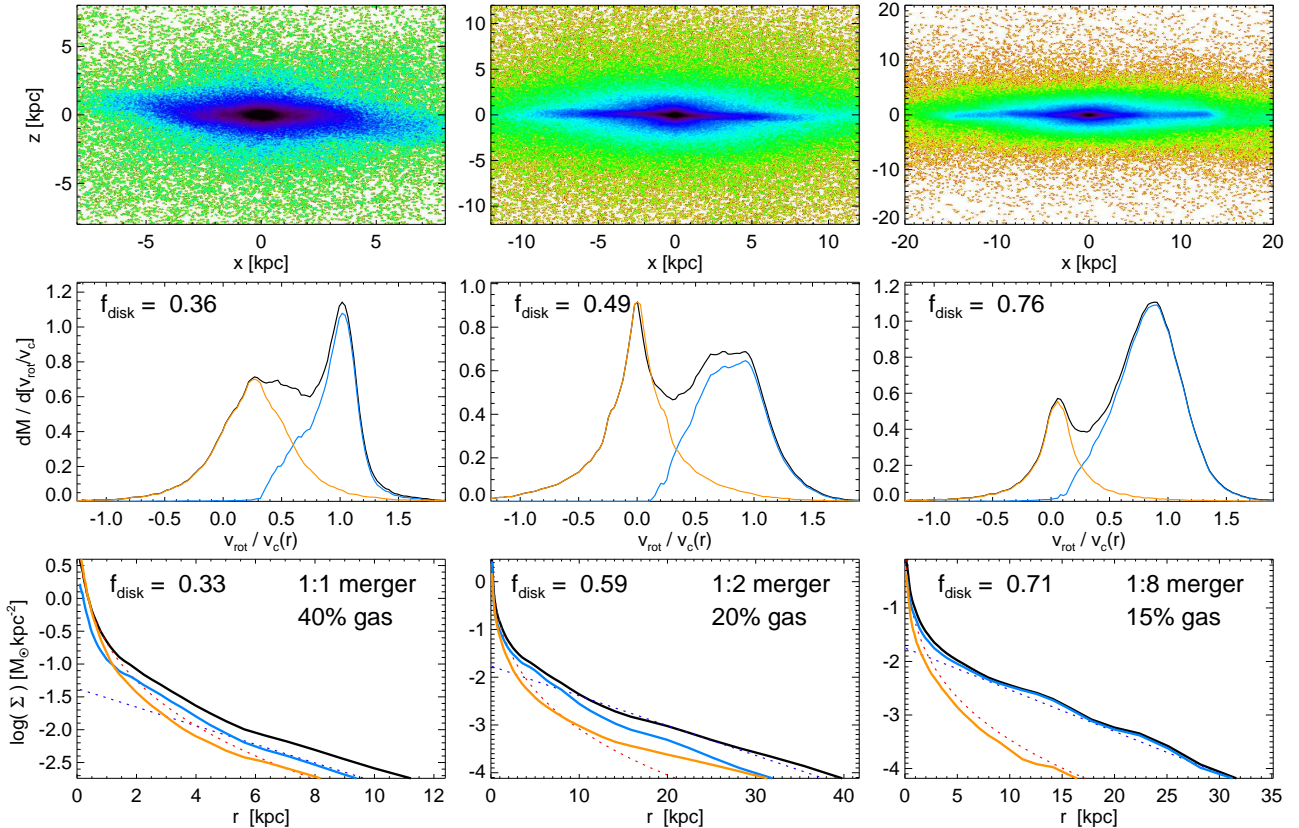}
    \caption{Examples of merger remnants with large disks. 
    {\em Top:} Edge-on projected stellar surface brightness 
    of the galaxy. {\em Middle:} the distribution of all stars in 
    their rotational support, $v_{\rm rot}/v_{c}(r)$, where 
    $v_{c}(r)$ is the circular velocity at $r$ and $v_{\rm rot}=v_{\phi}$ 
    in cylindrical coordinates is the rotational velocity about the 
    net stellar angular momentum axis. We decompose the clearly 
    bimodal distributions
    into bulge (orange, peak near $v_{\rm rot}/v_{c}(r)\sim0$) 
    and disk (blue, peak near $v_{\rm rot}/v_{c}(r)\sim1$) components, 
    with stellar mass fraction in the disk component ($f_{\rm disk}$) 
    labeled. {\em Bottom:} Azimuthally averaged face-on surface brightness 
    profile. We show the total profile (black) and profile of each of the 
    components separated by their rotational support (orange and blue respectively). 
    We fit the total profile to a standard bulge+disk decomposition, 
    and show the resulting fitted bulge (red dotted) and disk (blue dotted) 
    components and disk mass fraction. 
    The two methods recover similar decompositions in almost all cases: the ``disks'' 
    are rotationally supported with extended exponential profiles, 
    the ``bulges'' are dispersion supported with compact Sersic-law profiles. 
    We show three example remnants typical of our simulations: 
    {\em Left:} Equal mass (mass ratio $1:1$) merger remnant 
    with $\sim40\%$ gas at the time of merger. The remnant is 
    a bulge dominated elliptical/lenticular, but has a prominent 
    smooth stellar disk with $\sim30-40\%$ of the mass. 
    {\em Center:} Major (mass ratio $1:2$) merger remnant with 
    $\sim20\%$ gas at the time of merger. The remnant is a 
    marginally disk-dominated S0a-type galaxy, with some spiral structure in the disk. 
    {\em Right:} Minor (mass ratio $1:8$) merger remnant with 
    $\sim15\%$ gas at the time of merger. The remnant is a 
    Sb/Sc disk with a flattened, compact bulge. 
    \label{fig:jz.distrib}}
\end{figure*}

Figure~\ref{fig:jz.distrib} shows this for three simulations with 
large disks in the remnant. There is clearly a bimodal 
distribution in $\tilde{v}_{\rm rot}$, with one component having 
relatively little rotation (the bulge, with a peak near 
$\tilde{v}_{\rm rot}\approx0$), and 
one component being largely rotationally 
supported (the disk, with a peak near $\tilde{v}_{\rm rot}\sim1$). 
There are two stellar populations in these remnants, with a 
clean division in their rotational support. 

We can, from this plot alone, estimate a robust disk-bulge 
mass ratio, from fitting e.g.\ the sum of two Gaussian components 
(disk and bulge) 
to this distribution (or in a non-parametric sense, by 
assuming the bulge component has a symmetric 
rotation distribution about its peak, and mirroring 
the distribution about that, taking what remains to be disk). 
Our results are not sensitive to the exact details of our decomposition, 
but we experiment with a few different methods 
in order to estimate uncertainties on the bulge-disk decomposition 
which we refer to below. Again, we have repeated our entire 
analysis using alternative estimators of the disk-to-bulge mass 
(direct profile fits and velocity profile decompositions), and 
find that the same scalings apply in all cases (the uncertainties in 
the decomposition of a given simulation do, however, increase). 

\begin{figure*}
    \centering
    \plotone{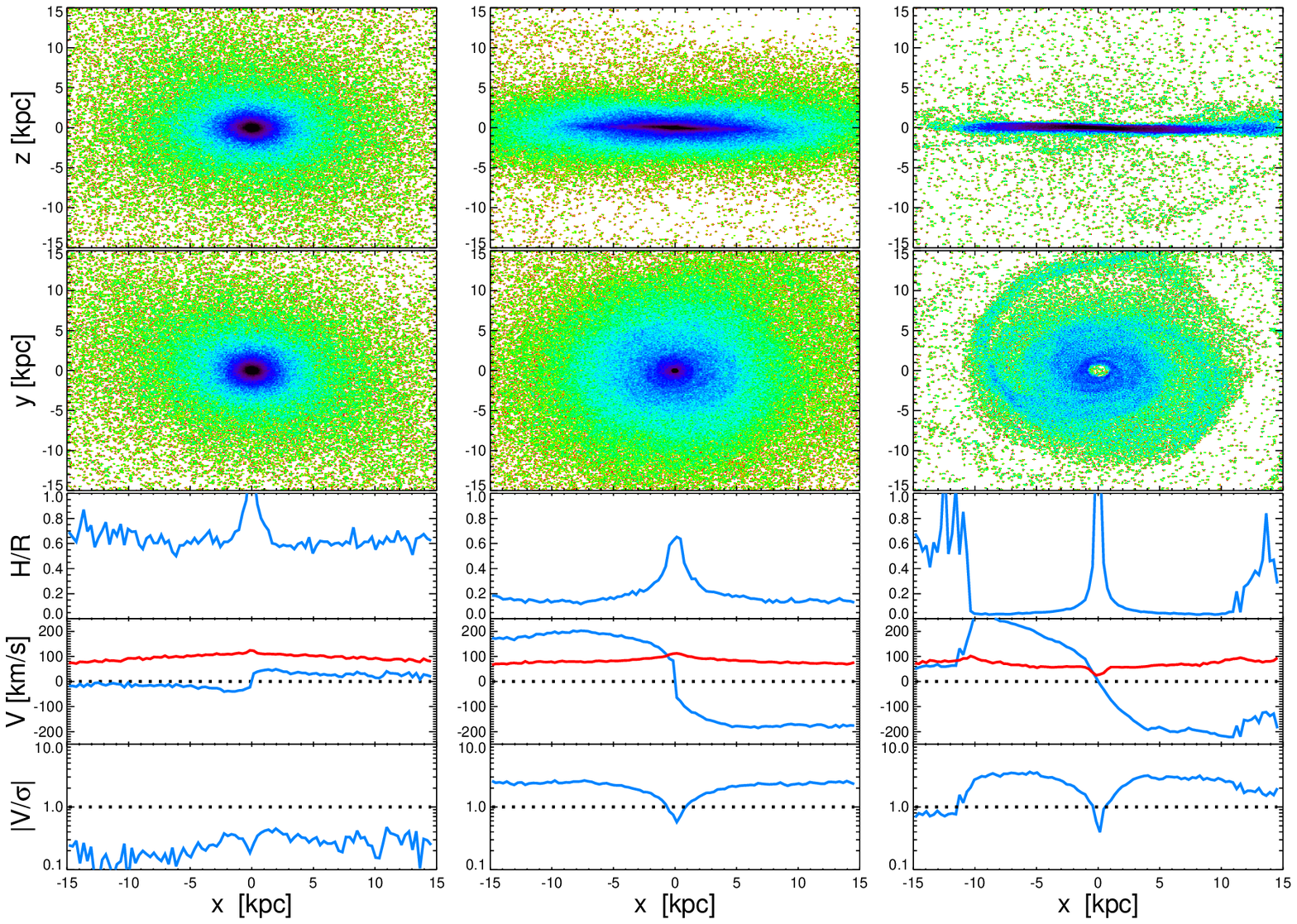}
    \caption{Bulge ({\em left}), stellar disk ({\em middle}), and 
    gas ({\em right}), in the remnant of a $1:2$ mass-ratio major merger 
    with $\sim20\%$ gas at the time of the merger, on a 
    typical random orbit (center panel in Figure~\ref{fig:jz.distrib}). 
    From top to bottom, panels show: 
    (a) Projected surface density (edge-on view). 
    (b) Face-on view. (c) Mean edge-on scale height 
    $H/R$ of the component as a function of circular radius $R=|x|$ (for 
    this projection). 
    (d) Edge-on velocity profile: mean velocity $v(r)$ (blue) and velocity 
    dispersion $\sigma(r)$ (red). (e) Rotational 
    support measure $v/\sigma$. 
    \label{fig:view.SbSc}}
\end{figure*}

Figure~\ref{fig:view.SbSc} shows one simulation from 
Figure~\ref{fig:jz.distrib} (S0 major-merger remnant), using this method to 
decompose the remnant into a stellar bulge and stellar disk. 
We also show the gas separately, which can cool and therefore forms an 
extremely thin disk. 
The properties are exactly what would be expected for a typical 
bulge-disk system: the ``bulge'' is a somewhat flattened 
ellipse with ellipticity $\epsilon\approx0.3-0.4$ ($H/R\approx0.6-0.7$), 
and is a pressure-supported system, with one-dimensional 
velocity dispersion $\sigma\sim120-150\,{\rm km\,s^{-1}}$ (depending 
on the sightline and slit width) and a 
rotation velocity $\sim30-50\,{\rm km\,s^{-1}}$. The resulting rotation parameter 
of the bulge itself 
($(V/\sigma)^{\ast}\sim0.4$) is typical of reasonably rapidly rotating 
bulges. It is also compact (as expected), with projected $R_{e}\sim 1-2$\,kpc. 
The stellar disk is like that of a combined thin-thick disk 
system, with $H/R\sim0.15-0.2$, and exhibits a flat rotation 
curve with $V_{\rm max}\sim200\,{\rm km\,s^{-1}}$. 
The disk is rotationally supported with typical values for a 
disk of similar mass and overall morphology, $V/\sigma\sim2-3$, 
and it is far more extended than the bulge ($R_{e}\sim10\,$kpc, 
putting it on the observed disk size-mass relation shown below). 
The properties of the gas disk are similar, with the obvious exception 
that, since the gas can cool, it forms a very thin disk 
($H/R\lesssim0.05$).

\begin{figure}
    \centering
    \plotone{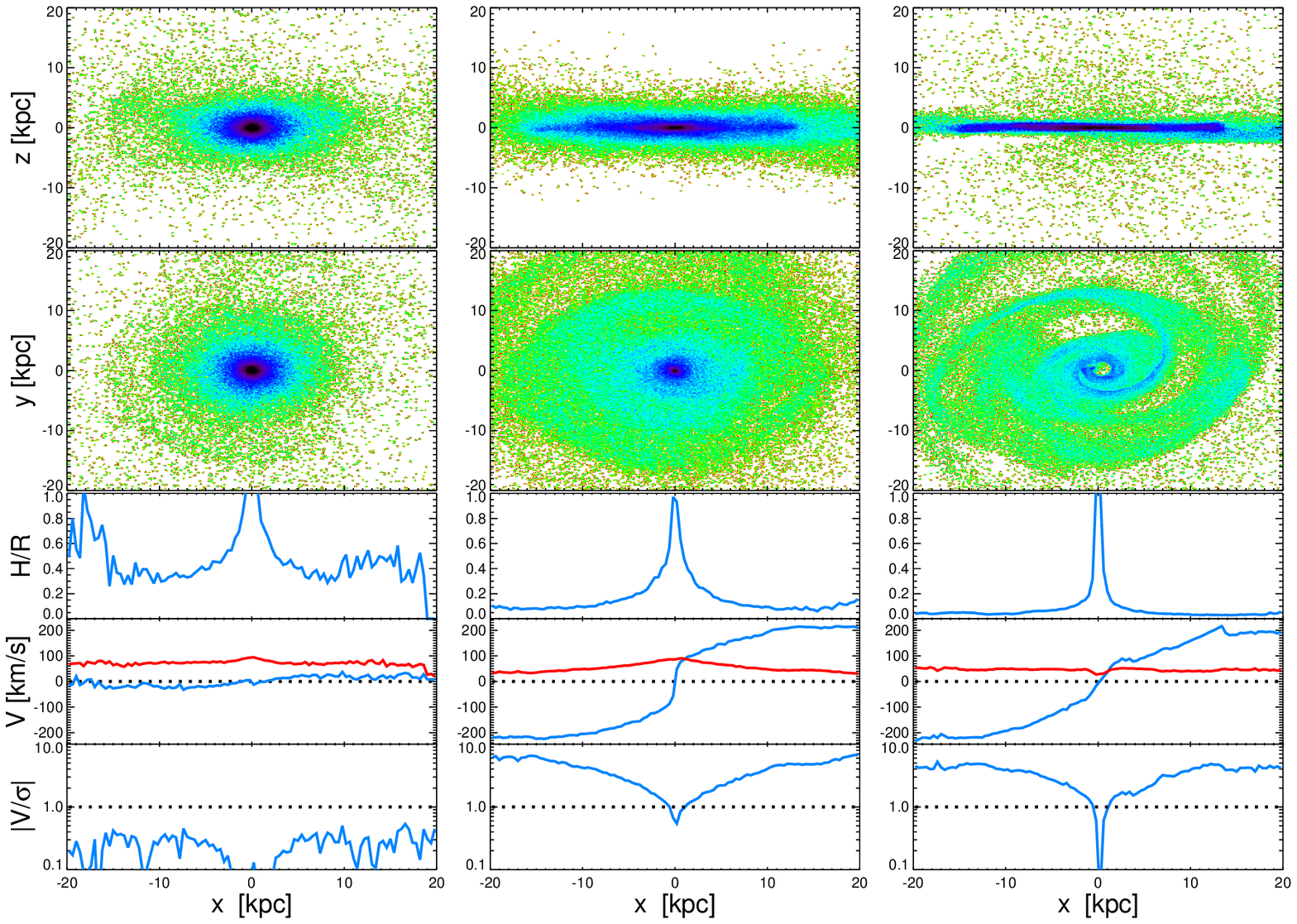}
    \caption{Bulge, disk, and gas, shown as in Figure~\ref{fig:view.SbSc}, 
    for the remnant of a $1:8$ mass-ratio minor merger with 
    $\sim15\%$ gas, on a polar orbit. 
    \label{fig:view.SbIm}}
\end{figure}

Figure~\ref{fig:view.SbIm} shows the components of 
another galaxy (Sbc minor-merger 
remnant) from Figure~\ref{fig:jz.distrib}, 
in the manner of Figure~\ref{fig:view.SbSc}. As expected given 
the smaller bulge-to-disk ratio in this case, the 
system is more flattened, with even larger rotational support 
($V/\sigma\sim5-10$ and $H/R\lesssim0.1$ even 
in the stellar disk). The bulge is clearly a distinct dispersion-supported 
component, despite being relatively flattened. 

\begin{figure}
    \centering
    \plotone{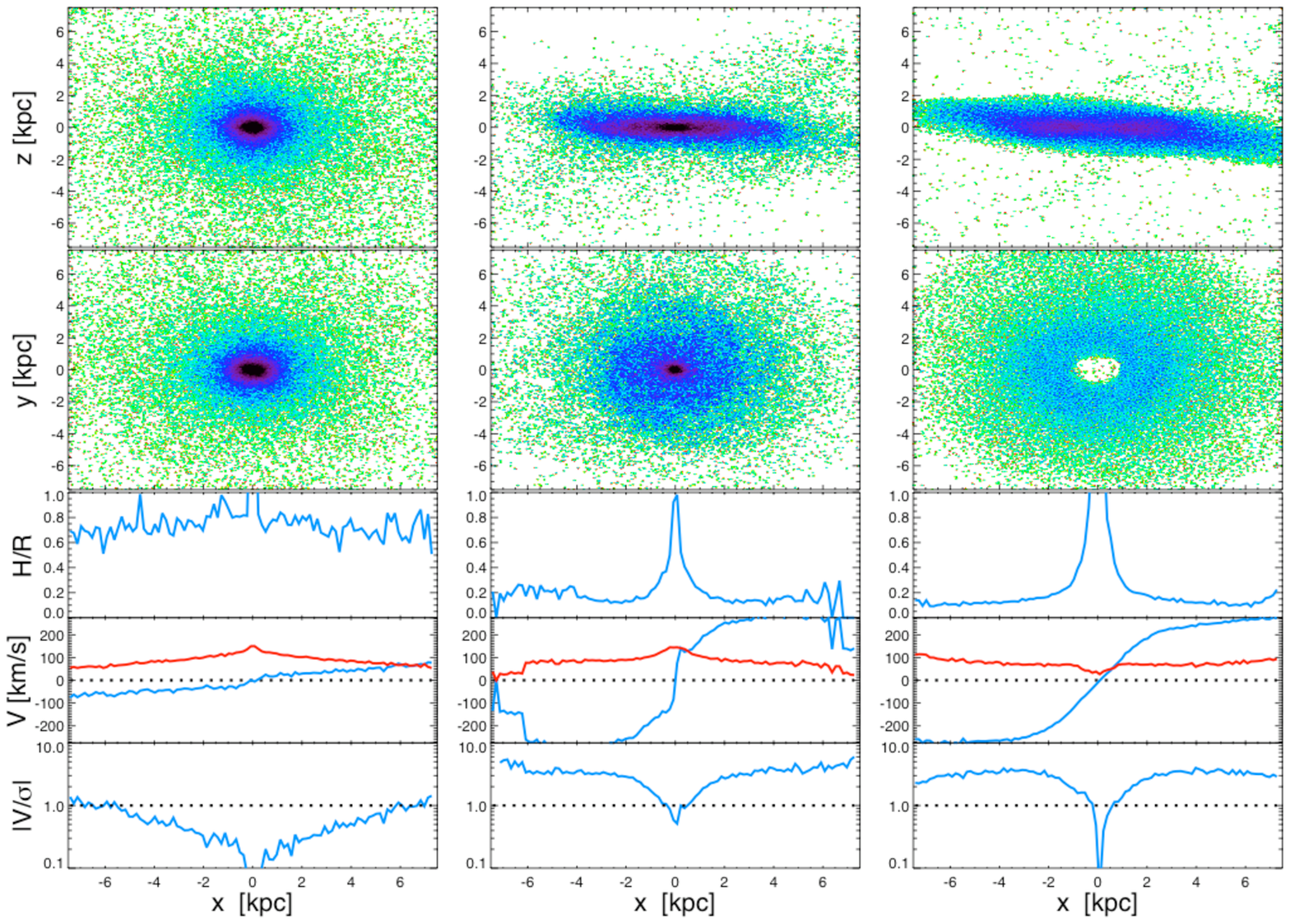}
    \caption{Bulge, disk, and gas, shown as in Figure~\ref{fig:view.SbSc}, 
    for the remnant of a $1:1$ mass-ratio major merger with 
    $\sim40\%$ gas, on an inclined polar orbit. 
    \label{fig:view.b3f}}
\end{figure}

Figure~\ref{fig:view.b3f} shows the components of 
the third galaxy (elliptical major-merger 
remnant) from Figure~\ref{fig:jz.distrib}, 
in the manner of Figure~\ref{fig:view.SbSc}. 
We show this case to demonstrate that embedded disks can be recovered 
reliably, and are indeed real, rotation supported ($V/\sigma\sim3-5$), 
and relatively thin ($H/R\lesssim0.2$ in the stellar disk, $\lesssim0.1$ in the gaseous disk)
kinematic objects even in simulations where they are not 
a majority of the mass (here, we find $B/T\sim0.7$).

\begin{figure}
    \centering
    \scaleup
    \plotone{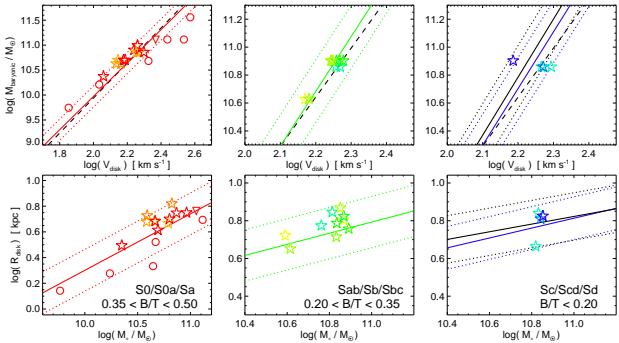}
    \caption{Simulated disk-dominated merger remnants on the observed baryonic 
    Tully-Fisher relation ({\em top}) and size-stellar mass relation ({\em bottom}). 
    We take $V_{\rm rot}$ from the rotation curves as in Figure~\ref{fig:view.SbSc} 
    and $R_{e}$ is the median projected half-mass radius. 
    We compare with the observed relations as a function of morphology 
    from \citet{courteau:disk.scalings} (solid lines in each panel; dotted lines 
    show the observed $\pm1\,\sigma$ scatter), for 
    S0-Sa ({\em left}; red), Sab-Sbc ({\em center}; green), and Sc-Sd ({\em right}; 
    blue and black, respectively) galaxies. For convenience, we assign our 
    simulations a ``morphology'' based on the bulge-to-disk ratio as labeled. 
    Most of our simulations are major (mass ratio $1:1$) mergers that yield significant, 
    but not dominant disks (and are therefore not shown here). Very late types 
    are only produced in our limited subset of small mass ratio ($\gtrsim1:8$) 
    mergers. In any case, the remnant disks all lie on the observed 
    Tully-Fisher and size-mass relations appropriate for their morphology -- 
    coupled with their rotation and scale heights, we can say they are real disks 
    in the observable sense. 
    \label{fig:TF}}
\end{figure}

As a further check that these are indeed real disks, 
Figure~\ref{fig:TF} plots the disks in disk-dominated simulation remnants on the 
baryonic Tully-Fisher and stellar size-mass relations observed 
for disks of similar morphology
\citep[see e.g.][]{belldejong:tf,mcgaugh:tf,shen:size.mass,courteau:disk.scalings}. 
For convenience, we will use the 
estimated bulge-to-disk ratio as a proxy for morphology throughout, 
with the values as labeled in Figure~\ref{fig:TF}. We take their 
velocities here from the projected disk rotation curves where they are flat, and 
take $R_{e}$ as the projected half-mass radius (note that this is different from the 
exponential disk scale length $h$; for a pure exponential disk 
$R_{e}=1.678\,h$, and we convert the observations where necessary accordingly). 

For each morphological class, our simulations agree well with the observed 
Tully-Fisher and size-mass relations. Given our limited sampling of very minor 
mergers with mass ratios $\gtrsim1:8$, we have only a few simulations with 
final $B/T<0.2$, but those nevertheless agree (as expected, since they have only been 
slightly modified from the original disk). We stress that we are {\em not} 
claiming to reproduce the Tully-Fisher or stellar size-mass relation of disks 
in an {\em a priori} manner: our (pre-merger) disks are constructed, by design, to 
more or less lie on the observed correlations. What we are saying is that, 
given progenitor disks that are similar to those observed, disks 
that form after or survive mergers (even 
major mergers) will remain on the appropriate correlations for their 
stellar mass and morphology. In short, when disks do survive mergers, 
they are ``real'' disks in the observable sense, not highly flattened bulges 
or unusual kinematic subcomponents.

\breaker
\section{Disk Formation in Major Mergers}
\label{sec:form.major}

Clearly, even major mergers can and do produce remnants with significant disks. 
We therefore ask how these disks form, and whether we can derive some 
analytic expectation for their masses as a function of progenitor and 
merger properties.

\subsection{Components of the Remnant: Surviving Gas Disks}
\label{sec:form.major:components}

For simplicity, let us begin with the case of an identical $1:1$ mass ratio 
merger (we will generalize to arbitrary mass ratios in \S~\ref{sec:model.massratio} below). 
Early in the merger, the galaxies experience a first passage and begin to 
lose angular momentum to the halos, rapidly coalescing on a timescale of 
order a couple orbital periods. In the final merger and coalescence of the 
galaxies, the stars which are initially ``cold'' (i.e.\ pre-merger disks) will 
scatter and violently relax \citep{lynden-bell67}, 
forming a \citet{devaucouleurs}-like quasi-spherical, 
dispersion supported profile. Of the gas available at the time of the final 
merger, some will lose its angular momentum, fall into the galaxy center, 
and (given the sudden rapid increase in density) rapidly transform into 
stars in a central starburst, forming a compact, central dissipational component of the 
remnant bulge \citep[for a detailed study of this 
component, see][]{hopkins:cusps.ell,hopkins:cusps.mergers}.
Gas that is at sufficiently large radii that it cannot efficiently 
fall in, or gas that for whatever reason cannot efficiently dissipate or lose 
angular momentum, will rapidly see the central potential relax (the equilibration 
timescale of the central bulge is only $\sim10^{8}\,$yr) and, 
having conserved its angular momentum, will rapidly cool and re-form 
a thin, rotationally supported disk. \citet{barneshernquist96} outline this process and show, in detail, 
how the cooling gas that survives the merger rapidly settles into a 
typical, rotationally supported exponential disk. This will then form stars, which 
constitute a new stellar disk. 

We emphasize these three components: 

{\bf Pre-Merger Stars:} These (along with the dark matter) constitute the 
collisionless (dissipationless) component of the merger. Because they are 
collisionless, the stars and dark matter distributions mix in the merger. A 
given star, as it moves through the merging galaxies on a random orbit, 
feels a rapidly fluctuating potential, which deflects its orbit and allows 
for the phase space distribution of the particles to uniformly mix. This violent 
relaxation process gives rise to a pressure supported system 
dominated by random velocities \citep{lynden-bell67} and transforms initially 
exponential disk into quasi-spherical \citet{devaucouleurs}-like
Sersic-law profiles. In the limit of a $1:1$ mass ratio merger, it is a 
good approximation to assume that all of the stars are violently relaxed -- the 
merger is sufficiently ``violent'' that no significant component of the 
pre-merger stellar disks will survive the merger. This is {\em not} 
necessarily true at lower mass ratios (see \S~\ref{sec:model.massratio}), but it simplifies our 
analysis to begin, while we consider such mergers. 

The remaining two components of the remnant can be identified 
with the gas supply available at the time of the final merger. 

{\bf Starburst Stars:} This is the remnant of a dissipational starburst, triggered in the 
merger. Some fraction of the gas will efficiently lose its angular momentum in the merger. 
Because gas can dissipate energy, it will then necessarily rapidly fall into the center 
of the merging system \citep[essentially free-falling to the center until the collapsing 
gas becomes self-gravitating; see][]{hopkins:cusps.mergers}. 
Collecting a large gas supply in the center, the result is a rapid, highly concentrated 
starburst -- in gas rich cases, this is analogous to that observed in 
e.g.\ nearby merging ULIRGs \citep{soifer84a,soifer84b,scoville86,sargent87,sargent89} 
and recent merger remnants \citep{kormendysanders92,hibbard.yun:excess.light,rj:profiles}. 
This builds up a dense, compact central stellar distribution, that raises the 
central phase space density and yields an effectively smaller, more baryon-dominated 
remnant. The starburst stars, being so concentrated (and typically 
having a more mixed orbital distribution owing to the random velocities of infalling gas 
in the starburst), are clearly part of the bulge (although they may have slightly 
different Sersic profiles and kinematics from the more extended bulge formed from 
violent relaxation of the pre-merger stars). This component is important for 
the structure and scalings of the bulge/spheroid component, and we study it in detail in 
\citet{hopkins:cusps.ell,hopkins:cores,hopkins:cusps.mergers}. It is essentially 
the dissipational component of the merger. For our purposes here, however, 
this is the gas ``lost,'' which becomes part of the bulge and no longer
contributes to the remnant disk. 

{\bf Surviving Gas/``Post-Merger'' Stars:} The gas that does {\em not} lose its angular 
momentum will, as described above, form a new disk as the remnant relaxes. 
For a $1:1$ merger, since (as noted above) the entire stellar distribution is violently 
relaxed, the post-merger disks can be entirely identified with gas that survives the merger. 
It is not, in this case, so much that the initial disks survive the merger intact, as it is that 
some of the gas remains at large radii/with significant angular momentum, which can 
rapidly re-form the disk after the merger. 
Essentially then (for major mergers), 
the question of how much of a disk will remain post-merger 
is a question of how much of the gas (at the time of the merger) 
will or will not lose its angular momentum.

\subsection{How Does the Gas Lose Its Angular Momentum?}
\label{sec:form.major:angloss}

How, then, does the gas lose angular momentum in a merger? 
The basic process has been understood since early simulations 
involving highly simplified models for gas dissipation in 
\citet{noguchi:merger.induced.bars.dissipationless,noguchi:merger.induced.bars.gas.forcing}, 
\citet{hernquist.89},
and \citet{barnes.hernquist.91}. With improved numerical models,
\citet{barneshernquist96} followed this process in detail, 
and showed that what happens 
in a typical major merger is as follows: 
the non-axisymmetric perturbation (owing to the companion) in the system 
induces (largely after first passage and on the final coalescence, since this is 
where the interaction is significant) a
non-axisymmetric response in
the disk.\footnote{In what follows, we will refer to this
non-axisymmetric response as a ``bar,'' for simplicity and 
because morphologically the induced feature, at least for 
some time during the merger resembles bars in isolated
barred spirals.  However, we caution that the formation 
mechanism which excites this response may be different
from that causing bars in isolated galaxies.  Furthermore,
while the non-axisymmetry is present throughout the merger,
it at times would not be classified as a bar morphologically,
particularly during the final coalescence of the galaxy
nuclei, when the resulting gas inflows are strong.}
A stellar bar and gas bar form, but because the gas is 
collisional and the stars are collisionless, 
the stellar bar will trail or lag behind the gas bar by a small offset 
(typically $\sim$a few degrees). The stellar bar therefore torques the gas bar, 
draining its angular momentum, and causing the gas to collapse to 
the center.

\begin{figure*}
    \centering
    \scaleup
    \plotterr{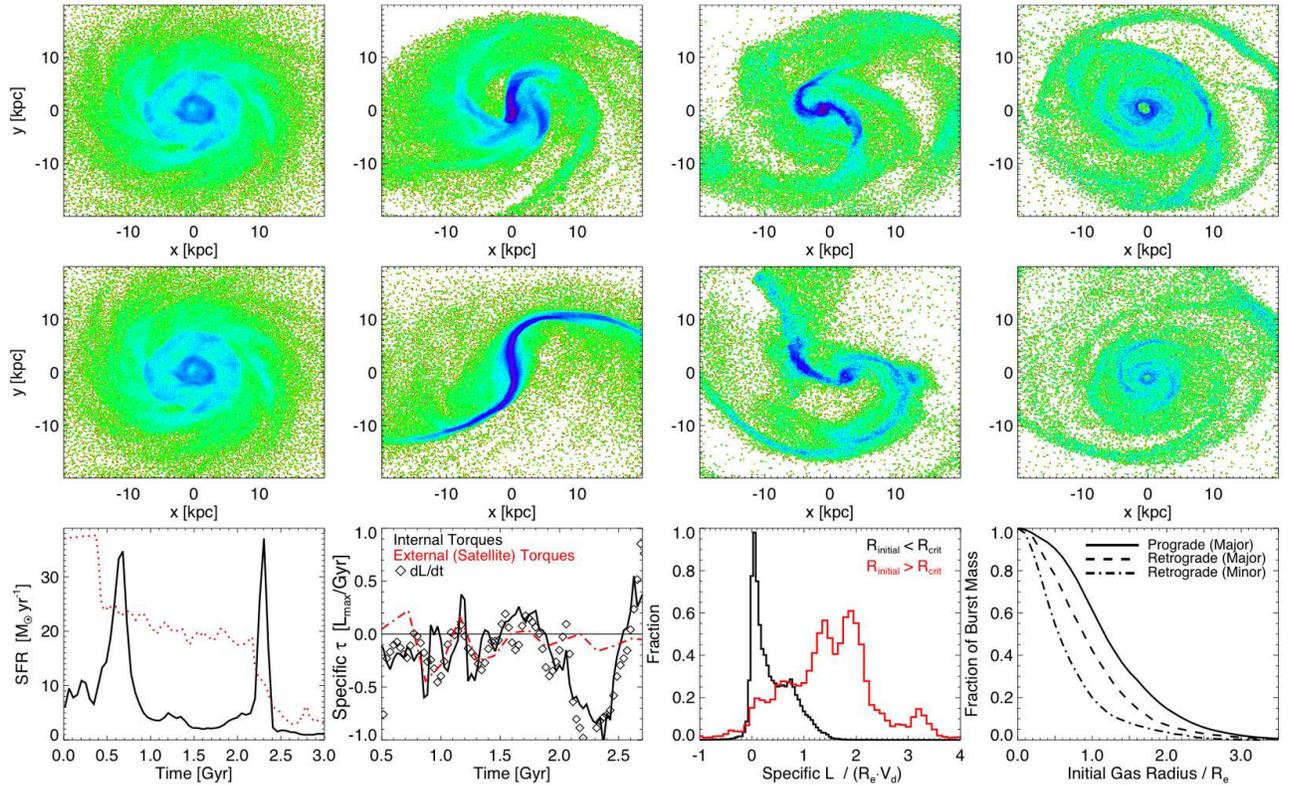}
    \caption{Illustration of the key processes that drive starbursts in a merger. 
    {\em Top:} Projected gas density (as in Figure~\ref{fig:view.SbSc}) in 
    the plane of the disk at representative times in a retrograde 1:3 merger 
    (left to right: before interaction, just after first passage, just after second passage/coalescence, 
    after relaxation). For clarity, just the gas from the primary is shown. 
    {\em Middle:} Same, for a prograde encounter. 
    The passage of the secondary induces a bar-like non-axisymmetric disturbance in the primary, 
    which survives after the short-lived passage and removes angular momentum from the 
    gas, leading to a starburst. The same process occurs on both passages, with a larger 
    (albeit less ``bar-like'') asymmetry on coalescence. The prograde encounters, being in 
    resonance, induce a stronger response that extends to larger radii. 
    {\em Bottom:} Quantities of interest in the merger. 
    {\em Left:} Star formation rate as a function of time in the prograde encounter (solid line; retrograde 
    is similar, but with a weaker enhancement in bursts). Red dotted line shows the specific 
    angular momentum of the gas that will participate in either 
    burst, in arbitrary units); as the gas rapidly loses angular momentum after the passages, 
    it drives a central starburst. 
    {\em Center Left:} Net specific torque on the primary gas that will participate in the final, 
    central starburst, as a function of time in the merger (in units of the initial 
    total angular momentum per Gyr). We compare the roughly numerically estimated net 
    torque (diamonds; from differentiating the specific angular momentum of the gas) 
    and the torque from two sources: stars in the same disk as the gas 
    (internal torques; black thick line), and the secondary galaxy and extended halos 
    (external torques; red dot-dashed line). The loss of angular momentum that 
    drives the secondary burst at $t\sim2.2-2.5$ is driven by internal torques from 
    the disturbed stellar disk; {\em not} the torque from the secondary galaxy itself. 
    {\em Center Right:} Final specific angular momentum content of 
    material (gas plus stars) that was originally gas at $R<R_{\rm crit}$ (our 
    predicted radius where merger-induced internal torques should be efficient 
    at removing angular momentum) or at $R>R_{\rm crit}$. 
    There is a strong division: gas inside a characteristic radius (corresponding 
    to where the internal asymmetry is strong; akin to the co-rotation resonance) 
    is mostly stripped of angular momentum. Gas at larger radii conserves 
    sufficient angular momentum to maintain a disk at similar $R_{e}$ and $V_{c}$. 
    {\em Right:} Original (cumulative) radial distribution of gas that participates 
    in the final starburst, relative to the initial disk effective radius (same in all cases) 
    for the prograde and retrograde cases shown and a 
    more minor retrograde merger. More resonant (prograde) and 
    more major encounters induce a 
    stronger response, with a larger co-rotation radius, and so torque gas out to larger 
    radii and efficiently strip more gas of angular momentum as predicted. 
    \label{fig:merger.demo}}
\end{figure*}

Figure~\ref{fig:merger.demo} illustrates this in a couple of
representative 1:3 mass ratio major merger simulations.  For a more
detailed description and illustration of the relevant physics, we
refer to \citet{barneshernquist96} (particularly their Figures~3-8);
but we briefly outline the scenario here.  We show the morphology of
the gas before the merger, when the disk is undisturbed, and shortly
after both first and second passages (the second passage leading, in
these cases, to a rapid coalescence), as well as in the relaxed
remnant.

The bar-like non-axisymmetric perturbation induced by the close
passages is clearly evident; the stars show a similar morphology at
each time, with a small phase offset in the bar pattern and (in the
remnant) a stellar bulge. As we discuss in \S~\ref{sec:model.orbit}, a
prograde encounter, being in resonance, produces a noticeably more
pronounced bar distortion (both in amplitude -- effectively ``bar
mass'' -- and spatial extent).  Shortly after each passage, this
double bar system efficiently removes angular momentum from the gas,
allowing it to fall into the center of the galaxy and participate in a
centrally concentrated starburst. 

Following \citet{barneshernquist96}, we track the gas in the primary
disk that will turn into stars in the final starburst, calculating the
net instantaneous gravitational torque decelerating the disk
rotation. We can coarsely infer what the total effective torque must
be by simply differentiating the specific angular momentum of this gas
at a given time, and compare this to the net torque from different
sources. Specifically, we separate the instantaneous gravitational
torques into the internal torques -- those from the stellar disk {\em
in the same galaxy} as the gas, chiefly from the bar (since the
axisymmetric disk, by definition, exerts no net torque) -- and the
external torques -- those from the gravity of the secondary galaxy
itself and the extended halos and their substructure. 

It is clear that, especially for the phases of interest shortly after
second passage and leading into the final starburst, when this gas
loses its angular momentum, the total torques are dominated by
internal torques from the stellar disk/bar system. The agreement
between these torques and the rate of change in the specific angular
momentum further argues that there are no other major sources of
angular momentum loss (specifically, both this comparison and direct
calculation demonstrate that the ``hydrodynamic torques'' defined by
pressure forces are not dominant).

As a result of these torques, gas within some critical radius where
the internal torques are strong (roughly inside the ``bar radius'' in
Figure~\ref{fig:merger.demo}) rapidly loses angular momentum. We
define this radius more precisely in \S~\ref{sec:model.overview}
below, but it is clear in the figure that at sufficiently large radius
the bar perturbation is weaker (and moreover, at larger radius the
potential of the disk, whether barred or unbarred, appears
increasingly axisymmetric); gas outside of these radii is relatively
unaffected. In general, then, the means for a more efficient encounter
to consume a larger fraction of the gas in the disk is to induce a
stronger bar disturbance, which is able to effectively exert internal
torques out to larger radii, stripping more gas of angular momentum
and bringing it into the central starburst (as evident in the stronger
prograde encounter in Figure~\ref{fig:merger.demo}).  Finally, the
system relaxes -- the gas that has not been subjected to strong
internal torques, having retained its angular momentum (at least in
large part), can rapidly re-form a disk. This may entail some
redistribution of that angular momentum (``filling in'' where the bar
depleted the gas of the disk), but does not lead to further
significant angular momentum loss.

\citet{barneshernquist96} and 
\citet{barnes:review} illustrate that this internal torquing 
is by far the dominant source of 
angular momentum loss, for typical orbits. This is 
because the stellar bar is: (a) more or less 
aligned in the plane with the gas bar, (b) trailing it 
by a small amount, and (c) relatively long-lived (it lives the rest of the duration of the 
merger, as opposed to the short time that is e.g.\ pericentric passage). 
The companion itself (either its baryonic mass or its halo), 
in most orbits, is not perfectly aligned with the 
gas disk, and the torque directly from it is much weaker (the tidal 
torquing drops by a factor $\sim(R_{\rm disk}/R_{\rm peri})^{3})$), 
and it can act only for a short duration on pericentric passage. 
There are some pathological orbits (e.g.\ perfectly coplanar prograde orbits) 
where this is not true, but these are exceptional cases, and we 
discuss them in \S~\ref{sec:model.exceptions}. 

At the final merger, one might image that mixing of random gas orbits or 
collisions and shocks 
would rapidly drain angular momentum, similar to what happens to 
the stars in violent relaxation. However, this is not possible, precisely because 
the gas is collisional: a Lagrangian gas element cannot go back and 
forth through the galaxy, but sticks to the other gas which has 
some net angular momentum. 
There could in principle be some net angular momentum cancellation, but this 
is inefficient -- the net angular momentum will almost always be 
comparable to the initial total. Even assuming random cancellation between 
two disks with comparable absolute angular momentum, the average 
change in net specific angular momentum is a factor $\sim2/3$; when 
one accounts for the angular momentum of the orbit -- typically comparable 
or even larger than that in the disks -- there is often no change or 
even a net {\em gain} in the gas specific angular momentum in a merger. 

A proper calculation 
shows that over the range in mass ratios $\mu\sim 0.1-1$, for a range of typical 
impact parameters $b\sim0.5-5\,\scalelen$, the expected final specific angular momentum 
after cancellation is approximately equal to the initial specific angular 
momentum of the primary (with $\sim20\%$ scatter). Cancellation is therefore 
inefficient. Even these cancellations, we find in detail, do not 
generally lead to a starburst in the same manner as a merger-induced bar, 
but simply lead to moderate disk contraction (and an equal number of mergers 
will scatter towards the opposite sense leading to disk expansion, keeping a mean 
specific angular momentum that is constant). They do not cause a starburst because, 
if two random parcels or streams of gas shock and lose angular momentum, 
the alignment and relative momenta would have to be near-perfect for them 
to lose, say $95\%$ of their angular momentum and fall all the way to the 
central $\sim 100$pc where a nuclear starburst would occur. Rather, they will lose 
some fraction of order unity of their angular momentum, fall in to a slightly smaller 
radius, and continue to orbit. 

Without the 
bar that can continuously drain angular momentum, the true burst is indeed 
inefficient. 
This initial bar-induced angular momentum loss scenario has been 
well-established in subsequent numerical studies 
\citep[see e.g.][]{noguchi:merger.induced.bars.dissipationless,
noguchi:merger.induced.bars.gas.forcing,
hernquist.89,hernquist:kinematic.subsystems,borderies:planetary.rings,
barnes.hernquist.91,barneshernquist96,barnes:review,
mihos:starbursts.96,springel:spiral.in.merger,robertson:disk.formation,
cox:kinematics,cox:massratio.starbursts,berentzen:gas.bar.interaction,
naab:gas}. 
We therefore can simplify our question to ask: how efficient will a given 
lagging stellar bar be at removing angular momentum from a 
leading gas bar?

\subsection{A Simple Model: Overview}
\label{sec:model.overview}

Consider a disk that contains a total gravitational mass $M$ (which 
can include a bulge and dark matter as well; the disk mass 
fraction we will denote $f_{\rm disk}$) within a 
characteristic scale length $\scalelen$. Some convenient dimensional variables are: 
\begin{eqnarray}
\nonumber & & \mdisk = f_{\rm disk}\,M\ ({\rm disk\ mass}) \\ 
\nonumber & & M_{\rm bar} = f_{\rm bar}\,M\ ({\rm stellar\ bar\ mass}) \\ 
\nonumber & & v_{c} = \sqrt{\frac{G\,M}{\scalelen}}\ ({\rm characteristic\ circular\ velocity}) \\ 
\nonumber & & \diskfreq = \frac{v_{c}}{\scalelen}\ ({\rm characteristic\ frequency}) \\ 
& & P = \frac{2\pi}{\diskfreq}\ ({\rm rotation\ period}) \, .
\end{eqnarray}
We also define 
the disk thickness according to a characteristic (assumed constant) 
relative scale height (height $H$ versus radius $R$; $H/R=$constant)
\begin{equation}
\tilde{H}\equiv H/R\ ({\rm disk\ scale\ height}). 
\label{eqn:scale.height}
\end{equation}
In these units, the circular velocity at a given 
cylindrical radius $R$ is given by 
\begin{equation}
v_{\rm circ}(r) = v_{c}\,{\tilde{v}(r)} \equiv v_{c}\,\sqrt{\frac{M_{\rm enc}(r)}{M}\,\frac{\scalelen}{R}}
\end{equation}
and dimensionless lengths are defined by 
\begin{eqnarray}
\nonumber & & \tilde{x} = x/\scalelen \\ 
\nonumber & & \tilde{y} = y/\scalelen \\ 
& & \tilde{R} \equiv \sqrt{x^{2}+y^{2}}/\scalelen. 
\end{eqnarray}
Throughout, we will use this notation: e.g.\ the dimensional variable $u$ 
is equal to the dimensionless variable $\tilde{u}$ times the appropriate combination 
of dimensional constants above. 

We will show that, in such a disk, a gas bar with a lagging stellar bar will 
efficiently cause gas to
lose its angular momentum and dissipate into a 
central starburst.
This will be the case for gas interior to a radius 
\begin{equation}
\frac{R_{\rm gas}}{\scalelen} \le \alpha\,(1-f_{\rm gas})\, f_{\rm disk} \, F(\theta,b)\, G(\mu) \, ,
\label{eqn:full.equation}
\end{equation}
where $\alpha\sim1$ is an appropriate 
integral constant (depending weakly on details of the stellar profile shape and 
bar dynamics), $f_{\rm gas}$ is the gas fraction in the disk and 
$f_{\rm disk}$ is the disk mass fraction. 
The factor 
\begin{equation}
G(\mu) \equiv \frac{2\mu}{(1+\mu)}
\end{equation}
contains the dependence on the merger mass ratio 
(where $\mu\le1\equiv M_{2}/M_{1}$). 
The term 
\begin{equation}
F(\theta,b) \equiv {\Bigl(}\frac{1}{1+[b/\scalelen]^{2}}{\Bigr)}^{3/2}\,\frac{1}{1-\orbitfreq/\diskfreq}
\label{eqn:full.eqn.orbit.1}
\end{equation}
accounts for the orbital parameters: $b$ is distance of 
pericentric passage on the relevant final passage before coalescence
($\sim1-$a couple $\scalelen$, for typical cosmological mergers) and 
$\orbitfreq$ is the orbital frequency at pericentric passage, 
\begin{eqnarray}
\nonumber \frac{\orbitfreq}{\diskfreq} &=& \frac{v_{\rm peri}}{v_{c}}\,
\frac{\scalelen}{b}\,\cos{(\theta)} \\
\nonumber &=& \sqrt{2\,(1+\mu)}\,[1+(b/\scalelen)^{2}]^{-3/4}\,\cos{(\theta)}\\
&\approx& 0.6\,\cos{(\theta)} \, ,
\label{eqn:full.eqn.orbit.2}
\end{eqnarray}
where $\theta$ is the inclination of the orbit relative to the disk, 
and the last equality comes from 
adopting typical cosmological orbits and mass ratios (but in any case, 
this is quite weakly dependent on the mass ratio). 

In the following sections, we derive this scaling piece by piece, and 
compare each aspect to the results from our library of hydrodynamic 
simulations. We show that it is robust and accurate as an approximation 
to the behavior in full numerical hydrodynamic experiments over a 
wide dynamic range of several orders of magnitude in surviving disk 
fraction (from systems with $\sim80-100\%$ of their disks surviving a 
merger to systems with $<1\%$ disk after a merger), as well 
as the entire dynamic range in mass, gas content, orbital properties, 
and different feedback prescriptions with which we experiment.

\subsubsection{Dependence on Disk Gas Content}
\label{sec:model.gas}

Let us consider an infinitely thin gas bar (a good approximation, owing to the 
efficiency of gas cooling) in a potential that is otherwise cylindrically symmetric 
except for the presence of a stellar bar. 
For simplicity, assume that the gas bar follows a fixed pattern speed 
$\Omega_{\rm b}$ ($\sim \orbitfreq$; we will derive the pattern speed later) 
in the disk (while there is not exactly a constant pattern speed in the outer regions of the disk, 
the torques are weak there in either case, and this approximation is globally quite good). 
We take $z=0$ to be the plane of the 
disk, and, without loss of generality, consider a 
frame rotating with the pattern speed of the gas bar, so that the 
bar lies along the $x$ axis. The material in the bar is rotationally supported, 
so it has instantaneous velocity ${\bf v}_{\rm \phi} =-v_{c}\,\tilde{v}(R)\,\hat{y}$, where 
$v_{c}\,\tilde{v}(R)$ (defined above) is the circular velocity at each point $x$. 

Now, consider a stellar bar of total mass $M_{\rm bar}$ also at fixed pattern speed, 
but offset by some instantaneous angle $\barangle$ 
from the gas bar (i.e.\ along the axis $y=\tan{(\barangle)}\,x$). 
The mass per unit length in the bar at a distance $R$ along the bar 
is ${\rm d}M_{\rm bar}/{\rm d}R = (M_{\rm bar}/\scalelen)\,\tilde{\Sigma}(R/\scalelen)$, 
where $\tilde{\Sigma}$ is the appropriate dimensionless mass profile 
and $\scalelen$ is some characteristic scale length (usually corresponding closely 
to the scale length of the unperturbed disk). If the initial disk is in equilibrium, 
(i.e.\ if the bar is some reasonable perturbation to the initial system), then 
the unperturbed net acceleration in the $x$ direction at some point $x$ in the gas bar 
will just be cancelled by the rotation of the system. Of interest here is the 
torque; if the stellar bar is also thin, then 
at a point $x=\tilde{x}\,\scalelen$ in the gas bar, the net torque per unit mass from the 
stellar bar will be 
\begin{equation}
\frac{{\rm d}j}{{\rm d}t}=\tilde{x}\,\scalelen\,\frac{{\rm d}v_{y}}{{\rm d}t}=
- \frac{G\,M_{\rm bar}}{\scalelen}\,
I_{0}(\barangle,\,\tilde{x}),
\label{eqn:bar1}
\end{equation}
where $I_{0}\sim1$ is a dimensionless integral which depends 
weakly on $\barangle$ and $\tilde{x}$ 
(at large $\barangle$, $I_{0}\rightarrow0$, 
reflecting the fact that the torque is dominated by times when the 
bars are close; since $\barangle\ll 1$ is expected, it is a good 
approximation to ignore the $\barangle$ dependence of $I_{0}$). 

If we assume the stellar bar is infinitely thin, there is 
a weak divergence in $I_{0}$ as $\barangle \rightarrow0$ 
(the accelerations become large when the bars nearly overlap). 
More accurately, we can allow for some finite height in the 
stellar disk/bar (it will always be thicker than the gas disk/bar). 
Let the stellar bar have a constant relative scale height $H/R$ given 
by Equation~(\ref{eqn:scale.height}), 
and for simplicity take its vertical profile to be constant 
density out to a height $\pm H$ (although assuming a 
more realistic vertical profile $\propto \exp{(-|z|/H)}$ or 
$\propto{\rm sech}^{2}{(z/H)}$ makes almost no difference to our calculation). 
The specific torque at $x$ in the gas bar now becomes 
\begin{equation}
\frac{{\rm d}j}{{\rm d}t}=
- \frac{G\,M_{\rm bar}}{\scalelen}\, 
\frac{1}{\sqrt{\sin^{2}{\barangle}+\tilde{H}^{2}}}\ 
I_{1}(\barangle,\,\tilde{x},\tilde{H}),
\label{eqn:bar2}
\end{equation}
where $I_{1}$ 
is an even weaker function of $\barangle$ and $\tilde{x}$ than $I_{0}$. 
The important behavior is entirely captured ignoring $I_{1}$, 
namely that the finite width of the stellar bar suppresses the 
numerical divergence seen earlier. 

The bar mass $M_{\rm bar}$ represents the stellar mass in the disk 
that is effectively part of the bar at the appropriate instant. We can therefore 
parameterize $f_{\rm bar} = M_{\rm bar}/M$ as 
\begin{equation}
f_{\rm bar} = (1-f_{\rm gas})\,f_{\rm disk}\,\Psi^{\prime}_{\rm bar}. 
\end{equation}
Here, $f_{\rm disk}$ is the disk mass fraction, and 
$f_{\rm gas}$ is the gas fraction in the disk (since we are interested 
in the cold gas, we explicitly ignore gas in e.g.\ a bulge or 
hot halo component). Therefore, the stellar mass of the 
disk is $(1-f_{\rm gas})\,f_{\rm disk}$ -- this defines the maximum 
mass that could be in the stellar bar. The parameter 
$\Psi^{\prime}_{\rm bar}$ thus defines the bar ``efficiency'' -- in 
an instantaneous sense as we have defined it here, 
$\Psi^{\prime}_{\rm bar}=0$ means there is no stellar bar, 
$\Psi^{\prime}_{\rm bar}=1$ implies the maximal stellar bar.

Already, we have one significant scaling -- the bar strength, 
and correspondingly the strength of the torques on the 
gas, scale with $(1-f_{\rm gas})$. In very gas rich 
systems where $f_{\rm gas}\rightarrow1$, there is no 
stellar mass to form a lagging bar and remove angular momentum 
from the gas. The gas itself may form a bar, but without a 
stellar bar to drag it, the angular momentum loss 
(over the timescales of relevance for a merger\footnote{At least, 
in this case, a major merger. The situation becomes more complicated 
in the limit of minor mergers with mass ratios $\sim$1:10; see 
\S~\ref{sec:model.secular}}) is  
inefficient. This is well known in e.g.\ dynamical studies of 
pure gas and stellar bars \citep[e.g.][]{schwarz:disk-bar,
athanassoula:bar.orbits,pfenniger:bar.dynamics,
combes:pseudobulges,friedli:gas.stellar.bar.evol,oniell:bar.obs}. There might be 
some angular momentum loss in such a case, between e.g.\ bar 
and halo \citep[e.g.][]{hernquistweinberg92},
but it will be small -- certainly nowhere near the 
efficient stripping of angular momentum needed to 
induce a significant starburst. 

If we consider a Lagrangian gas element at some initial 
radius $x_{0}$, then its orbit will decay as it loses angular 
momentum. The instantaneous rate of change in 
the radius $R=|x|$ will be given by 
${\rm d}R/{\rm d}t = (R/v_{\phi})\,{\rm d}v_{\phi}/{\rm d}t$. 
The characteristic timescale for the system to evolve is 
given by $2\pi/\diskfreq$, where 
$\diskfreq\sim v_{c}/h$ is the characteristic 
frequency of the disk. Define 
the timescale 
\begin{equation}
\tau \equiv \frac{\diskfreq}{2\pi}\,t = \frac{v_{c}}{\scalelen}\,t = \sqrt{\frac{G\,M}{\scalelen^{3}}}\,t, 
\end{equation}
where $M$ is the total effective gravitational mass of the 
disk and $\scalelen$ is again a characteristic scale length. 
We now have: 
\begin{equation}
\frac{{\rm d}\tilde{x}}{{\rm d}\tau} = 
\frac{2\pi}{\diskfreq}\,\frac{1}{v_{\phi}}\,\frac{G\,M_{\rm bar}}{\scalelen^{2}}\,I_{1}
\equiv 2\pi\,f_{\rm bar}\, I_{2}(\barangle,\ \tilde{x}) \, .
\end{equation}

Because the merger occurs on a couple of dynamical timescales, 
i.e.\ a time $\Delta\tau\sim1$, 
to lowest order (ignoring e.g.\ the complications of different orbital 
parameters) we expect that gas 
within a radius 
\begin{equation}
\tilde{x}\ll \Delta\tau\,\frac{{\rm d}\tilde{x}}{{\rm d}\tau}
\end{equation}
will efficiently lose angular momentum and fall to the center, becoming 
part of the central starburst, while gas at 
$\tilde{x}\gg \Delta\tau\,\frac{{\rm d}\tilde{x}}{{\rm d}\tau}$ will avoid 
the starburst. 
This defines a 
scale
\begin{equation}
\frac{R_{\rm gas}}{\scalelen} \lesssim (1-f_{\rm gas})\, f_{\rm disk} \,
\Psi_{\rm bar}(\barangle,\tilde{H},...)
\label{eqn:rtemp1}
\end{equation}
within which the gas will lose angular momentum. 
For convenience, we have collected all of the dimensionless 
integral factors, including $\Psi^{\prime}_{\rm bar}$ (the efficiency of forming 
the stellar bar) and the dynamical integral factors (e.g.\ $I_{1}$, $I_{2}$) from above, 
into the term $\Psi_{\rm bar}$ that represents the full solution. 
We write $\Psi_{\rm bar}(\barangle,\tilde{H}...)$ because, as we will show, 
this quantity (at present) encapsulates our ignorance of e.g.\ the orbital 
parameters and merger mass ratio; for a $1:1$ merger on a typical orbit, 
however, $\Psi_{\rm bar}\sim1$. 

The total gas mass 
which will lose angular momentum and 
fall into the center of the galaxy will be 
$f_{\rm gas}\times f(<R_{\rm gas})$, where $f(<R_{\rm gas})$ is the 
mass fraction within the characteristic radius above, according to the 
details of the mass profiles and dimensionless integral above. 
We consider solutions for a variety of 
profiles, including e.g.\ an exponential, isothermal sphere, and a 
\citet{mestel:disk.profile} $1/R$ disk profile. In general, we find that there is little difference
between the predictions for these various profiles -- the differences in the 
mass profile shapes tend to cancel out and leave only weak corrections to the 
simple dimensional scaling. 
An exponential disk 
with $\Sigma\propto \exp(-R/\scalelen)$ contains a 
mass fraction 
\begin{equation}
1-(1+R/\scalelen)\,\exp{(-R/\scalelen)}, 
\label{eqn:fburst.1}
\end{equation}
within a radius R; here $\Psi_{\rm bar}$ must be solved numerically. 
We obtain nearly identical predictions, however, 
assuming a $1/R$ disk or an isothermal sphere profile for the gas, 
which allows us to analytically solve the relevant equations and 
write the predicted gas fraction consumed in the form: 
\begin{equation}
f_{\rm burst} = f_{\rm gas}\,(1-f_{\rm gas})\, f_{\rm disk}\, \Psi_{\rm bar}(\barangle,\tilde{H},...) \, ,
\label{eqn:fburst.2}
\end{equation}
where $\Psi_{\rm bar}\sim 1$ can be analytically calculated for 
these profiles (with the equations above) under certain conditions: 
if the dependence on the orbital parameters is separable 
and we define $\Psi_{\rm bar}$ 
by the requirement that the radius in Equation~(\ref{eqn:rtemp1}) satisfy 
$\tilde{x}=\Delta\tau\,\frac{{\rm d}\tilde{x}}{{\rm d}\tau}$, then for 
instantaneous bar lag of $\barangle\sim$a few degrees in a thick disk of 
height $\tilde{H}\sim0.2$, $\Psi_{\rm bar}$ is given by
\begin{equation}
\Psi_{\rm bar} \approx F(...)\,{\Bigl\{}1 - 
\exp{{\Bigl[}-\frac{\sin(2\,\barangle)}
{\sin^{2}(\barangle)+\tilde{H}^{2}}{\Bigr]}}{\Bigr\}} \sim 1 \, ,
\label{eqn:fburst.3}
\end{equation}
where we explicitly show $F(...)$ 
as we have suppressed our ignorance of the orbital parameters. 
Nevertheless, this simple scaling alone provides a remarkably successful 
description of many of our simulations.

\begin{figure}
    \centering
    \scaleup
    \plotter{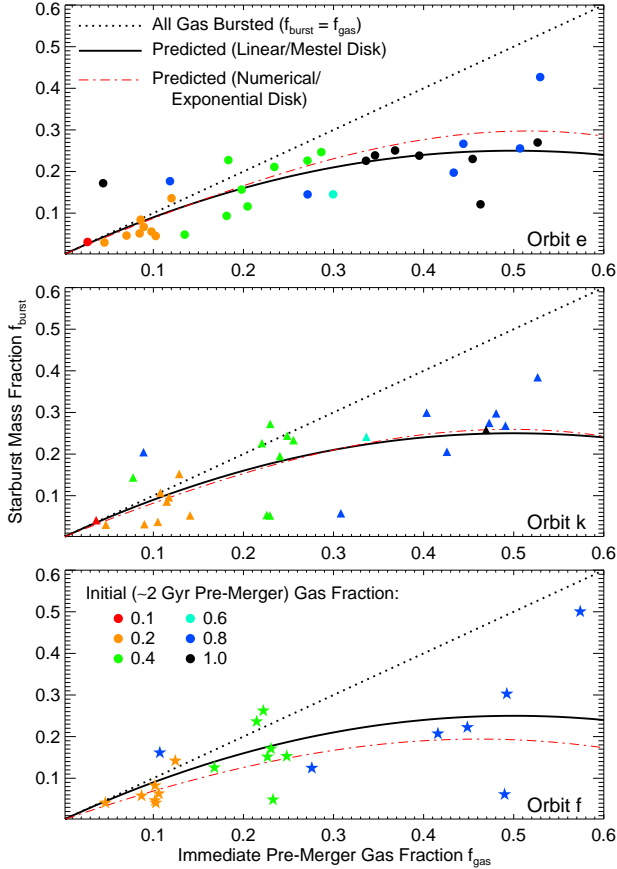}
    \caption{Mass fraction formed in the central, dissipational starburst 
    as a function of gas mass fraction at the time just before the starburst, 
    in a suite of major $1:1$ mass-ratio mergers. 
    Solid lines are our theoretical predictions ($f_{\rm burst}=f_{\rm gas}(1-f_{\rm gas})\,\Psi$, 
    see Equation~\ref{eqn:fburst.2}), 
    dotted line corresponds to bursting all the available gas 
    ($f_{\rm burst}=f_{\rm gas}$). 
    We show results here for several orbits from Table~\ref{tbl:orbits}: a typical random orbit 
    with both disks inclined ({\bf e:} {\em top}), an inclined polar-prograde orbit ({\bf k:} {\em middle}), 
    and a polar-polar orbit ({\bf f:} {\em bottom}). 
    The simulations agree well with our analytic predictions: more gas-rich mergers 
    are less efficient at torquing angular momentum away from the gas 
    and funneling it into the starburst (efficiency $\sim(1-f_{\rm gas})$). 
    \label{fig:fgas.fsb}}
\end{figure}

Figure~\ref{fig:fgas.fsb} tests this simple prediction. For a suite of merger simulations, 
we compare the mass fraction in the central starburst, $f_{\rm burst}$, 
to the gas content of the 
(immediately pre-merger) disks, $f_{\rm gas}$.
We can either determine the starburst mass fraction 
by directly measuring the gas mass that loses its angular momentum and participates 
in the brief nuclear starburst, or by measuring the gas content that survives and 
forms a disk (described in \S~\ref{sec:form.major}) and assuming the gas that did not survive 
(relative to that available just before the final merger) was part of the burst. 
In either case, we obtain a nearly identical answer for each simulation. For now, 
we consider only simulations with a $1:1$ mass ratio -- we will generalize to arbitrary 
mass ratios below. In all these simulations, the 
pre-merger $f_{\rm disk}\approx1$, and measuring 
$\barangle$ and $\tilde{H}$ just before the merger we 
expect $\Psi_{\rm bar}\approx1$. We consider one set of orbits at a time -- i.e.\ compare 
only systems with the same orbital parameters, so that we can temporarily suppress 
the dependence on them (this yields a systematic offset between each set of 
orbital parameters -- the solutions plotted account for that following our solution in the 
next section). At a fixed orbit, for these mergers, then, the only parameter 
that matters should be $f_{\rm gas}$. The simulations at each orbit span a wide 
range in $f_{\rm gas}$, from $\sim0.01-1$. 

We compare the relation between $f_{\rm burst}$ and $f_{\rm gas}$ resulting 
from the full numerical experiments to the simple scalings predicted by 
Equations~(\ref{eqn:rtemp1})-(\ref{eqn:fburst.2}). In detail, we show two solutions -- 
first, the scaling given by Equation~(\ref{eqn:fburst.2}), 
$f_{\rm burst}\propto f_{\rm gas}(1-f_{\rm gas})$, appropriate for an isothermal 
sphere or \citet{mestel:disk.profile} disk profile with 
$\Phi_{\rm bar}\approx1$; and second, the appropriate numerical solution 
(following Equations~\ref{eqn:rtemp1}-\ref{eqn:fburst.1}) for an 
exponential disk. In either case the analytic solutions are similar, and agree 
well with the trend seen in the simulations. 
It is clear that the efficiency of the burst in simulations is -- as we predict -- 
{\em not} constant. It is not the case that the entire gas supply is always stripped 
of angular momentum and consumed in the final merger (which would yield 
$f_{\rm bust}=f_{\rm gas}$). Rather, when $f_{\rm gas}$ is sufficiently high, 
only a fraction $\sim(1-f_{\rm gas})$ of the available gas is able to efficiently lose 
its angular momentum and participate in the starburst.

\begin{figure}
    \centering
    \scaleup
    \plotone{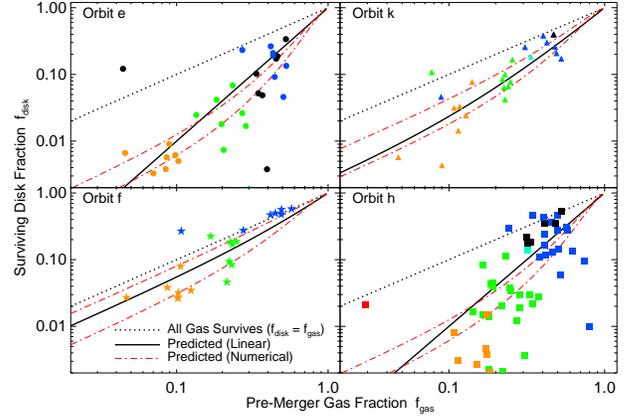}
    \caption{Relaxed post-merger remnant disk mass fraction versus gas fraction 
    just before the merger, for $1:1$ major mass-ratio mergers. In this case essentially 
    all the pre-merger stellar mass is transformed (violently relaxed) into bulge -- the 
    disk is formed from the gas that survives the merger. Panels consider different orbits, 
    with points as Figure~\ref{fig:fgas.fsb}. Solid lines are our theoretical predictions 
    ($f_{\rm disk}=f_{\rm gas}[1-(1-f_{\rm gas})\,\Psi]$, see Equation~\ref{eqn:fburst.2}), 
    dotted lines correspond to all the gas surviving and forming a disk 
    ($f_{\rm disk}=f_{\rm gas}$). Again, the simulations agree well with our 
    analytic predictions; gas-rich mergers are inefficient at stripping angular momentum 
    from the gas, leaving significant gas content that rapidly re-forms a post-merger disk. 
    \label{fig:fgas.fdisk}}
\end{figure}

Figure~\ref{fig:fgas.fdisk} repeats this comparison in terms of the surviving disk mass. 
We argued that the gas that does not lose angular momentum in the merger will 
survive to re-form a disk. Because these are $1:1$ mergers where we can safely assume 
the entire stellar disks are destroyed, we expect then that 
the disk mass fraction will be $f_{\rm disk} = f_{\rm gas} - f_{\rm burst}$. 
Using the method described in \S~\ref{sec:form.major} to estimate the 
remnant disk mass fractions, we plot $f_{\rm disk}$ versus $f_{\rm gas}$ for 
each of several orbital parameter sets. Again, the exact details of the predictions 
depend on orbital parameters in a manner we derive below, but for 
now we are interested in whether or not they obey 
the predicted scaling with $f_{\rm gas}$. 
Indeed, they do. Over $2-3$ orders of magnitude in fractional disk mass 
(and $\sim5-6$ in absolute disk mass), the simple scaling here agrees well with 
full numerical experiments. It is clear that some of $f_{\rm gas}$ is consumed, as 
expected (if all the gas survived, we would obtain $f_{\rm disk} = f_{\rm gas}$; but 
in fact, especially at low $f_{\rm gas}$, the efficiency of angular momentum loss 
is high as predicted and the gas participates in the starburst).

\begin{figure}
    \centering
    \plotone{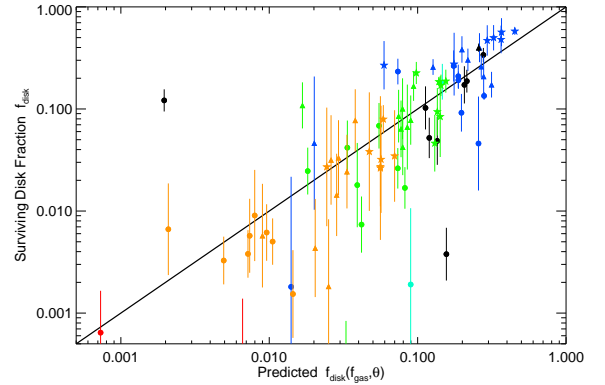}
    \caption{Relaxed post-merger remnant disk mass fraction versus our 
    analytic predictions as a function of gas fraction and orbital parameters, 
    for $1:1$ mass-ratio mergers. Error bars correspond to variation using 
    different methods to estimate the disk-bulge decomposition. 
    \label{fig:fgas.fdisk.exp}}
\end{figure}

Figure~\ref{fig:fgas.fdisk.exp} simplifies this -- we again 
compare $f_{\rm disk}$ and $f_{\rm gas}$, but effectively 
put all the orbits on the same footing
by plotting $f_{\rm disk}$ versus the predicted $f_{\rm disk}(f_{\rm gas},...)$ (i.e.\ including 
the orbital parameters according to the predictions in \S~\ref{sec:model.orbit} below). 
Essentially this 
amounts to implicitly including $F(...)$ in Equation~(\ref{eqn:fburst.2}) above. The 
remaining scaling should just represent the predicted 
$f_{\rm burst}\propto f_{\rm gas}(1-f_{\rm gas})$. For each simulation, we show 
an error bar corresponding to the range of $f_{\rm disk}$ estimated using 
different methods (e.g.\ a full three-dimensional kinematic decomposition, 
one and two-dimensional kinematic modeling, and surface brightness profile 
fits, as described in \S~\ref{sec:form.major}). The agreement is surprisingly good, 
given the simplicity of our derivation. Moreover, the scatter is quite small -- a 
factor $\sim2-3$ at very low $f_{\rm gas}$ and considerably smaller ($\lesssim50\%$) 
at high $f_{\rm gas}$. It seems that our simple scaling indeed 
captures the most important physics of angular momentum loss -- namely that 
with less fractional stellar material in the disk, there is less mass 
available to torque on the 
gas bar in a merger, therefore less angular momentum loss in the gas.

\subsubsection{Dependence on Orbital Parameters}
\label{sec:model.orbit}

We now turn to how the details of the orbit affect the loss of 
angular momentum. Before, we made the simplifying assumptions 
that the stellar bar lagged by some constant angle $\barangle$ 
and that the characteristic time for the perturbation to act 
was of order the disk rotational period/dynamical time. While 
these turn out to give reasonable scalings, we can improve upon 
them. 

The secondary (``perturber'') galaxy will have some characteristic 
orbital frequency $\orbitfreq$, 
approximately given by 
\begin{equation}
\orbitfreq \sim \frac{v_{\rm peri}}{b} \, ,
\end{equation}
where $v_{\rm peri}$ is the velocity at pericentric passage and 
$b$ is the impact parameter or pericentric passage distance. 
Because the behavior we are interested in is relatively short-lived, this 
is reasonable even for first passages or ``flyby'' encounters -- 
we are interested in the orbital frequency at pericentric passage because this 
is when the forcing is strongest and the bar is driven. 

Now, consider the frame rotating with the 
disk/bar at a frequency $\sim\diskfreq$.  In this frame, 
the secondary galaxy will have an 
apparent frequency for an orbit projected into the disk plane of 
\begin{equation}
\Omega_{\rm eff} = \orbitfreq\,\cos{\theta}-\diskfreq, 
\end{equation}
where $\theta$ is the inclination of the orbit relative to the plane of the 
disk (in standard parlance for orbital parameters as described in \S~\ref{sec:sims}). 
Note that we are interested in the 
component of the orbital motion in the plane of the disk: in terms of the 
standard orbital parameters $\theta$ and $\phi$ of the primary 
galaxy (angle of the angular momentum vector of the disk relative to the 
plane of the orbit), this is $\orbitfreq\,\cos{\theta}$. 
For a case with e.g.\ $\theta=0$ (prograde) and 
a parabolic orbit with small impact parameter ($\orbitfreq\sim\diskfreq$), 
the system is maximally prograde -- in the frame of the rotating disk 
there is almost no net circular motion of the secondary. 
For the same orbit but $\theta=180\degree$ (retrograde), 
the secondary completes a circular orbit around the disk 
in just half a disk dynamical time ($\Omega_{\rm eff}\sim -2\,\diskfreq$). 
The time required for the secondary to complete a revolution in 
this frame is therefore (in our dimensionless units $\tau = t / (2\pi/\diskfreq)$)
\begin{equation}
\tau_{\rm circ} = \frac{1} {1-\frac{\orbitfreq}{\diskfreq}\,\cos{\theta}}. 
\end{equation}

The timescale for a gas element to lose its angular momentum and 
fall to the center of the galaxy is given by our earlier 
estimate of the torque, as $\tau_{\rm loss}\sim \tilde{x} / ({\rm d}\tilde{x}/{\rm d}\tau)$. 
If $\tau_{\rm loss}\ll \tau_{\rm circ}$ at a given radius, then 
the derivation we have obtained is essentially valid: the system 
sees a quasi-static perturbation to the potential, loses its angular 
momentum, and collapses before the perturbation can damp out 
or circularize. However, if $\tau_{\rm loss}\gg \tau_{\rm circ}$, 
then the system has not lost much angular momentum by the time 
the secondary completes a revolution, and will gain 
some of those losses back as the system comes around the other 
side. In the limit where 
$\tau_{\rm circ}$ is short (much shorter than the local 
dynamical time), for example, then the potential is effectively circularized -- 
the gas at these radii may undergo oscillatory motion and even have e.g.\ 
spiral waves driven by this external forcing, but there is no means by 
which the system can introduce a strong net asymmetry to drive inflows. 

We can therefore improve our previous estimate: instead of taking 
$\Delta\tau\sim1$ (i.e.\ a disk rotation period) as 
the only characteristic timescale, we 
argue that gas with 
\begin{equation}
\frac{\tilde{x}}{{\rm d}\tilde{x}/{\rm d}\tau} \lesssim 
\tau_{\rm circ} = \frac{1}{1-\frac{\orbitfreq}{\diskfreq}\,\cos{\theta}}
\label{eqn:circ.x}
\end{equation} 
will lose its angular momentum, while gas at larger radii will not. 

In detail, we can integrate the equations from 
\ref{sec:model.gas} for a parcel of gas at some initial radius 
$x_{0}$, in a time-dependent potential of this nature. For simplicity, we assume that 
the secondary drives a circular perturbation in the potential with frequency 
$\omega=\orbitfreq$ and calculate the bar response using the 
gaseous disk (assuming, again, that it 
is infinitely cold) and stellar disk (assuming that the scale height $\tilde{H}$ 
translates into a corresponding velocity dispersion $\sigma/v_{c}$) 
wave dispersion relations from 
\citet{binneytremaine}. In practice, we find this is not much different 
from assuming that the lag in the stellar bar grows with 
time $\propto \tau_{\rm circ}$ (i.e.\ that the stellar bar can keep up or reverse 
sense tracking the perturbation without significant energy loss; 
or, more or less equivalently, that the two bars 
are only in phase when the perturbation is strong, and then rapidly fall out of 
phase -- at the $\gtrsim5-10\degree$ level, once the perturbation is weak or 
reverses its sense as the phase of the secondary reaches $\gtrsim \pi/2$). 
In principle, we now have a physically motived and fully time-dependent 
model for $\barangle(t)$ and the response of the gas bar. This allows us 
to properly integrate out the dependence of $\Psi_{\rm bar}$ on $\barangle$ and 
instantaneous conditions and replace it with the appropriate integral dependence 
on orbital parameters and disk gas content and structure. 

We find that there is a strong division in expected behavior, at more or less 
exactly the characteristic radius implied by Equation~(\ref{eqn:circ.x}). Within 
this radius, gas (in our simple numerical calculations) is effectively torqued 
efficiently as it enters the gas bar near resonance (but slightly leading 
the stellar bar), and plunges to the center. Gas outside this radius begins to 
feel a perturbation, but then the phase of the secondary cycles around and the sense of the 
torques begin to weaken or reverse (depending on the details of the orbit), 
and generate wave motion in the gas but no significant angular momentum loss 
or infall. Not only is the transition between these two regimes predicted by the simple 
scalings above, but we find in more detailed numerical calculations that the 
width of the transition region (where behavior is more sensitive to the details of 
e.g.\ the profile shapes and assumptions about the bars) is quite narrow, 
$\sim20\%$ of that radius. 

This should not be surprising. Essentially, what we have derived is a rough 
equivalent of the co-rotation condition, but for forced bars as opposed to 
isolated self-generating (swing-amplified) bar instabilities. It is well-known 
from studies of idealized bars \citep[see e.g.][]{schwarz:disk-bar,
pfenniger:bar.dynamics,noguchi:merger.induced.bars.gas.forcing,binneytremaine,
berentzen:gas.bar.interaction}
that gas can be efficiently torqued inwards 
inside of the co-rotation resonance (in the language above, given the forcing 
with pattern speed $\orbitfreq$, this is interior to the radius where the relative motion of 
the secondary is slow relative to the dynamical time, and so the perturbation 
does not circularize). Moreover, the resonant structure around these radii 
is known to be sharp; if we 
follow a derivation similar to \citet{borderies:planetary.rings} 
(their derivation is intended to apply to planetary disks with satellites, but the 
relevant physics is similar) it is straightforward to show that the 
detailed numerical prefactors will be swamped for all but a narrow range of 
radii around this resonance by the strong dependence of the resonant forcing 
on radius (roughly going as some large power 
of $(r/r_{\rm crit})$ -- such that the forcing is strong inside the resonance 
and rapidly weakens outside). 

This gives us confidence that we can adopt the scalings above and 
robustly assume that there is indeed a characteristic radius (depending 
in detail upon orbital parameters) interior to which the gas will lose its angular 
momentum. 
This resonant structure of the angular momentum loss is actually 
quite convenient from an analytical perspective, as it means that more subtle 
issues of e.g.\ the thermal pressure and state of the ISM, stellar and AGN 
feedback, and the exact mix of e.g.\ gas and stars or density structure of the 
gas will not contribute significantly to determining which gas can or cannot 
lose angular momentum. Unlike e.g.\ a self-generating bar in an unstable disk, 
there is no issue of stability analysis -- the torques inside this critical radius (and 
the inducing perturbation) are sufficiently strong such that all the material 
therein loses angular momentum in a very short time (much less than a single 
orbital time, in practice). 

For example, it is well known that in isolated cases, a 
pure gas disk is more unstable to gravitational perturbation than a stellar disk 
\citep[see e.g.][]{christodoulou:bar.crit.1,christodoulou:bar.crit.2,
mayer:lsb.disk.bars}, however in the driven case this is not applicable: the 
distortion in the local stellar/gas distribution is caused by the secondary, not by 
e.g.\ orbital ``pileup'' or instability in the primary. The location of the resonant 
radius is not determined by the internal structure of the primary (unlike in an 
isolated case, where it is determined by how e.g.\ those orbits can overlap and 
where various stability criteria are satisfied), but rather by the orbital motion 
of the secondary (relative to the internal motion of the primary), and therefore knows 
nothing about e.g.\ the gas to stars ratio, phase structure, and feedback 
situation in the primary. Inside this radius, the distortion is sufficiently strong that it 
does not matter whether one configuration or another is more or less prone to 
gravitational instability -- the driving force (and therefore angular momentum loss) 
is large in any case. 

Exactly what the pressure support of the gas inside 
this radius is may effect e.g.\ how far it free-falls after losing angular momentum 
before shocking and forming a central starburst, but it will not change the fact 
that the angular momentum loss is efficient. Quantitatively, the torque is 
$\gg j_{\rm disk}\,\Omega_{d}$ (as it must be in order for the gas to lose its 
angular momentum in much less than an orbital period); but e.g.\ the pressure 
gradients resisting gas collapse cannot be larger than (in energetic terms)
$\tilde{H}\,M_{d}\,V_{d}^{2}\sim\tilde{H}\,j_{\rm disk}\,\Omega_{d}\ll j_{\rm disk}\,\Omega_{d}$ 
(or else the disk could not be thin) -- therefore whether or not there is even considerable 
pressure support or e.g.\ thermal feedback or a modified ISM equation of state makes a 
negligible correction to the behavior seen in the simulations. 

Before moving on, we would like to translate the general 
scaling above in terms of $\orbitfreq$ and $\diskfreq$ into 
more convenient parameters. As noted above, $\orbitfreq\sim v_{\rm peri}/b$. 
We expect $b\sim1-3\,\scalelen$ for common parabolic cosmological orbits -- 
as we discuss below, for orbits with larger $b$ that will eventually merge, all 
that matters in terms of the end product is the impact parameter of 
the final passage or two when the most dramatic forcing occurs, so even for 
initially larger passages, angular momentum transfer to the halo will ensure 
a value in this range towards the final stages of the merger. 

Assuming a parabolic orbit, $v_{\rm peri}$ will be given by 
the infall velocity from infinity, $\sqrt{(G\,M\,(1+\mu)/b)}$ (where 
$\mu$ is the merger mass ratio, discussed below). 
Because the merging systems are extended, as $b\rightarrow0$ 
these expressions should be replaced by a more complicated 
function of $b/\scalelen$ (for the case $b=0$, the infall velocity asymptotes 
to the escape velocity from the center of the primary $\sim\sqrt{G\,M/\scalelen}$), 
which requires a numerical solution for an arbitrary density profile.  
In practice we find that we can interpolate between the limits $b=0$ and 
$b\gg \scalelen$ quite accurately by replacing $b$ 
with $\sqrt{b^{2}+\scalelen^{2}}$ (which also happens to be an exact solution 
for e.g.\ a Plummer sphere density profile). 
Combining these 
factors, we find that (for the regime of typical interest) 
\begin{eqnarray}
\nonumber \frac{\orbitfreq}{\diskfreq} &\sim& \frac{v_{\rm peri}}{v_{c}}\,
\frac{\scalelen}{b}\,\cos{(\theta)} \\
\nonumber &=& \sqrt{2\,(1+\mu)}\,[1+(b/\scalelen)^{2}]^{-3/4}\,\cos{(\theta)}\\
&\approx& 0.6\,\cos{(\theta)} \, ,
\end{eqnarray}
where the last term comes from inserting a typical major merger 
mass ratio and $b\sim2\,\scalelen$. 
The orbital dependence is then -- as we would expect -- largely a function of 
the inclination angle $\theta$. Prograde orbits induce a strong bar 
response -- despite the fact that in these mergers the 
orbital angular momenta are all aligned, we actually expect the most 
angular momentum loss and least efficient disk formation. Retrograde 
and polar mergers, on the other hand, despite having completely 
un-aligned or cancelling total angular momentum, should most efficiently 
form disks. 

Inserting this dependence on orbital parameters into our previous 
derivation in Equation~(\ref{eqn:rtemp1}) allows us to effectively 
replace the part of  $\Psi_{\rm bar}$ which parameterized our ignorance of 
orbital parameters ($F(...)$ in Equation~\ref{eqn:fburst.2}), giving
\begin{equation}
\Psi_{\rm bar}(\theta,\tilde{H},...) \propto \frac{1}{1-\frac{\orbitfreq}{\diskfreq}\,\cos{\theta}}. 
\label{eqn:orbit.dept.1}
\end{equation}
In short, our previous derivation applies, but the orbital dependence is now 
explicit in $\Psi_{\rm bar}$. 

Revisiting Figures~\ref{fig:fgas.fsb} \&\ \ref{fig:fgas.fdisk}, recall that we included 
this orbital dependence in the predicted curves therein. For each orbit, the 
predicted curve is given by the solution for the gas mass within the critical 
$R_{\rm gas}/\scalelen$ (Equation~\ref{eqn:rtemp1}) with 
the dependence on orbital parameters as in Equation~\ref{eqn:orbit.dept.1} -- 
we insert the appropriate orbital inclination $\theta$ and impact parameter 
$b$ for the two disks in the orbit and sum their expected $f_{\rm burst}$ or $f_{\rm disk}$ 
to derive the model prediction. 
The difference between the different orbits does not appear dramatic 
in Figure~\ref{fig:fgas.fsb} -- but this is because the burst fraction $f_{\rm burst}$ is plotted 
on a linear scale, suppressing the dependence on $\theta$ at small $f_{\rm gas}$ 
(most of the visible dynamic range in the plot is at large $f_{\rm gas}$ -- in this regime, however, 
the stellar bar is weak in any case because there is not much stellar mass in the disk -- so 
the result is that much of the gas survives and becomes part of the disk, regardless of orbital 
parameters). 

However, the difference between different orbits is 
much more clear in Figure~\ref{fig:fgas.fdisk}, displayed on a logarithmic scale. At 
low $f_{\rm gas}$, there is much more stellar mass in the disk than gas mass, so in principle 
the stellar bar could (if maximal) easily torque away all the angular momentum of the 
gas. Here, however, the orbital parameters become important in determining 
just how efficient this process should actually be. For the orbits close to retrograde 
($\cos{\theta}\approx-1$), 
the scaling we have just derived suggests that $\Psi_{\rm bar}$ should be 
suppressed by a factor $\sim2$. 
But for orbits close to coplanar prograde ($\cos{\theta}\approx1$), 
$\Psi_{\rm bar}$ is enhanced by a factor $\sim2-3$ -- in other words, because the orbit 
is nearly resonant, the effective co-rotation resonance (the orbit interior to which the gas 
can efficiently lose angular momentum to the induced stellar bar) is moved out 
by a substantial factor, including a larger fraction of the disk gas (in those extremes, 
only the gas at very large radii survives the merger). 

Again, Figures~\ref{fig:fgas.fsb}-\ref{fig:fgas.fdisk} demonstrate that 
the simple scalings based on our model provide an accurate description of the 
behavior in the full numerical experiments. Figure~\ref{fig:fgas.fdisk.exp} 
combines these into a single plot -- we compare the disk fractions in our 
simulations to the full expectation based on our derivation thus far as a function of 
both gas fraction and orbital parameters. As noted above, the agreement is 
good, with a reasonably small scatter. 

\begin{figure}
    \centering
    \scaleup
    \plotterr{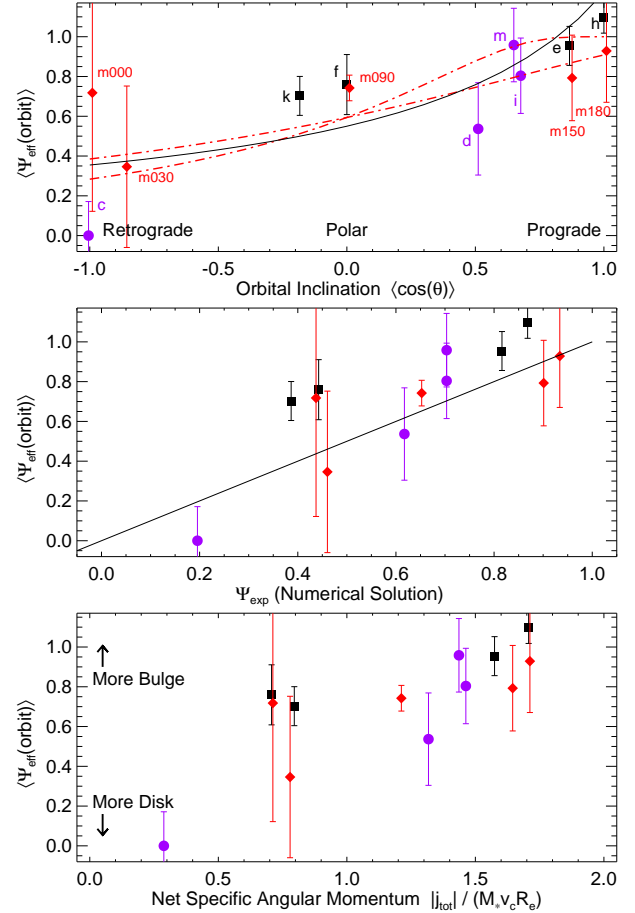}
    \caption{Effective efficiency of bars (the parameter $\Psi$, 
    efficiency at torquing gas into a starburst, removing 
    its angular momentum and destroying the disk), as a function of 
    the effective orbital parameter. Each point represents the effective constraint on 
    $\Psi$ from fitting a correlation of the form in Equation~(\ref{eqn:fburst.2})
    to a suite of simulations over a range of gas fractions, masses, and 
    mass ratios (since we are interested in a comparison of orbital parameters 
    here and not e.g.\ gas fractions, we choose to normalize so that $0<\Psi<1$). 
    Black points are the most well-sampled orbits (e, h, k, f), 
    shown in Figure~\ref{fig:fgas.fdisk}, purple points are 
    more limited studies of orbits (c, i, m, d), red points are 
    a study of major and minor mergers with $\phi=0$, 
    $\theta=0,\ 30,\ 90,\ 150,\ 180\degree$. 
    {\em Top:} $\Psi_{\rm eff}$ versus mean orbital inclination 
    $\langle\cos{(\theta)}\rangle$ (for 1:1 mergers, we average the two 
    inclinations). Black solid line is the simple linear scaling 
    $\Psi_{\rm eff}\propto1/(1 - 0.6\,\langle\cos{\theta}\rangle)$ 
    from Equation~(\ref{eqn:orbit.dept.1}); 
    red dot-dashed lines are the numerical solutions for the appropriate 
    $\langle\cos{\theta}\rangle$, as in Figure~\ref{fig:fgas.fdisk}. 
    {\em Middle:} $\Psi_{\rm eff}$ versus the full numerical 
    expectation (properly solving for the effective bar strength in each disk 
    and then adding the burst fractions, rather than just taking $\Psi(\langle\cos{\theta}\rangle)$. 
    The efficiency of disk destruction 
    and angular momentum loss scales with orbital parameters in the 
    simple manner predicted in Equation~(\ref{eqn:orbit.dept.1}). 
    Over a typical random 
    cosmological ensemble of orbits, we expect values similar to those 
    between our typical {\bf e} and {\bf f} orbits. 
    {\em Bottom:} $\Psi_{\rm eff}$ versus net specific angular momentum 
    of the system (adding/cancelling the initial disk plus 
    orbital angular momenta and dividing by the final baryonic mass). 
    The two are actually {\em anti}-correlated, demonstrating that disks do 
    not arise after or survive mergers owing to co-addition of angular 
    momentum (and cancellation is inefficient at destroying disks) -- rather, 
    systems with aligned angular momentum vectors are in greater resonance, 
    triggering stronger internal asymmetries in the primary that drain more 
    angular momentum from the gas.     
    \label{fig:psi.theta}}
\end{figure}

These results are for four representative orbits spanning a reasonable range in 
orbital parameters -- those for which we 
have a large number of simulations covering a wide range in the
space of other parameters. We consider them first because this allows us 
to robustly determine that the predicted orbital scalings do not depend on e.g.\ stellar 
mass, halo properties, feedback prescriptions, or other varied physics in the 
simulations. Having done so, we consider a more limited sampling of a much 
broader range in orbits given by Table~\ref{tbl:orbits} in order to survey the 
full dynamic range of orbital parameters. 

Figure~\ref{fig:psi.theta} shows the results of this. For a given suite of simulations 
with some particular orbital parameters, we first construct the correlation 
$f_{\rm disk}(f_{\rm gas})$ as in Figure~\ref{fig:fgas.fdisk}. Rather than adopt some 
{\em a priori} model for the orbital dependence, we then fit the points in that 
correlation to a function of the form in Equation~(\ref{eqn:fburst.2}) -- i.e.\ 
effectively fit for the normalization or ``efficiency'' of angular momentum 
removal, which we define as $\langle\Psi_{\rm eff}({\rm orbit})\rangle$. 
We compare this, for our ensemble of orbits, to our analytic expectation 
from the simple scaling in Equation~(\ref{eqn:orbit.dept.1}) and to 
a full numerical solution (technically, for 1:1 mergers, we want to solve this separately 
for each disk and add the two, although just considering the primary is a 
good approximation for less major mergers). 
The agreement with our analytic model is quite good across the entire range 
of orbital parameters, implying that we have captured the most important 
physics of resonant interactions in this simple scaling. 

We also compare this effective efficiency of disk destruction with the 
net specific angular momentum of the merger remnant (assuming 
pure addition/cancellation of the initial baryonic angular momenta of the disks 
and the orbital angular momentum). The result is actually an anti-correlation: 
systems with aligned angular momentum vectors, e.g.\ coplanar prograde mergers 
being the extreme case, induce the most efficient bars and remove angular 
momentum most efficiently from the gas. Systems where the angular momentum 
vectors are misaligned (e.g.\ polar orbits) or anti-aligned (retrograde) 
actually leave the largest disks in place. This clearly emphasizes that it 
is not, in fact, any direct addition/cancellation of angular momentum that 
determines or enables disks to form in and survive mergers. Rather, the cases with the 
largest net angular momentum are most resonant, inducing the strongest resonant 
asymmetries in the merging pair, which most effectively drains angular momentum 
from the gas and leaves a compact, bulge-dominated remnant.

\subsubsection{Dependence on Mass Ratio}
\label{sec:model.massratio}

The major remaining parameter to study is the merger mass ratio. 
Thus far, we have restricted our attention to equal mass 
$1:1$ mergers, which allowed us to make several convenient simplifying assumptions. 
Nevertheless, most of our previous derivation applies. None of the 
scalings that we have explicitly derived up to now are dependent upon 
mass ratio. However, we have quantified the strength of the induced stellar and 
gas bars with the parameter $\Psi_{\rm bar}$, which we expect should 
scale with mass ratio. Moreover, we have made the assumption that the 
pre-merger disk stars are entirely violently relaxed by the merger. While this 
is a good assumption for $1:1$ mergers, it is not true for minor mergers 
(a $1:10$ mass ratio merger, even with no gas, will clearly not transform the entire 
primary stellar disk into bulge). 

First, consider this stellar component: there are a number of ways to derive 
the disturbance of the stellar component in the merger. The simple expectation 
is that the mass in galaxy $M_{1}$ which can be violently relaxed by 
collision with galaxy $M_{2}$ is proportional to $M_{2}/M_{1}$ -- the net energy deposit, 
tidal forces, and the mass fraction brought in from a potentially disrupted satellite 
all scale in this manner. For simplicity, consider the case where the secondary 
$M_{2}$ is much smaller and more dense than the primary, and falls in on a nearly radial orbit 
in the final encounter (which is a good approximation, given the efficiency of 
angular momentum transfer from the orbit to the halo). Since we are assuming 
$M_{2}\ll M_{1}$, treat $M_{2}$ as a point mass, and consider its final orbital 
decay, where it oscillates with rapidly decaying amplitude through the center of the primary 
with initial impact velocity $v_{\rm i}\approx v_{c}$ and damping spatial amplitude 
$\ell_{\rm max}\lesssim R_{d}$. 
At some instant, then, the secondary is at 
location $(R^{\prime},\phi^{\prime},\,z^{\prime})=\ell\,(\cos{\theta},\,0,\,\sin{\theta})$ 
(we rotate such that the secondary orbit defines $\phi=0$ without loss of generality). 
A star in the primary disk at $(R,\phi,z=0)$ then feels some potential from the 
secondary ($\equiv \Phi_{2}$) and experiences
a vertical deflection out of the 
disk $\partial\Phi_{2}/\partial z = (G\,M_{2}/R^{3})\,\ell\,\sin{\theta}\,f(\ell/R)$, 
where $f(u)=[1+u^{2}-2\,u\cos{\theta}\,\cos{\phi}]^{3/2}\sim1$. We are only interested 
in the time the secondary spends at $\ell \sim R$ (when its much closer to the 
disk or further away, the vertical perturbation is weak), so it effectively 
acts for a time $\delta t \sim R/v_{\rm i}$ as it passes through $\ell \sim R$ 
in its ringing about the center. If we know the full potential, we can 
solve for the deflection as a function of time and calculate the full acceleration of 
the disk stars at $(R,\phi)$, which yields an effective 
net velocity deflection while the secondary is on one side of the galaxy 
of $\delta{v} = (G\,M_{2}/R\,v_{\rm i})\,(\ell_{\rm max}\sin{\theta}/R)\,f(\ell_{\rm max}/R)$. 
Deflection occurs when $\delta{v} \sim v$ or larger, so if 
$v=\tilde{v}(r)\,v_{c}$ (where $\tilde{v}$ depends weakly on $r$)
and we substitute for $v_{c}\equiv\sqrt{G\,M_{1}/R_{d}}$ here and in $v_{\rm i}$ we obtain 
the criterion $G\,M_{2}/R \gtrsim v_{c}\,v_{i} \sim v_{c}^{2} \sim G\,M_{1}/R_{d}$, 
i.e.\ (rearranging) $R/R_{d} \lesssim M_{2}/M_{1}\,\sin{\theta} = \mu\,\sin{\theta}$. 
The $\sin{\theta}$ dependence comes because we considered only vertical 
deflection of stars (i.e.\ some heating to $v_{z}^{2}$)-- 
a coplanar orbit (in this limit) will obviously induce no such 
heating, but will introduce deflections in the radial direction (heating 
$v_{R}^{2}$). We can repeat 
our derivation considering where these deflections are significant, and find 
(as one would expect) $R/R_{d} \lesssim \mu\,\cos{\theta}$. So, the absolute 
mass fraction scattered should be more or less angle-dependent, although 
the orbital anisotropy $\beta_{z}\equiv 1 - \bar{v_{z}^{2}}/\bar{v_{R}^{2}}$ 
will depend significantly on  the orbital inclination $\theta$. 

For the case of a thin 
\citet{mestel:disk.profile} disk with no bulge, we can solve these equations exactly and obtain 
the simple solution that 
a merger with a secondary $M_{2}$ scatters exactly $M_{2}$ worth of stars in the primary, 
completely independent of the inclination $\theta$, (but with an 
anisotropy 
$\beta_{z}(\theta) \sim 1 - \frac{2\,\sin^{2}{\theta}}{(1+2\,\cos^{2}{\theta})}$ -- 
although this ignores a proper treatment of further mixing as the perturbed stars 
interact with each other, and thus does not reproduce orbits quite as radial as seen 
in simulations). 
The full numerical solutions for arbitrary cases yield the general result that, 
when inside a radius that encloses a mass $\sim M_{2}$ in the primary, 
then the presence of the mass $M_{2}$ is a significant perturbation, which 
scatters those stars in the primary -- i.e.\ deflections occur rapidly, so the stars 
violently relax. At larger radii, where $M_{\rm enc} \rightarrow M_{1}>M_{2}$, 
the motion of the $M_{2}$ secondary at the center is a small perturbation. The 
disk at these radii is perturbed adiabatically by the motion of the secondary, which 
can induce some warps and/or disk heating, but will not violently relax the stars. 
Reversing this derivation for the secondary, it is trivial that essentially all 
the mass in the secondary (we ignore stripping of the tightly bound stellar mass) 
will be violently relaxed. So the total mass violently relaxed will 
be $\sim M_{2}$ (in the primary) plus $M_{2}$ (the secondary), out of a total 
mass $M_{1}+M_{2}$ -- i.e.\ in terms of the mass ratio $\mu\equiv M_{2}/M_{1}$, the 
fraction of the pre-merger {\em stellar disk} mass which is destroyed and turned into 
bulge is
\begin{equation}
f_{\ast,\rm disk}(\rm destroyed) = \frac{2\,\mu}{1+\mu}. 
\label{eqn:minor.scaling}
\end{equation}
Technically this assumes the systems are initially pure disk, but the corrections 
if they have pre-existing bulges are not large (generally smaller than the 
simulation-to-simulation variation; although we discuss them in 
more detail in \S~\ref{sec:prescriptions}), so this is a reasonable approximation 
for general cases. 

Now, consider the gas. It turns out that a similar 
linear scaling in Equation~(\ref{eqn:minor.scaling})
is found for how the gas mass in the starburst (i.e.\ the fraction 
that loses its angular momentum) scales with mass ratio, as one might expect. 
In detail, though, the derivation must be revisited (and will include additional terms 
depending on orbital parameters): because the gas is collisional, 
even a large vertical deflection of 
gas at some $R$ does not translate to a loss of that gas disk, since the gas can 
dissipate the vertical energy and no loss of rotational angular momentum 
has occurred. Deflections in the $R$ direction will be resisted by hydrodynamic forces. 
So, for a proper derivation, we return to our model of the stellar bar torquing 
the gas bar. The essential question is how the amplitude of the induced stellar bar 
(our term $\Psi_{\rm bar}$) should scale with mass ratio. 

Take the thin disk limit (this is just for convenience, the final scaling is unchanged if we 
allow for a finite stellar disk thickness); the 
disk surface density is linear in the potential according to Poisson's equation, 
\begin{equation}
\nabla^{2}\Phi = 4\pi\,\Sigma(R,\phi)\,\delta{(z)}. 
\end{equation}
So, since the non-axisymmetric 
potential of the secondary, at some distance $b$ (the impact parameter), 
must scale as roughly $\Phi \sim G\,M_{2} / b^{3}$, 
we expect the amplitude of the induced bar (perturbation in $\Sigma$) 
should also scale as $M_{2}/b^{3}$. Fractionally, this yields 
$\Psi_{\rm bar}\propto M_{\rm bar}/M_{1} \sim \mu\,(h/b)^{3}$. 

We can show this more formally: 
if $\Phi_{0}$ is the (azimuthally symmetric) potential of the primary and 
$\Phi_{1}$ is the perturbative potential of the secondary, which induces the 
surface density perturbation $\Sigma_{1}\propto f_{\rm bar}\propto \Psi_{\rm bar}$ 
that defines the bar, we 
have $\nabla^{2}\Phi_{1} = 4\pi\,\Sigma_{1}(R,\phi)\,\delta{(z)}$. 
We can expand any potential $\Phi_{1}$ 
as $\Phi_{a}(k\,R)\,\exp{[\imath\,(m\,\phi-\omega\,t)-k\,|z|]}$, which gives the 
trivial solution 
$\Sigma_{1} = \Sigma_{a}(k\,R)\,\exp{[\imath\,(m\,\phi-\omega\,t)-k\,|z|]}$ 
where $\Sigma_{a} = -|k|/(2\pi\,G)\,\partial^{2}\Phi_{a}/\partial r^{2}$. 
We expect $\Phi_{a}\sim -G\,M_{2}/r$, so 
we obtain $\partial^{2}\Phi_{a}/\partial r^{2} \sim -2\,G\,M_{2}/r^{3}$ (note 
that we can generalize this to extended distributions for the secondary, 
relevant for e.g.\ more major mergers, and the change is, for reasonable 
profiles, equivalent to replacing $r$ with $\sqrt{r^{2}+a^{2}}$). 
The details of the mode structure turn out not to be important, since 
the behavior we are interested in is dominated by modes with $|k|\sim 1/\scalelen$; 
but we can, for example, treat $\Phi_{a}(k\,R)$ as the potential 
generated by a point source (appropriate for e.g.\ the small mass-ratio limit) 
at the impact parameter $b$
and expand the wave modes appropriately, then integrate over the modes 
in the disk to determine the bar strength (i.e.\ the total 
mass effectively contributing to the bar). In any case, up to a numerical 
constant that is weakly sensitive to the mode structure, 
we obtain 
$M_{\rm bar} \propto \frac{\scalelen^{3}}{(b^{2}+\scalelen^{2})^{3/2}}\,M_{2}$. 
In terms of $\Psi_{\rm bar} = M_{\rm bar}/M_{1}$, this gives 
\begin{equation}
\Psi_{\rm bar}\propto \mu\,  (1+[b/\scalelen]^{2})^{-3/2} \, ,
\label{eqn:massratio.b.forcing}
\end{equation}
where again $b$ is the impact parameter and the 
$1+[b/\scalelen]^{2}$ term effectively allows for the interpolation between 
the case of an orbiting point mass and a penetrating encounter 
\citep[see e.g.][]{binneytremaine}. 

As noted in \S~\ref{sec:model.orbit}, 
if we are just interested in the end product of a merger -- i.e.\ we do not 
care what happens on each passage separately as the companions lose 
angular momentum, but only in the surviving disk fraction and total 
burst fraction -- then we are not interested in some initial impact parameter 
$b$ but only in the impact parameter on the final passages close to 
coalescence, when angular momentum loss has made the orbits 
nearly radial and the forcing is strong. We can see directly from 
Equation~(\ref{eqn:massratio.b.forcing}) 
that the forcing is dramatically suppressed by a factor $\sim(b/\scalelen)^{3}$ 
on earlier, large-impact parameter passages, so these can be effectively 
ignored in calculating the remnant properties (we confirm this is 
true in a sample of simulations with much larger $R_{\rm peri}$). 
Eventually, for any systems which are destined to merge, 
angular momentum transfer yields a nearly radial orbit with $b\sim\scalelen$, 
and this is where most of the forcing occurs, so the remnant solution is 
effectively given by ignoring the $b$ dependence above (technically 
summing over each passage with the appropriate $b$ is possible, 
but in practice we obtain the same result to within the
simulation-to-simulation scatter by assuming $b\rightarrow0$ in 
Equation~\ref{eqn:massratio.b.forcing}). The final dependence on any 
initial impact parameter $b$ is therefore weak, so long as the 
systems are bound to merge. The dependence on mass ratio $\mu$, however,
is fixed. 

If the mass enclosed is linear in $\Psi_{\rm bar}$ (the case 
for e.g.\ the \citet{mestel:disk.profile} disk and an isothermal sphere, and not a 
bad approximation for the regime of interest for an exponential 
disk), we then have a similar result for the gas as the stellar distribution: 
the secondary 
induces a burst of mass $\propto M_{2}$ in the primary $M_{1}$. 
Reversing the derivation, the primary (since it is larger, so 
$M_{1}/M_{2}>1$ induces a burst (assuming the two have similar initial 
gas fractions) $\propto M_{2}$ in the secondary (i.e.\ bursting all its gas). 
The net burst mass $\propto 2\,M_{2}$ relative to the remnant 
mass $M_{1}+M_{2}$ is then 
\begin{equation}
f_{\rm bust}\propto \frac{2\,\mu}{1+\mu}. 
\end{equation}
This is generally applicable for mergers; however we will note below 
that, because they do not coalesce (and therefore do not 
eventually come in with $b\rightarrow0$ or brake their orbital energy 
interior to the stellar distribution of the primary), this is not exactly 
applicable to e.g.\ fly-by or first passage scenarios. 

\begin{figure*}
    \centering
    \scaleup
    \plotone{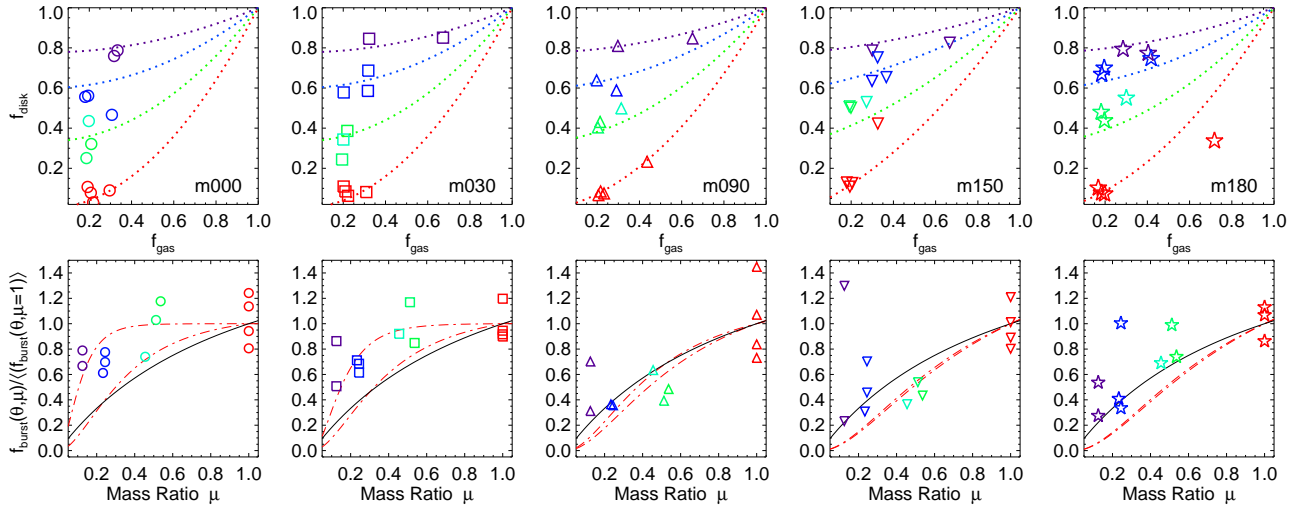}
    \caption{{\em Top:} Surviving disk mass fraction $f_{\rm disk}$ as a function of  
    pre-merger gas fraction $f_{\rm gas}$, for a series of 
    mergers of varying mass ratios (symbols). Each panel shows a series of mergers with 
    different orbital parameters (orbits {\bf m000-m180}). Color encodes 
    mass ratio of the merger: 1:1 (red), 1:2 (green/cyan), 1:4 (blue), 1:8 (purple). 
    Dotted lines (of corresponding color) show our prediction (Equation~\ref{eqn:full.equation}) 
    for the given orbital parameters and mass ratio. Note that for minor 
    mergers, $f_{\rm disk}>f_{\rm gas}$ is allowed, because some of the original 
    stellar disks are predicted to survive the merger as well as some of the gas 
    which does not lose angular momentum.  
    {\em Bottom:} Starburst mass fraction in mergers of a given mass ratio $\mu$, 
    relative to our model prediction for 1:1 mass ratio mergers with 
    the same orbit and pre-merger gas content (symbols, as 
    top panels). Lines show our simple linear model (solid black; this does 
    well for typical orbits but the bursts in nearly prograde orbits -- m000 -- 
    are somewhat more efficient than predicted owing to the effects described in 
    \S~\ref{sec:model.exceptions}) and full numerical calculation (dot-dashed red; 
    two lines correspond to different mass profiles), as in Figure~\ref{fig:fgas.fdisk}. 
    Minor mergers induce less efficient bursts, and do not completely destroy the 
    primary disk: the scaling of these efficiencies with mass ratio agrees well 
    with our dynamical model predictions. 
    \label{fig:massratio.fdisk}}
\end{figure*}

\begin{figure}
    \centering
    \plotter{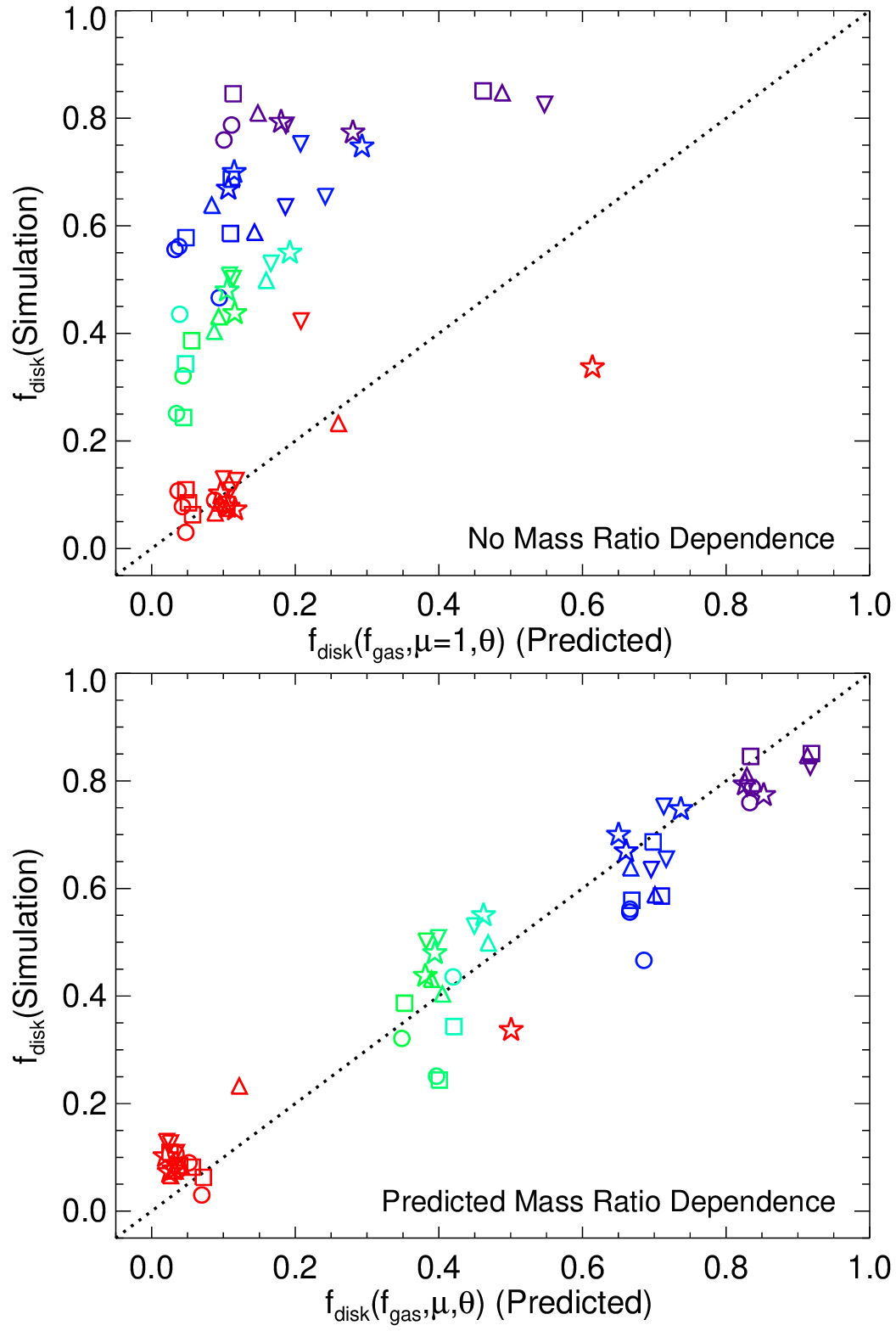}
    \caption{Predicted (as a function of 
    orbital parameters, pre-merger gas content, and 
    mass ratio) and actual post-merger surviving disk fraction for the 
    simulations in Figure~\ref{fig:massratio.fdisk} (symbol type 
    and color denote orbit and mass ratio in the same style). 
    {\em Top:} Comparison assuming there is no dependence on 
    mass ratio (i.e.\ treating all cases as $\mu=1$). Clearly, this 
    is inappropriate for minor mergers, but it is also inappropriate 
    for even intermediate major mergers (mass ratios $\mu=0.3-0.5$). 
    {\em Bottom:} Comparison including the predicted 
    dependence on mass ratio 
    of both destruction of the stellar disk and 
    angular momentum loss in the gas. 
    Our predictions as a function of mass ratio, orbital parameters, 
    and gas fraction are accurate in the simulations to $\sim0.1$ in $f_{\rm disk}$. 
    \label{fig:massratio.pred}}
\end{figure}

Figure~\ref{fig:massratio.pred} tests this prediction in an ensemble of 
simulations spanning a range in mass ratio from 
$\mu=0.1-1$. For a given set of orbital parameters (fixed), we plot 
the disk and burst fractions ($f_{\rm disk}$ and $f_{\rm burst}$) of the 
remnant, as a function of the immediate pre-merger 
gas fraction $f_{\rm gas}$, as in Figures~\ref{fig:fgas.fsb}-\ref{fig:fgas.fdisk}. 
For the $1:1$ mergers, we plot our expectation based on the 
simple scaling in Equations~(\ref{eqn:rtemp1})-(\ref{eqn:orbit.dept.1}), 
including the dependence on $f_{\rm gas}$ and orbital parameters ($\theta$) 
following \S~\ref{sec:model.gas}-\ref{sec:model.orbit}. 
We then show the prediction for mass ratios $1:2$, $1:4$, and $1:8$, according 
to our derivations here. 

This includes two important corrections: 
instead of assuming the entire stellar disk is turned into bulge (which 
was a good approximation for the $1:1$ mergers), we allow the fraction 
of the stellar disk that is destroyed (turned into bulge) to 
depend on mass ratio following Equation~(\ref{eqn:minor.scaling}) -- so 
some (considerable) fraction of the disk is assumed to survive in 
higher mass-ratio mergers. We also include the scaling with mass ratio 
in $\Psi_{\rm bar}$, used as before to calculate how much of the 
gas participates in the starburst. So, in the high mass-ratio cases, 
both the fraction of the gas that loses its angular momentum 
(fraction of $f_{\rm gas}$) and fraction of the pre-merger primary stellar 
disk turned into bulge (fraction of $(1-f_{\rm gas})$) are suppressed 
by a factor $\sim\mu$. 

For each of the orbits surveyed (and the range in e.g.\ absolute masses, 
gas fractions, and feedback prescriptions in our minor merger 
simulations), this simple rescaling according to the merger mass 
ratio provides a good approximation to the behavior in the 
full hydrodynamic experiments. Both the total surviving disk fraction 
(which reflects both the ability of the pre-merger stellar disks 
and the pre-merger gas to survive the merger) and the burst 
fractions (which reflect only how much of the gas survives/loses angular 
momentum) are accurately predicted, suggesting that our derivations 
are reasonable for both the dissipational and dissipationless components 
of the galaxy.

Figure~\ref{fig:massratio.pred} summarizes these results. We first compare 
the final disk fraction in the simulations to our prediction including e.g.\ 
the dependence on gas content and orbital parameters but {\em without} 
any accounting for mass ratio (assuming all mergers are just as efficient 
as a $1:1$ merger). Unsurprisingly, this works for the $1:1$ mergers, 
but is a terrible approximation to mergers of very different mass ratios. 
We then compare allowing for the same scalings but including the 
predicted mass ratio dependence. The agreement between full simulation 
and our simple analytic expectations is good -- with a scatter for 
the high disk fractions typical of intermediate and minor merger remnants 
as low as $\sim20\%$. 

One important caveat here is that, for mergers of increasingly
small mass ratio $\mu$,
the merger timescales become long. At the smallest mass ratios we 
consider, $\sim$1:10, this timescale may become sufficiently long that 
the secular (i.e.\ self-amplifying) instability/response of the disk 
may become important over the duration of the merger. It is not entirely 
clear what the response of such an (initially driven) system will be; whether or 
not, for example, the driven non-axisymmetric modes will remain locked to 
their driver (the secondary orbit) or de-couple and move at the 
pattern speed dictated by the internal stability properties of the disk. 
This competition between secular processes (more sensitive to e.g.\ the 
detailed structure, rotation, and pressure support of the disk) and 
merger-driving in this regime probably contributes to 
some of the increased scatter in burst fractions seen in Figure~\ref{fig:massratio.fdisk} 
at the lowest $\mu$. For this reason, it is reasonable to restrict a 
definition of ``mergers'' to this mass ratio and more major interactions: at smaller mass ratios, 
secular/internal processes (even if initially driven by interactions) may 
be more important than the direct driving from the interactions themselves 
(or at least operate on comparable timescales).

\subsubsection{First Passage and Fly-By Encounters}
\label{sec:model.flyby}

We have derived a general equation for the disk mass that should be lost in 
mergers, and demonstrated that it is robust to variations in a wide range of 
galaxy properties. Most of our derivation is completely generalizable as 
well to encounters where the systems will not merge (or at least are not 
immediately merging). Two cases of interest (which are, in the short term, essentially 
equivalent) arise: first passages and ``fly-by'' encounters (in which 
there is a close encounter but the velocities are sufficiently large to 
delay or prevent a merger). 

In such a passage, there is of course no violent relaxation and mixing of stars, 
so we assume the stellar disk is left intact (excepting the bar response). 
The same physics will govern bar formation and 
loss of gas angular momentum. The primary difference is the suppression 
by the appropriate impact parameter $b$ in 
Equation~(\ref{eqn:massratio.b.forcing}). 
We argued before that the term 
$\propto [1+(b/\scalelen)^{2}]^{3/2}$ should ultimately be neglected for 
mergers because in the final passage(s) that dominate, the condition of merging 
more or less guarantees $b\rightarrow0$. However, clearly this is not the case 
on a non-merging passage. 

This introduces a non-trivial uncertainty -- 
we quote $[1+(b/\scalelen)^{2}]^{3/2}$ where $b$ is the impact parameter 
and $\scalelen$ is some characteristic scale length of the system. But in detail, 
the appropriate ``impact parameter'' is really the {\em actual} distance of closest 
approach, which is usually somewhat smaller than the distance of approach 
estimated from infinity (the formal impact parameter definition), as some angular 
momentum is already lost. Moreover, in detail, is 
the appropriate $\scalelen$ the exponential scale length? The half-mass radius? 
Any such radii are of course closely related, and all of these uncertainties in the 
exact definition change the term $b/\scalelen$ only at the factor $\sim2$ level, 
but since the dependence $\sim(b/\scalelen)^{3}$ is fairly strong, this is important 
on a quantitative level for these fly-by situations. 

\begin{figure*}
    \centering
    \scaledown
    \plotterr{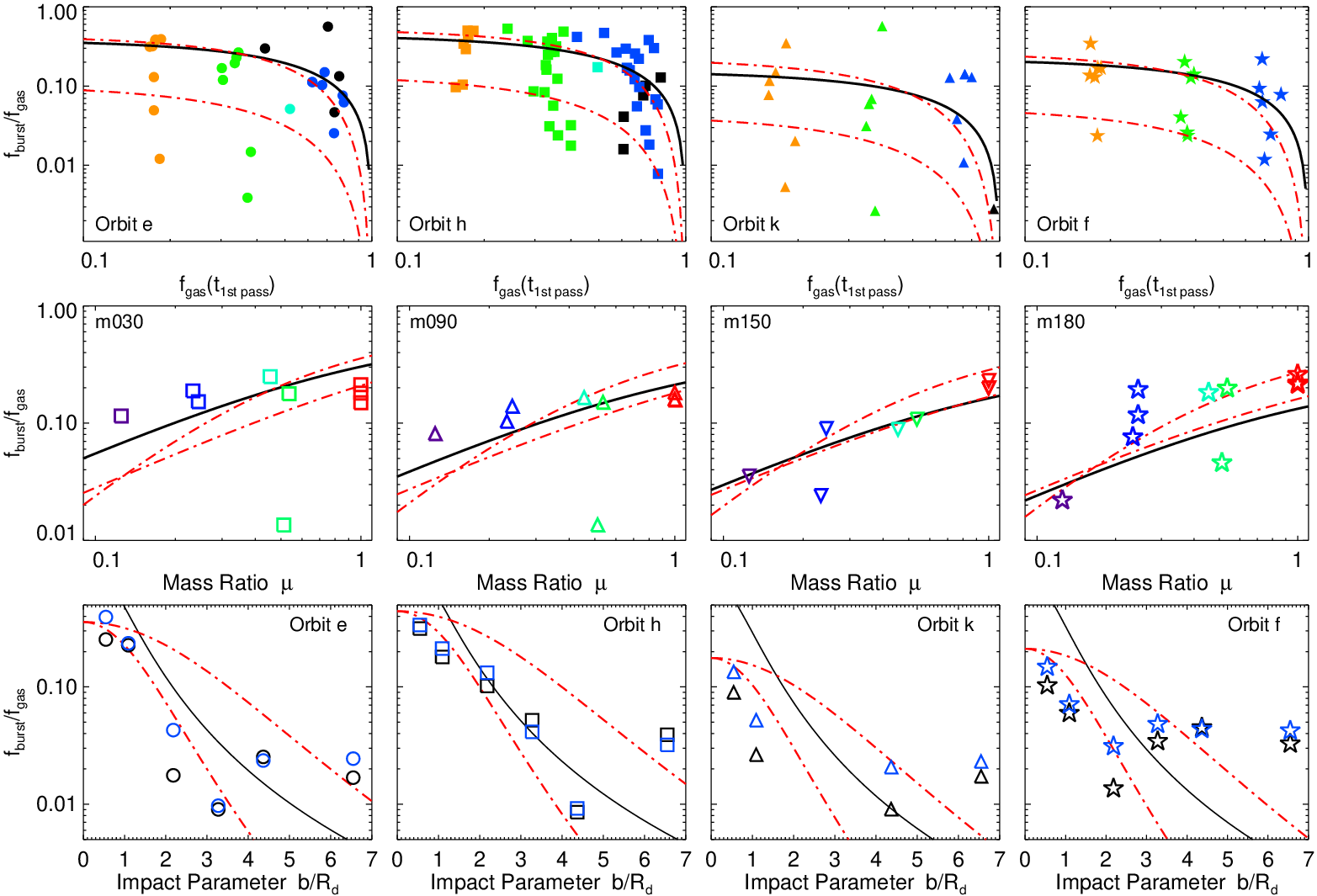}
    \caption{Comparison of the burst fractions in single fly-by (or equivalently, first-passage) 
    scenarios and the general application of our model scalings. In these cases, there is 
    no significant violent relaxation (no stars merge), so the stellar disk is left completely 
    intact. Some pseudobulge may result from the induced bar and disk heating, but 
    we are not modeling that here. Some burst still results from the same induced 
    non-axisymmetry in the primary, which should be described by 
    our same scaling (Equation~\ref{eqn:full.equation}). The important differences from 
    a case that will merge are: {\bf (1)} the suppression of the induced burst 
    by a factor $\sim [1+(b/\scalelen)^{2}]^{3/2}$ (where $b$ is the impact parameter and 
    we find decent agreement with our simulations when $\scalelen$ is the half-mass 
    disk radius), whereas in cases that will merge $b\rightarrow0$ is appropriate, 
    {\bf (2)} the lack of violent relaxation of the stellar disk, and {\bf (3)} an expected 
    increased scatter, as the details of the approach are more important (and there is no 
    merger/in-spiral, which tends to average out the exact details of the approach). 
    {\em Top:} Burst fraction (relative to gas supply at the time of the passage) versus 
    gas fraction. Our simple linear model prediction (black solid) and 
    numerical predictions (red dot-dashed) are shown, with the results from 
    the first passages and fly-by encounters of the simulations in Figure~\ref{fig:fgas.fdisk}, 
    appropriate for each set of orbital parameters shown. These cases had 
    $b/\scalelen\approx1$ (and that was used in the predictions -- the 
    curves assuming $b=0$, as we used for the post-merger systems, would be a 
    factor $\sim3$ higher, in conflict with the simulations). 
    {\em Middle:} Same, but as a function of mass ratio for systems in 
    Figure~\ref{fig:massratio.fdisk} with otherwise equal orbital parameters and 
    gas fractions at the time of passage. 
    {\em Bottom:} Same, but as a function of impact parameter for 1:1 mergers 
    with $\fgas\approx0.2$ (black) and $0.4$ (blue). Note that ``burst'' fractions 
    $\lesssim1\%$ of $\fgas$ are essentially equivalent to zero (equivalent to 
    random fluctuations in isolated disks). 
    Our predictions describe first passages and 
    fly-by encounters reasonably well, although there is larger scatter about them 
    owing to differences in the details of how the passage proceeds. 
    \label{fig:firstpass}}
\end{figure*}

In practice, we find that using the impact parameter $b$ defined as the halos approach 
(i.e.\ neglecting detailed resonant loss of angular momentum) and 
taking $\scalelen$ to be the half-mass radius of the system works well in a mean 
sense. The results of this exercise are shown in Figure~\ref{fig:firstpass}. 
We plot the fraction of the gas available at the time of a first passage or fly-by 
encounter which is consumed in the induced burst \citep[we define the strength of the 
induced burst by integrating the star formation excess over the interpolation between 
the pre- and post-flyby star formation rates; see e.g.][for details]{cox:massratio.starbursts}, 
as a function of the gas content, for different orbits as in Figure~\ref{fig:fgas.fdisk}. 
We predict that the efficiency of channeling gas into the burst should 
scale as $\sim (1-\fgas)$, as before, and that the scaling with orbital parameters 
should be similar. 
We also show, for cases with otherwise identical gas fraction at the time of 
first passage and the same orbits as Figure~\ref{fig:massratio.fdisk}, 
how this scales with merger mass ratio (again, expected to be the same as 
that we derived above). Altogether, adopting our previous estimates, but 
re-normalizing appropriately for the impact parameter of the passage 
(here $b/\scalelen\approx1$, as we defined it above) yields a good approximation 
to the typical behavior in our simulations. 
We also test the behavior as a function of impact parameter, and find 
that our simple scaling is a reasonable approximation, yielding rapidly 
diminishing bursts as the impact parameter is increased to 
$b\gg\scalelen$ (at some point here, our estimates from the simulations become 
ambiguous, as a $\sim1\%$ enhancement in star formation is 
below the level of random fluctuations in isolated disks). 
We have also checked whether or not the pre-flyby stellar disks 
are destroyed -- as expected, they are left more or less intact by
fly-by encounters. The disks may be heated, and in fact some ``pseudobulge'' 
can form from the buckling of the bar induced in the stars, but we are 
not attempting to predict or study pseudobulge formation here (rather 
considering it, as is often the case in observations, to be fundamentally 
still part of the stellar disk rather than part of a violently relaxed ``classical'' 
bulge). In an average sense, then, our derived scaling is generally 
applicable. 

However, the details of exactly how the approach proceeds will 
introduce considerable scatter in the amount of burst triggered on 
first passages and in fly-by encounters. This is plain in the large (factor $\sim$ a few) 
scatter in Figure~\ref{fig:firstpass}. Further, details such as the 
structure of the bulge are increasingly important in the limit of weak interactions, 
where distortions in the potential of the primary that would trigger gas inflows 
can be suppressed by the presence of a larger bulge (and note the 
caveat from \S~\ref{sec:model.massratio}, that the secular/internal response 
of the disk will become relatively more important in weaker interactions 
with smaller mass ratios and larger impact parameters). 
We therefore expect in general that our predictions can be quite broadly 
applied, but are less robust for any specific case if it is a single fly-by 
as opposed to an integration over a full merger. 
Fortunately, in the case of systems that will actually merge, these details 
tend to average out or be unimportant, yielding the relatively 
small scatter we have seen in our previous predictions. In those cases, we do not need to 
be too concerned with the exact details of the impact approach, 
nor the structural details of the galaxy (in particular because our predictions 
are for integral quantities at the end of a merger, various effects will 
tend to cancel out -- for example retaining more gas on first passage will 
yield a larger supply for the second burst, etc.).

\subsubsection{Independence from ``Feedback'' Physics}
\label{sec:model.feedback}

Our derivation of the torques causing gas to lose angular 
momentum in mergers is purely dynamical. All else being 
equal (i.e.\ for systems with the same gas content and dynamical structure 
at the time of the final merger), we therefore expect that the detailed 
physics of e.g.\ ``feedback'' from supernovae, stellar winds, and AGN activity 
should make little difference. 

\begin{figure}
    \centering
    \scaleup
    \plotone{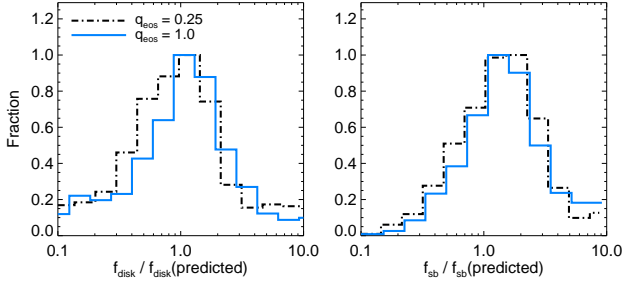}
    \caption{The effects of feedback on disk survival in mergers. 
    We show the distribution in $f_{\rm burst}/f_{\rm burst,\ pred}$, 
    i.e.\ the burst mass fraction, relative to that predicted (or equivalently, the 
    mean in our simulations) for the given pre-merger gas fraction and 
    orbital parameters, for simulations with two different ISM gas 
    feedback prescriptions (different effective equations of state $q_{\rm eos}$). 
    We also show the corresponding (but measured differently) 
    disk mass fractions $f_{\rm disk}/f_{\rm disk,\ pred}$. 
    There is perhaps a small offset in the sense expected (a stiffer, higher-feedback 
    equation of state for the ISM suppresses bursts by an average factor $\sim1.1-1.2$), 
    but this is much smaller than the simulation-to-simulation scatter. For a given 
    gas content at the time of the merger, then, feedback makes almost no 
    difference (true for AGN feedback and starburst winds as well). 
    \label{fig:qeos}}
\end{figure}

Figure~\ref{fig:qeos} demonstrates that this is indeed the case. We compare 
the starburst and surviving disk gas fractions 
of merger remnants, relative to those predicted by our simple dynamical model 
as a function of the merger mass ratio, orbital parameters, and 
gas content at the time of the merger, for suites of simulations with two different 
prescriptions for supernovae feedback and the effective equation of state of 
the ISM. In terms of our $\qeos$ parameter (see \S~\ref{sec:sims}), 
we compare $\qeos=0.25$ simulations (a nearly isothermal equation of state 
with effective temperature $\sim10^{4}\,$K) to $\qeos=1$ simulations 
(the ``full'' stiff \citet{springel:multiphase} equation of state, with effective temperature 
$\gtrsim10^{5}\,$K at the densities of interest here). 
There is no significant systematic offset between either the median 
result or the scatter about our simple analytic expectation. 
At most, there may be a $\sim20\%$ systematic offset, in the sense that 
more highly pressurized systems ($\qeos=1$) have slightly more 
gas survive -- a small offset like this is expected
because the bars in these 
cases are slightly more ``puffy'' (essentially the same as a slightly 
thicker disk -- for which we derive an analytic expectation in Equation~\ref{eqn:fburst.3} 
that yields an expected $\sim10-20\%$ difference at most based on 
the full possible range of $\qeos$). In any case, such an offset is small 
relative to other systematic uncertainties in disk structure and the scatter 
about the median predictions. 

\begin{figure}
    \centering
    \scaleup
    \plotone{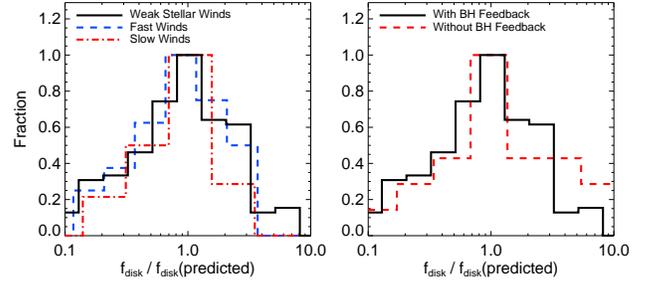}
    \caption{As Figure~\ref{fig:qeos}, but comparing the distribution of 
    disk fractions in merger remnants relative to our simple predictions as a 
    function of stellar wind and quasar feedback prescriptions. We plot 
    the distribution of disk fraction $f_{\rm disk}$ in simulations 
    relative to our predicted $f_{\rm disk}(f_{\rm gas},\mu,\theta)$ 
    (i.e.\ our calculation as a function of immediate pre-merger 
    gas fraction, orbital parameters, and merger mass ratio). 
    {\em Left:} Varying starburst-driven wind prescriptions. We compare 
    our usual weak stellar wind scenario (winds self-consistently can 
    generate from the hot gas, but additional mass loading is only 
    $\sim1\%$ of the star formation rate) to a fast winds scenario 
    (with additional mass loading $\sim0.5\,\dot{M}_{\ast}$ and 
    wind launch velocity $\sim800\,{\rm km\,s^{-1}}$) and a slow 
    winds scenario (additional mass loading $\sim2\,\dot{M}_{\ast}$ 
    and launch velocity $\sim200\,{\rm km\,s^{-1}}$). 
    {\em Right:} Simulations with and without feedback from 
    accreting black holes. 
    For otherwise fixed merger parameters (orbit, mass ratio) 
    and disk properties (cold gas content, mass profiles), 
    these feedback prescriptions make no different to the starburst or 
    surviving disk gas fractions. They will, however, change the 
    the gas content, consumption, and distribution leading into the merger. 
    \label{fig:sbw}}
\end{figure}

In Figure~\ref{fig:sbw}, we perform 
a similar exercise for cases with and without 
central supermassive black holes (we have also 
examined initial BHs with varying 
initial masses from $\lesssim 10^{5}\,\msun$ to $\sim10^{7}\,\msun$), 
and cases with or without a simple 
implementation of starburst-driven winds where winds are launched 
(in addition 
to the stellar feedback implicit in our multi-phase ISM model)
with a mass-loading efficiency $\dot{M}_{\rm wind} = \eta_{w}\,\dot{M}_{\ast}$ 
relative to the star formation rate $\dot{M}_{\ast}$ and energy loading 
efficiency $\epsilon_{w}$ relative to the total energy (for a \citet{salpeter:imf} 
IMF) available for supernovae \citep[sampling the range $\eta_{w}\sim 0.01-10$ 
and $\epsilon_{w}\sim 0.1-1$; see][]{cox:winds}. Shown in 
Figure~\ref{fig:sbw} are our fiducial weak winds ($\eta_{w}\sim0.01$, 
$\epsilon_{w}\sim0.0025$), cases with moderate mass loading 
into very fast winds ($\eta_{w}=0.5$, $\epsilon_{w}=0.25$, yielding a wind 
launch speed $\sim800\,{\rm km\,s^{-1}}$), and cases 
with high mass loading but correspondingly slower wind velocities 
($\eta_{w}=2.0$, $\epsilon_{w}=0.0625$, yielding a wind 
launch speed $\sim200\,{\rm km\,s^{-1}}$). In all these cases we find 
a similar result: at otherwise fixed properties at the time of merger, 
feedback makes no difference to our conclusions.

The reasons for this are described in \S~\ref{sec:model.orbit} and 
in more detail below (\S~\ref{sec:model.secular}). Recall, the 
distortion in the primary is driven by the secondary and as such depends only 
on the gravitational physics of the merger. Given this distortion, the 
gravitational torques within some characteristic radius are sufficiently strong to remove 
the angular momentum from the gas in much less than an orbital time. 
Feedback, then, insofar as it changes the effective pressurization or equation of 
state of the gas or drives a wind, is largely irrelevant: because the angular momentum is removed 
in a timescale much shorter than the orbital time, the gas (regardless of the 
strength of feedback) cannot dynamically respond with these hydrodynamic forces, but 
must essentially free-fall into the center of the galaxy where the starburst is triggered. 
The radius interior to which the torques are strong is not a function of e.g.\ the stability of the 
galaxy to perturbation, because it is not an instability in the first place, but a driven 
distortion in the system. Moreover, the entire system is strongly in the non-linear regime 
for the time of interest (when we consider interactions 
of the magnitude simulated: mass ratios $\sim$1:8 and more major 
mergers) -- no amount of 
making the system more robust against linear instability would be sufficient to avoid 
a strong gravitational distortion in the violent coalescence surrounding 
the actual merger (this must be so, because the distortion occurs where the disturbances in 
the potential are greater than order unity -- for hydrodynamic forces to resist distortion, 
they would have to be stronger than large-scale gravitational forces in the equilibrium system, 
negating the concept of a rotationally supported thin disk). 
So what matters is instead where that coalescence occurs and how long 
it introduces such a strong distortion, relative to e.g.\ the local dynamical or orbital time 
of some disk element, giving rise to the simple dynamical criteria for angular 
momentum loss developed here.

That is {\em not} to say that for fixed {\em initial} conditions (significantly pre-merger 
or e.g.\ at first passage), feedback will not change the result. 
There are two primary means by which feedback can indirectly have a strong 
influence on disk survival: 

{\bf (1) Retaining Gas (Lowering the Stellar Mass Fraction):} 
As has been demonstrated 
in a number of works \citep{weil98:cooling.suppression.key.to.disks,
sommerlarsen99:disk.sne.fb,sommerlarsen03:disk.sne.fb,
thackercouchman00,thackercouchman01,governato:disk.formation,
robertson:disk.formation,springel:models,springel:spiral.in.merger,
okamoto:feedback.vs.disk.morphology,scannapieco:fb.disk.sims}, these forms of feedback can have 
dramatic implications, in even a short time period, for the rates at which cooling of 
new cold gas from the halo and consumption of existing gas by star formation proceed. 
In cases with no feedback, star formation may exhaust gas efficiently, leading to 
predicted systems that are much more gas-poor at the interesting time 
of the final merger -- according to our model, then, these will 
not be able to form disks as efficiently as more gas-rich systems. 
In cases with strong feedback from e.g.\ star formation to lower 
the effective star formation efficiency and recycle gas, the predicted gas fractions 
at the time of merger (from some gas-rich initial conditions) could be much higher. 
Inclusion of stellar and supernovae feedback responsible for injecting 
energy and turbulent pressure into the ISM may also be necessary to 
prevent the onset of clumping and disk fragmentation in 
isolated gas-rich cases, enabling the stable existence and evolution of 
quiescent gas-rich disks \citep[see e.g.][]{springel:multiphase,
robertson:cosmological.disk.formation}. 
In short, feedback may be critical to give rise to high gas fractions in the first place, 
which we have shown have dramatic implications for the survival of disks -- but for a 
given gas fraction (however that comes about in the first place), the results of the 
merger will be (in the short term) independent of feedback. 

{\bf (2) Changing the Spatial Distribution of Gas (``Kicking Gas Out'' of $R_{\rm max}$):} 
Recall, our derivations demonstrate that it is not necessarily a fixed fraction of 
gas that loses its angular momentum: rather (see Equation~\ref{eqn:rtemp1} 
and \S~\ref{sec:model.gas}-\ref{sec:model.orbit}) 
it is the mass inside some radius $R_{\rm max}/\scalelen$ relative to that of the 
stellar disk (characteristic radius $\scalelen$) which will lose its angular momentum. 
If some form of feedback can change the spatial gas distribution, then, it could have 
dramatic implications for disk survival. We have used the radius $R_{\rm max}$ to 
estimate the mass fraction that will burst by assuming the gas density profile 
is broadly similar to that of the stars (which is 
true in our simulations, given their feedback 
prescriptions). But one could easily imagine the extreme limit, where some 
strong feedback keeps all the gas at large radii $r\gg R_{\rm max} \sim \scalelen$ (i.e.\ 
a case in which there is a large hole in the gas distribution, 
or in which the gas is at least much more extended 
than the stellar distribution) -- 
the stellar disk torques only act effectively within $R_{\rm max}$, so only a tiny 
fraction of 
the gas in such a case would 
lose its angular momentum. Especially at high redshift, 
this may be important in avoiding overcooling and the 
formation of too much bulge mass in many systems 
\citep[see e.g.][]{robertson:cosmological.disk.formation,
governato:disk.formation, donghia:disk.ang.mom.loss,
ceverino:cosmo.sne.fb,zavala:cosmo.disk.vs.fb}. Again, we stress that for a 
given gas density profile at the time of merger, our calculations are independent of 
feedback; but if feedback alters the gas profile -- keeping the 
gas at radii $\gg R_{\rm max}$, then it will largely survive the merger.

\subsubsection{Exceptions and Pathological Cases}
\label{sec:model.exceptions}

We have derived a general model for how disks are destroyed in mergers 
and shown that it applies to a wide range of gas fractions, orbital parameters, 
galaxy mass ratios, and prescriptions for feedback and gas physics. 
However, there are some pathological cases of more than academic interest, 
as these can explain some small differences with previous results 
as well as illustrate the important physics in our model. 

For example, consider the starburst mass fraction and surviving disks in 
our ``{\bf h}'' orbits: i.e.\ a prograde-prograde, coplanar merger of two 
disks. In this case, the angular momentum vectors of both disks 
and the orbital angular momentum are all perfectly aligned. Naively, 
one might then expect that these unique cases would create the largest disks. 
In fact, the {\em opposite} is true. This is largely for the reasons we 
outline in \S~\ref{sec:model.orbit} -- the alignment of angular momentum vectors 
means that the system is in near-perfect resonance, so it excites the largest 
tidal and bar asymmetries that rapidly drain the gas of all angular momentum. 
As we have shown, the much larger space of less-aligned orbits is in 
fact more favorable to disk survival. 

However, while the perfect resonance means that the bar 
efficiency $\Psi_{\rm bar}$ is large, the amount of mass in the stellar bar still 
scales as $1-\fgas$, so the burst fraction should vary as 
$\fgas\,(1-\fgas)$ in our simple model (i.e.\ we would still expect 
that a $100\%$ gas disk would have 
no stellar bar, hence no burst, as we have seen for more representative orbits
in Figure~\ref{fig:fgas.fsb}). In fact, though, we typically find 
in these cases that the 
burst fraction seems to scale as $f_{\rm burst}=\fgas$ all the way to high values of $\fgas$ -- 
in short, almost all the gas always bursts -- 
there is no suppression by a $1-\fgas$ factor as would be expected 
if the stellar bar were doing the torquing. 

The reason for this is simple -- again, the orbits here are perfectly coplanar 
and in resonance; so this is the one case where the secondary 
galaxy as a whole can directly act as an efficient torque on the gas. In short, 
because the systems are perfectly coplanar and in a resonant orbit, the 
entire secondary galaxy (all baryons and dark matter within the 
stellar $R_{e}$) acts directly to introduce a non-axisymmetric potential 
perturbation (the secondary itself plays the role of the bar). 
So because of this, to an even greater extreme than our scalings for 
more general orbits would predict, this narrow range of orbits is pathological and biased 
{\em against} disk formation. However, understanding why this is the 
case, we can check and explicitly show that it is not so for  
more general orbits, even nearly prograde-prograde orbits (such as case {\bf e}) -- 
in all those cases, even those just slightly out of coplanar resonance, the 
stellar bar is indeed the primary source of torque, and our assumptions 
are justified. This example therefore nicely illustrates what the consequences would 
be if our fundamental assumptions were not true, as well as showing why 
they are in fact true for non-pathological cases. 

Another pathological case of interest is one in which the disks are $100\%$ 
gas at the time of merger. Here, as we have said, our simple model 
predicts no starburst or angular momentum loss. In practice, there will still 
be some loss of angular momentum owing to direct cancellation in e.g.\ shocks 
between the disks; but as discussed in \S~\ref{sec:form.major:angloss}, there will also be 
the possibility of some gain owing to the angular momentum of the merger. 
In fact, over the range in mass ratios $\mu\sim 0.1-1$, for a range of typical 
impact parameters $b\sim0.5-5$, the expected final specific angular momentum 
from after cancellation is approximately equal to the initial specific angular 
momentum of the primary (with $\sim20\%$ scatter). Cancellation is therefore 
inefficient. A random distribution of orbits might negate $\sim20\%$ of 
the angular momentum in $\sim$ half the systems merging, but will leave 
$\sim 80-100\%$ of the disk intact. Even these cancellations, we find in detail, do not 
generally yield a starburst in the same manner as a merger-induced bar, 
but lead to moderate disk contraction (and an equal number of mergers 
will scatter towards the opposite sense leading to disk expansion, keeping a mean 
specific angular momentum that is constant). They do not cause a starburst because, 
if two random parcels or streams of gas shock and lose angular momentum, 
the alignment and relative momenta would have to be near-perfect for them 
to lose, say $95\%$ of the angular momentum and fall all the way to the 
central $\sim 100$pc where a nuclear starburst would occur. Rather, they will lose 
some fraction of order unity of their angular momentum, fall in to a radius smaller 
by a factor $\sim2-3$ (but not to very small radii), and continue to orbit. Without the 
bar that can continuously drain angular momentum, the true burst is indeed 
inefficient. 

Although we show in \S~\ref{sec:model.feedback} that the physics of interest 
are generally independent of feedback prescriptions, 
there are some pathological feedback regimes. These are discussed in 
detail in \citet{cox:winds}; here, we outline the pathological behavior. 
If e.g.\ starburst-driven winds are implemented with extreme efficiencies 
$\dot{M}_{\rm wind}\gg \dot{M}_{\ast}$ and with 
moderate to large velocities $\gtrsim200\,{\rm km\,s^{-1}}$, then there is no definable 
``starburst'' in the simulations any more, even when the 
gas loses angular momentum -- indeed, it becomes almost impossible 
to trigger starbursts by {\em any} mechanism. This is because the feedback 
is so extreme that any parcel of gas that begins forming stars above some 
threshold rate is immediately blown apart and drives away all the surrounding 
gas. However, observations suggest that these cases are almost certainly not relevant -- 
observationally inferred mass-loading factors of winds are well below 
the predicted threshold where we see this behavior \citep[see e.g.][]{veilleux:winds,
martin99:outflow.vs.m,martin06:outflow.extend.origin,
erb:lbg.metallicity-winds,sato:outflow.hosts}, 
and moreover the ubiquity of starbursts and recent starburst remnants in 
observed gas-rich major mergers 
\citep[e.g.][]{soifer84a,soifer84b,scoville86,sargent87,sargent89} 
implies that feedback, 
while still potentially efficient, is not able to ``self-terminate'' a starburst 
before it even begins \citep[this is in fact directly confirmed in 
observations of outflows in ongoing massive, merger-induced 
starbursts; see e.g.][]{martin05:outflows.in.ulirgs}. 
A similar pathology can appear if we include extreme 
coupling of black hole feedback to the galaxy gas (e.g.\ allowing 
$100\%$ of the BH accretion energy to couple efficiently), but this is also 
ruled out observationally, both for the arguments above \citep[starbursts 
exist, and the winds seen are not so enormous; 
see the discussion in][]{cox:winds,
hopkins:lifetimes.methods,
hopkins:lifetimes.interp,hopkins:lifetimes.obscuration,
hopkins:lifetimes.letter,hopkins:qso.all}, 
and because such a prescription 
yields black hole masses orders-of-magnitude discrepant from the 
observed \citep{FM00,Gebhardt00} $M_{\rm BH}-\sigma$ relation 
\citep[see e.g.][]{hopkins:bhfp.theory,hopkins:bhfp.obs}

\subsubsection{Longer-Lived Perturbations: Relation to Secular Evolution}
\label{sec:model.secular}

Thus far, we have focused on activity during the merger, roughly defined 
as the short timescale $\sim10^{8}$\,yr following first passage and 
coalescence. In this regime, we have shown that (for typical conditions), 
the dominant source of angular momentum loss is the torque on 
gas from stars in the same disk. However, it is well known from 
studies of isolated barred galaxies \citep[e.g.][]{weinberg:bar.dynfric,
hernquist:bar.spheroid.interaction,
friedli:gas.stellar.bar.evol,
friedli:gas.bar.ssp.gradients,
athanassoula:bar.halo.growth,
athanassoula:bar.vs.concentration,weinberg:bar.res.requirements,
kaufmann:gas.bar.evol,foyle:two.component.disk.evol.from.bars} that 
a long-lived bar (regardless of whether the bar is purely 
stellar or purely gaseous) will exchange angular momentum with 
itself (or e.g.\ gas/stars further out in the disk) and the 
dark matter halo, allowing for further angular momentum loss 
and building a central bulge or ``pseudo''-bulge \citep{patsis:gas.flow.in.bars,
athanassoula:bar.evol.in.int,
mayer:lsb.disk.bars,berentzen:gas.bar.interaction}. 
Here, we discuss 
the relation of this process to what we have described in 
our merger simulations: in general, we find that it (while potentially 
very important for the long-term evolution of the disk and bulge 
masses and structure) is a second-order effect within the merger 
itself, and on longer timescales is more appropriately considered 
an independent, secular evolution process (despite being initially 
triggered by a merger), whose study is better described in 
simulations of idealized and long-lived bars. 

As discussed in \S~\ref{sec:model.orbit}, there is a limit to how far the analogy to 
barred galaxies can be drawn. Recall, we use the term ``bar'' more generally to 
represent a quadrupole moment or non-axisymmetric distortion 
in the stellar disk: it does not necessarily (and, especially after second passage, 
usually does not) morphologically resemble isolated barred spirals 
and may not even have an $m=2$ mode structure. Critically, the distortion is 
driven externally by the gravitational perturbation of the secondary orbit -- it is 
not the result of an instability within the primary. As we note in \S~\ref{sec:model.orbit}, 
this already gives rise to a couple of important distinctions: because the distortion is 
driven by the orbital motion of the secondary, it has a characteristic frequency 
(and corresponding radius) internal to which angular momentum loss is very 
efficient (determined entirely by the gravitational properties and relative 
motions of the systems, not 
the subtleties of their internal orbital structure), 
{\em regardless} of properties of the primary (e.g.\ gas phase structure, 
feedback, etc.) that might otherwise make the system more or less stable to 
the development of internal instabilities. 

It is straightforward to estimate the relative importance (over a 
short timescale after the merger) of angular momentum loss from 
the gas to the shared stellar bar/distortion induced by the merger, versus 
that to itself and the dark matter halo (the standard secular scenario, which 
other than the initial driving in the merger, will {\em not} be driven by the 
relative gravitational motions but by the more standard bar stability and 
spin-down criteria). 
Approximating the gas as a rigid, thin bar of mass 
$M_{\rm bar,\, gas}\approx f_{\rm gas}\,M_{\rm bar}$ and 
radius $R_{\rm bar}\sim R_{d}$, we can estimate 
the specific torque from the remaining gas disk 
and halo in the dynamical friction limit, following 
\citet{weinberg:bar.dynfric} \citep[for more detailed solutions, which ultimately 
give similar results, see][]{hernquist:bar.spheroid.interaction,
athanassoula:bar.slowdown,weinberg:bar.res.requirements}: 
${\rm d}j/{\rm d}t = -4\,\pi\,\alpha\,G^{2}\,M_{\rm bar,\, gas}\,\rho(R_{d})\,
v_{\rm bar}^{-2}$, where $\rho(R_{d})$ is the background density and 
$\alpha\sim1$ is a numerical constant (depending on the exact shape of the bar, potential, 
and phase-space distribution of the background). If the ``background'' is 
a \citet{mestel:disk.profile} disk or isothermal sphere, this becomes 
$-2\,\alpha\,G\,M_{\rm bar,\, gas}/R_{d}$. Compare this to our 
Equation~(\ref{eqn:bar2}) for the instantaneous 
torque on the gas bar from the stellar bar: 
$-G\,M_{\rm bar,\, \ast}/(R_{d}\,\sqrt{\sin^{2}{\barangle}+\tilde{H}^{2}})$. 
Removing the common factors, the gas/halo torque 
goes as $\sim f_{\rm gas}$, whereas that from the 
stellar bar goes as $\sim (1-\fgas)/\sqrt{\sin^{2}{\barangle}+\tilde{H}^{2}} 
\sim (1-\fgas)/\barangle$ (because $\barangle\sim\tilde{H}\ll 1$). In short, 
the torque from the gas disk and halo goes as $f_{\rm gas}$ because it is a 
second-order resonance effect (amplified 
and trading off with the gas bar), whereas the stellar bar strength goes 
as the stellar mass fraction $(1-\fgas)$, but boosted by a factor 
$\sim1/\barangle$ representing the small angle of offset between the two 
bars -- i.e.\ the stellar bar is in much closer spatial proximity (in particular 
in spatial alignment in the disk plane) to the gas bar. 

This simple comparison gives a reasonable quantitative prediction of 
the relative torques exerted by the halo and stellar disk in our 
simulations. 
Essentially, we have just re-derived the 
well-known fact that the timescale for a bar to damp its own angular momentum 
via resonant interactions with itself and/or the halo is some number ($\sim$ a few) bar 
rotational periods \citep{athanassoula:bar.vs.concentration,
athanassoula:bar.slowdown, 
weinberg:bar.res.requirements,
kaufmann:gas.bar.evol} 
(each bar rotational period being $\sim1-2$ times 
the disk rotational period), whereas in the typical mergers the gas is drained of 
angular momentum by the much stronger local torques on a timescale much 
shorter than an orbital time, allowing it to more or less free-fall into the 
galactic center. Comparing these timescales gives a similar ratio of 
torque strengths. 
Obviously, 
as $\fgas\rightarrow1$, the torque from the halo must eventually 
dominate, but this will not happen until 
$f_{\ast}=(1-\fgas)\lesssim \barangle\sim 0.1$ (given typical bar lags of 
$\sim$ a few degrees or the ratio of the timescales above). 

In practice, such a situation is somewhat contrived (it is very difficult to 
maintain a disk with a true $\gtrsim90\%$ gas fraction), and unlikely to 
be of broad cosmological relevance (we have no simulations in this regime 
with which to compare, in fact, because even initially $100\%$ gas disks with 
low star formation efficiencies will be $\lesssim 80\%$ gas by the time of 
the actual merger). However, our Equation~(\ref{eqn:full.equation}) can be trivially 
modified to include these effects: the $(1-\fgas)$ term should be replaced 
with a more appropriate $(1-\fgas+\epsilon_{h})$, where 
$\epsilon_{h}\sim \barangle \sim0.1$ represents the contribution of 
angular momentum 
loss to the halo and outer gas disk during the merger. This exchange of angular momentum, 
therefore, sets some minimum 
bulge mass (with mass fraction $\sim10\%$) that would form even in a 
pure-gaseous disk merger. 

By comparing the relative {\em instantaneous} amplitude of the torques from 
the halo/gas and stellar disk, we are comparing how important they each are 
in the loss of angular momentum from the gas over the same (relatively 
short) merger timescale. More important is the fact that, in a gas-rich case 
where the distortion to the stellar distribution may be an inefficient torque, 
the gas bar could be long-lived and continue to lose angular momentum 
over longer timescales. It is not necessarily clear that this would 
happen, however -- a number of studies suggest that 
gas and stellar bars become self-damping once a central mass concentration 
(i.e.\ a nuclear starburst triggered by gas inflows, in this case) 
is in place with a mass fraction larger than a few percent 
\citep{bournaud:gas.bar.renewal.destruction,
berentzen:bar.destruction.in.int,
berentzen:self.damping.tidal.bar.generation,
berentzen:gas.bar.interaction,
athanassoula:bar.vs.cmc} 
\citep[but see also][]{kaufmann:gas.bar.evol}. In such a case, we again arrive at the conclusion 
that these processes set a minimum bulge mass from a large 
bar-inducing perturbation, but do not dominate the 
creation of much larger bulges in mergers. 

Regardless of this effect, it is 
not clear that a bar can survive a substantial merger: 
recall, the distortion following second passage 
and coalescence resembles a bar only in that it introduces a rotating 
quadrupole distortion in the disk potential (allowing us to describe it as a 
``bar'' for analytic convenience), not necessarily in its structure or longevity (it 
does not necessarily share the orbital ``pileup'' that allows a bar to survive), 
and moreover the actual coalescence of the galactic nuclei will disturb any 
bar structure that may be present. Quantitatively, we find our remnants 
rarely have significant long-lived 
$m=2$ modes in the stars or gas -- the mode amplitude tends to damp 
after merger on a timescale $\lesssim10^{8}$\,yr (i.e.\ the free-fall or dynamical 
time, much slower than the typical significant number of orbital times 
for standard bar self-braking). This is similar to the conclusions in the bar 
studies of e.g.\ \citet{bournaud:gas.bar.renewal.destruction} and 
\citet{berentzen:gas.bar.interaction}, who find that the combination of the 
formation of a bulge/small central mass concentration from gas inflows 
and the disturbance/heating to the bar itself in interactions prevents even 
gas-rich systems from maintaining or very rapidly re-forming a bar 
after a significant merger \citep[as opposed to a fly-by passage, which 
may more efficiently induce long-lived bars; see e.g.][]{berentzen:self.damping.tidal.bar.generation}. 
That is not to say a bar 
may not form in the re-formed remnant disk, but such a bar would arise in a 
standard secular fashion, and should be considered in the context of 
the long-term secular evolution of the merger remnant. 

If, however, the potential distortion survives the merger to form a 
stable bar, it can certainly be important to the long-term evolution of the system 
and buildup of the bulge. However, this case is outside the scope of this paper, 
and should be more appropriately considered as subsequent evolution of 
the remnant (albeit with an initially merger-induced bar). This is because the 
timescale for the bar to lose angular momentum and contract is some number 
of rotational periods -- so the gas losing angular momentum will slowly 
spiral inwards in some number of orbital periods (turning into stars and 
possibly being ejected by feedback as it does so), rather than free-falling 
into a central burst in a time much less than an orbital period. 
The end result of such angular momentum loss can resemble a bulge 
\citep{mayer:lsb.disk.bars,debattista:pseudobulges.a}, 
although the expectation of rotational support and ``disky-ness'' 
in the material lead to it more likely being a ``pseudo-bulge'' typical 
of secular processes \citep{combes:pseudobulges,kuijken:pseudobulges.obs,oniell:bar.obs,
kormendy.kennicutt:pseudobulge.review,athanassoula:peanuts}. 
Depending on e.g.\ details of the equation of 
state, feedback, and rotational support of the gas disk, it may also amount to 
steady disk contraction \citep{debattista:pseudobulges.b} or emergence of a two-component disk 
\citep{kaufmann:gas.bar.evol,
foyle:two.component.disk.evol.from.bars}. A number of effects will be important 
in this regime, including the effects of feedback in pressurizing the disk and 
smoothing out substructure, and the role of accretion and mergers 
in rebuilding the disk as such evolution continues (since it is occurring on 
timescales $\sim$ several Gyr, comparable to the characteristic timescales 
for new accretion and mergers). 

These effects make it difficult to predict the net effect of 
such evolution. For example, if in a pure gas merger 
of mass ratio $\mu$ the specific angular momentum (on average) is 
increased (by addition of specific angular momenta plus orbital angular momentum) 
by an amount $\sim \epsilon_{m}\,\mu\,j_{\rm disk}$, but then the induced 
bar (of amplitude $\sim \mu$) loses its angular momentum 
($\sim \mu j_{\rm disk}$) on a timescale $\sim N\,t_{\rm rot}$, 
the sequence of mergers and induced bars compete (given the 
cosmologically expected timescale $\approx \mu\,t_{H}$ between 
mergers of mass ratio $\mu$ in \citet{fakhouri:halo.merger.rates}, and 
that for a disk of mass fraction $m_{d}$ relative to the halo, 
$t_{\rm rot}\sim m_{d}\,t_{H}$, one obtains 
${\rm d}j/{\rm d}t\sim j_{\rm disk}/t_{H}\,[\epsilon_{m} - \mu/N\,m_{d}]$ -- i.e.\ 
more major events will tend to lead to angular momentum loss 
in gas, whereas the net effect of very minor mergers and smooth 
gas accretion, even where it induces instabilities, may be to 
``spin up'' the disks). 

As discussed in \S~\ref{sec:model.massratio}, there is also an interesting 
regime of parameter space, namely minor mergers with 
mass ratios $\sim$1:20-1:10 or so, in which the characteristic merger 
timescales and secular/internal evolution timescales are
comparable. The secondary may be large enough to induce a significant 
bar/non-axisymmetric response, but the merger/dynamical friction time 
may be sufficiently long that the primary could respond almost as if in isolation for 
several orbital periods. 
In such a case it becomes less clear whether the 
merger or the secular response of the disk is ultimately the dominant 
driver of evolution (and the answer probably depends on e.g.\ the exact orbital 
parameters and stability properties of the disk, and may be
sensitive to feedback, 
the gas phase structure and pressure support, and detailed halo structure).  
In any event, it is clear that these processes 
require study in a more complete cosmological context, and 
can contribute significantly to the bulge population (especially 
in less bulge-dominated galaxies, below the typical thresholds 
we simulate) over a Hubble time of evolution. However, 
although the bar itself may be triggered in the merger, the nature of the 
relative strength of the interaction and characteristic timescale for angular momentum 
loss make it not a violent process associated with the merger itself, but rather a 
secular process that should be considered more analogous to bars in non-merging 
systems.

\breaker
\section{Application to Semi-Analytic Models}
\label{sec:prescriptions}

Our results clearly have potential uses as prescriptions for 
analytic and semi-analytic models of galaxy formation. 
Here, we summarize and give some simple 
recommendations for these applications. 

When a merger is identified in a semi-analytic model, the two key quantities 
we can predict here are the mass fraction of the disk that is destroyed 
(violently relaxed into a bulge) and the 
fraction of the cold gas in the disk that will lose angular momentum 
and contribute to the bulge by forming a compact starburst. 

First, the stellar disks: in a merger of secondary mass $M_{2}$ with 
primary mass $M_{1}$, the secondary is destroyed (adding $M_{2}$ 
to the bulge) and the mass within a radius enclosing $\approx M_{2}$ 
in the primary is violently relaxed. If the primary were pure 
disk, this would add $2\,M_{2}$ to the bulge. However, one can imagine 
the limit where the primary is entirely bulge-dominated inside that 
radius (with the stellar disk dominant only at much larger radii) 
-- then the violent relaxation of the merger will act primarily to 
heat existing bulge stars, and only a mass $1\,M_{2}$ will be added to the bulge. 
Obviously, its also true that if the total disk mass of $M_{1}$ is less than 
$M_{2}$, then that is a maximum to how much can be added to the bulge 
(i.e.\ really ${\rm MIN}(M_{2},\,f_{\rm disk}\,M_{1})$ is added). 
For most purposes, this factor $2$ possible range is not critical 
in the semi-analytic models, and picking a constant (effective mean) 
fraction $(0-1) \times M_{2}$ to violently relax in the primary in all mergers 
is acceptable. However, if more detail is desired, an estimate of the 
mass profiles of bulge plus disk components in the primary can be used to 
determine the total primary 
disk mass within a radius enclosing a mass $\approx M_{2}$, and 
then that will be the fraction violently relaxed. For a \citet{hernquist:profile} 
bulge and exponential disk obeying roughly the observed size-mass 
relations from \citet{shen:size.mass}, the primary disk mass that should be 
violently relaxed in a merger with mass $M_{2}$ can be 
approximated as $f_{\rm disk,\,\ast}\,M_{2}/(1+[M_{1}/M_{2}]^{\alpha})$, 
where $f_{\rm disk,\,\ast} \equiv (1-f_{\rm gas})\,(1 - f_{\rm bulge})$ is the  
mass fraction of the stellar disk (relative to the baryonic galaxy) and 
the term $(1+[M_{1}/M_{2}]^{\alpha})$ is a correction for e.g.\ the 
relative sizes of the two components as a function of mass ratio and 
other properties (for the assumptions above, 
$\alpha\approx 0.3-0.6$, depending on the details of the disk 
mass profile). 

Two clarifications should be emphasized. First, these derivations only 
apply to cases where the secondary is sufficiently massive that it survives 
to merge with the center of the primary. If the secondary is destroyed or 
shredded by tidal forces before merger, then it will not add either its own 
mass or any violently relaxed mass to the bulge. This generally occurs in 
the limits of smaller mass ratios ($\lesssim1:10$, which we have 
considered), but is included in some models. Second, for most applications, 
the masses $M_{1}$ and $M_{2}$ should be taken to be the 
{\em baryonic} masses {\em within} the galaxies (stars in the galaxy and cold gas -- not 
diffuse stellar halo or pressure-supported hot gas in the extended halo). 
This is how we have defined our models and fits to our simulations (although those 
simulations do include dark matter and extended halo gas and stars) throughout. 
The halos are much more extended, and much lower density, so they merge and mix 
more efficiently, and do not strongly participate in the central 
violent relaxation process that defines the bulge. 
Moreover, there can be a wide range in halo masses for galaxies of similar mass -- 
but most of these halos are large and often independent substructures that should 
not be used to define e.g.\ the mass ratios of merging encounters. 
What dark matter is 
carried in with the galaxies is that enclosed in their stellar 
effective radii $R_{e}$, which tends to track the baryonic mass much more closely 
than, say, the total halo mass, so it is not a bad proxy to still define mass ratios, 
etc.\ in terms of the baryonic masses. 

Next, in such an encounter, our analysis provides a means to estimate 
the fraction of the cold gas mass in the pre-merger stellar disks that should 
lose angular momentum and be funneled into a nuclear starburst. 
The cold gas inside some radius $R_{\rm gas}/\scalelen$ will participate in this starburst, 
where $R_{\rm gas}$ is given by Equation~(\ref{eqn:full.equation}). 
There are five variables that go into this equation: (1)
$f_{\rm gas}$, which we define as the mass fraction {\em of the disk} 
that is in cold (rotationally supported) gas (i.e.\ if the disk is $50\%$ cold gas, then regardless of the 
bulge fraction of the galaxy, $f_{\rm gas}=0.5$. Note that we only care about 
cold, rotationally supported gas. Hot gas in the galactic halos can cool, of course, 
and form new stars, but that process is relatively independent of the merger, 
and is not related to angular momentum loss (also because the hot gas is 
pressure-supported, it is fairly resistance to significant redistribution in the merger, 
and if anything will tend to be shocked to even higher temperatures rather than 
forming stars in the short-lived merger). 
(2) $f_{\rm disk}=(1-f_{\rm bulge})$, the total (gas plus stellar) baryonic mass 
fraction of the disk. (3) $\mu\equiv M_{2}/M_{1}$, the mass ratio of the 
merger (defined as above). (4) $\theta$ and $b$, equivalently the 
orbital parameters of the merger. As discussed in \S~\ref{sec:model.orbit}, 
for cases that will merge the appropriate limit is $b\rightarrow0$, since 
most of the action will occur on the final merging passages after the 
angular momentum is removed. We discuss what should be adopted 
for the orbital inclination $\theta$ below. 
(5) $\scalelen$, the scale length of the disk stars. 

In any semi-analytic or analytic model, variables (1)-(3) should be well-known 
beforehand. Given some choice of orbital parameters and an assumed 
mass distribution of the disk, it is trivial then to translate 
Equation~(\ref{eqn:full.equation}) into a fraction of the gas that will 
burst. Because orbital parameters are generally undetermined in these 
models, there are two choices for the assumed orbital inclination $\theta$. 
First, one could draw a random value of $\theta$ for each merger 
(uniformly sampling in $\cos(\theta)$ as appropriate for an isotropic 
orbit distribution), and use 
Equations~(\ref{eqn:full.eqn.orbit.1})-(\ref{eqn:full.eqn.orbit.2})
for each merger. Alternatively, we can average over 
a random distribution of orbits and quote an ``effective'' 
orbital dependence $F(\theta,b)$ for Equation~(\ref{eqn:full.eqn.orbit.1}). 
Note that this is only strictly appropriate if all disks have the same mass 
profiles and those are such that the enclosed mass is linear in $R/\scalelen$ 
(otherwise the appropriate average would have to be weighted by 
other terms such as $(1-\fgas)$ in Equation~\ref{eqn:full.equation}). In 
any case doing so yields an effective mean orbital dependence 
$F(\theta,b)\approx1.2$. 

The only remaining issue is the assumed mass profile of the disk. Here, models 
have some freedom. As we have emphasized, the exact profile (e.g.\ choice 
of exponential disk or some other profile) does not have a dramatic effect. 
What is important, however, is the assumption of how the gas is distributed 
relative to the stars. Recall, Equation~(\ref{eqn:full.equation}), with the variables 
above inserted, gives that the gas inside some radius 
$R_{\rm gas} = x\,\scalelen$ (where $x$ is a constant depending on those 
variables, and $\scalelen$ is a characteristic scale length of the {\em stellar} disk) 
should lose angular momentum and participate in the burst. 
Given a gas mass profile $M_{\rm gas}(R/R_{e,\, {\rm gas}})$, in terms of a 
characteristic gas disk scale length $R_{e,\, {\rm gas}}$, this gives 
the gas mass that bursts, $M_{\rm gas}(x\,\scalelen/R_{e,\, {\rm gas}})$. 
For our simulations, we have generally assumed 
(and can see that it is a good approximation) that the gas and stellar disks 
initially trace one another ($\scalelen\approx R_{e,\, {\rm gas}}$). 
However, since our derivation and Equation~(\ref{eqn:full.equation})
show that it is the gas inside some fraction of the {\em stellar} disk half-mass 
radius $\scalelen$ that loses angular momentum, then if the gas is e.g.\ much 
more extended than the stars, a lower gas fraction will end up in the burst. 
We discuss this in \S~\ref{sec:model.feedback}, and consider how such situations may 
in fact arise owing to e.g.\ supernova feedback blowing gas out to large radii. 
Semi-analytic models therefore have some freedom in adopting these prescriptions 
based on their implicit assumptions about feedback and disk formation, encapsulated 
effectively in our prescriptions as the ratio of the 
stellar to gas disk scale lengths $\scalelen/R_{e,\, {\rm gas}}$. Lacking 
some detailed model for both values in the semi-analytic models, 
a constant value $\sim1$ is probably a good choice (with the exact choice 
reflecting implicit assumptions about feedback and outer disk formation). 

Those prescriptions define both the violently relaxed and starburst components 
induced in mergers of arbitrary mass ratios, gas content, and orbital parameters. 
If desired, appropriate scatter (a factor $\sim2$) can be added to both 
quantities, reflecting the scatter we see between various numerical 
realizations (although it should still be ensured that, with scatter, the implied 
violently relaxed and burst fractions are within the sensible physical limits). 

Although not discussed here, in \citet{hopkins:cusps.ell,
hopkins:cores,hopkins:cusps.fp,hopkins:cusps.evol,
hopkins:cusps.mergers} we consider 
how the sizes and velocity dispersions of these components should scale, 
and we refer to those papers for detailed analysis of those results. Briefly, 
we note that in the absence of dissipation, it is straightforward to calculate the 
size of the dissipationless component (the violently relaxed stars from the 
pre-merger stellar disk), given phase space and energy conservation. 
Roughly, this implies that the component will have the same (modulo 
projection effects since it transforms from a disk to a sphere) 
scale radius as the disk (or radius within the disk) from which it forms. 
Again, conservation of energy in subsequent dissipationless re-mergers, 
along with the assumption of preserved profile shape 
\citep[which we demonstrate is reasonable in][]{hopkins:cores} yields 
the evolution in subsequent events of these radii (in a 
re-merger of masses $M_{1}$ and $M_{2}$, the dissipationless 
bulge component will have final size 
$R_{f}/R_{1} \approx (1+\mu)^{2}/(1+\mu^{2}\,R_{1}/R_{2})$). 
Dissipation complicates this -- it is possible to solve separately for the 
size of the dissipational component by allowing for energy loss 
in the collision followed by (after angular momentum loss) collapse 
to a self-gravitating limit, and then subsequently evolve the 
component as a dissipationless body, added with the violently relaxed 
components to give a total bulge effective radius. Fortunately, 
\citet{covington:diss.size.expectation} 
perform such an exercise and we show in \citet{hopkins:cusps.ell} 
that their results can be conveniently 
approximated (in both an analytic manner and as a fit to the results of 
numerical simulations) by the scaling: 
$R_{e}({\rm bulge}) = R_{e}(f_{\rm sb}=0)/(1+f_{\rm sb}/f_{0})$, 
where $f_{\rm sb}$ is the total mass fraction of the bulge/spheroid which 
originally formed dissipationally (as opposed to being violently relaxed), 
$R_{e}(f_{\rm sb}=0)$ is the radius the system would have if purely 
dissipationless (calculated as described above), and $f_{0}\approx 0.25-0.35$ 
is a constant. 

Our modeling could also be applied in the manner described in \S~\ref{sec:model.flyby} 
to fly-by (non-merging) encounters, but we caution that these 
are usually ill-defined in semi-analytic models (and if adopted, the 
cautions in \S~\ref{sec:model.flyby} about the appropriate meaning of the impact 
parameter adopted should be borne in mind). In any case, the rapid suppression 
of bursts with increasing impact parameter means that such cases should 
be relatively unimportant in a representative cosmological ensemble.

\breaker
\section{Discussion and Conclusions}
\label{sec:discussion}

We have derived a general physical model for how 
disks survive and/or are destroyed in mergers and interactions. 
Our model describes both the dissipational and dissipationless 
components of the merger, and allows us to predict, for a 
given arbitrary encounter, the stellar and gas content of the 
system 
that will be dissipationlessly violently relaxed, dissipationally lose 
angular momentum and form a compact central starburst, or 
survive (without significant angular momentum loss or violent relaxation) 
to re-form a disk. 
We show that, in an immediate (short-term) sense, the amount of stellar 
or gaseous disk that survives or re-forms 
following a given interaction can be understood purely 
in terms of simple, well-understood gravitational physics. 
Knowing these physics, our model allows us to accurately predict the 
behavior in full hydrodynamic numerical simulations across as a function of 
the merger mass ratio, orbital parameters, pre-merger cold gas 
fraction, and mass distribution of the gas and stars, 
in simulations which span a wide range of parameter space 
in these properties as well as prescriptions for gas physics, 
stellar and AGN feedback, halo and initial disk structural 
properties, redshift, and absolute galaxy masses. 

The fact that we can understand the complex, nonlinear behavior 
in mergers with this analytic model, and moreover that (for given conditions 
at the time of merger) our results are independent of the details of 
prescriptions for gas physics, star formation, and feedback, owes 
to the fact that the processes that strip angular momentum 
from gas disks and violently relax stellar disks are fundamentally 
{\em dynamical}. 

Gas, in mergers, primarily loses angular momentum to 
internal gravitational torques (from the stars in the same disk) 
owing to asymmetries in 
the galaxy induced by the merger (on the close passages 
and final coalescence of the secondary, during which phase 
the potential also rapidly changes, scattering and violently relaxing the 
central stellar populations of the stellar disk).\footnote{We note again 
that although we have described these asymmetries as ``bars'' or 
``bar-like'' at certain points in this paper, there are a number of properties 
of the non-axisymmetric distortions induced in mergers  
(discussed in \S~\ref{sec:model.orbit} and \S~\ref{sec:model.secular}) that make them -- 
at least over the short relaxation timescale of the merger -- dynamically 
distinct from traditional bar instabilities in isolated systems.} 
Hydrodynamic torques and 
the direct torquing of the secondary are second-order effects, 
and inefficient for all but pathological orbits. 

Once gas is efficiently drained of angular momentum, 
there is little alternative but for it to fall to the center of the galaxy and 
form stars, regardless of the details of the prescriptions for star 
formation and feedback -- we show that even strong supernova-driven 
winds (with mass loading efficiencies several times the star formation rate 
and wind mass-loading velocities well above the halo escape velocity) 
do not significantly effect our conclusions. Such processes, after all, 
can blow out some of the gas, but they cannot fundamentally alter the 
fact that cold gas with no angular momentum will be largely 
unable to form any sort of disk, or the fact that a galaxy's worth of 
gas compressed to high densities and small radii 
will inevitably form a large mass in stars. 

For these reasons, many processes and details that are important
cosmologically (systematically changing e.g.\ the pre-merger 
disk gas fractions) -- in some sense setting the initial conditions for 
our idealized study of what happens in mergers -- 
do not alter the basic dynamical behavior within the mergers themselves, 
and therefore do not change our conclusions. 

\begin{figure*}
    \centering
    \scaleup
    \plotone{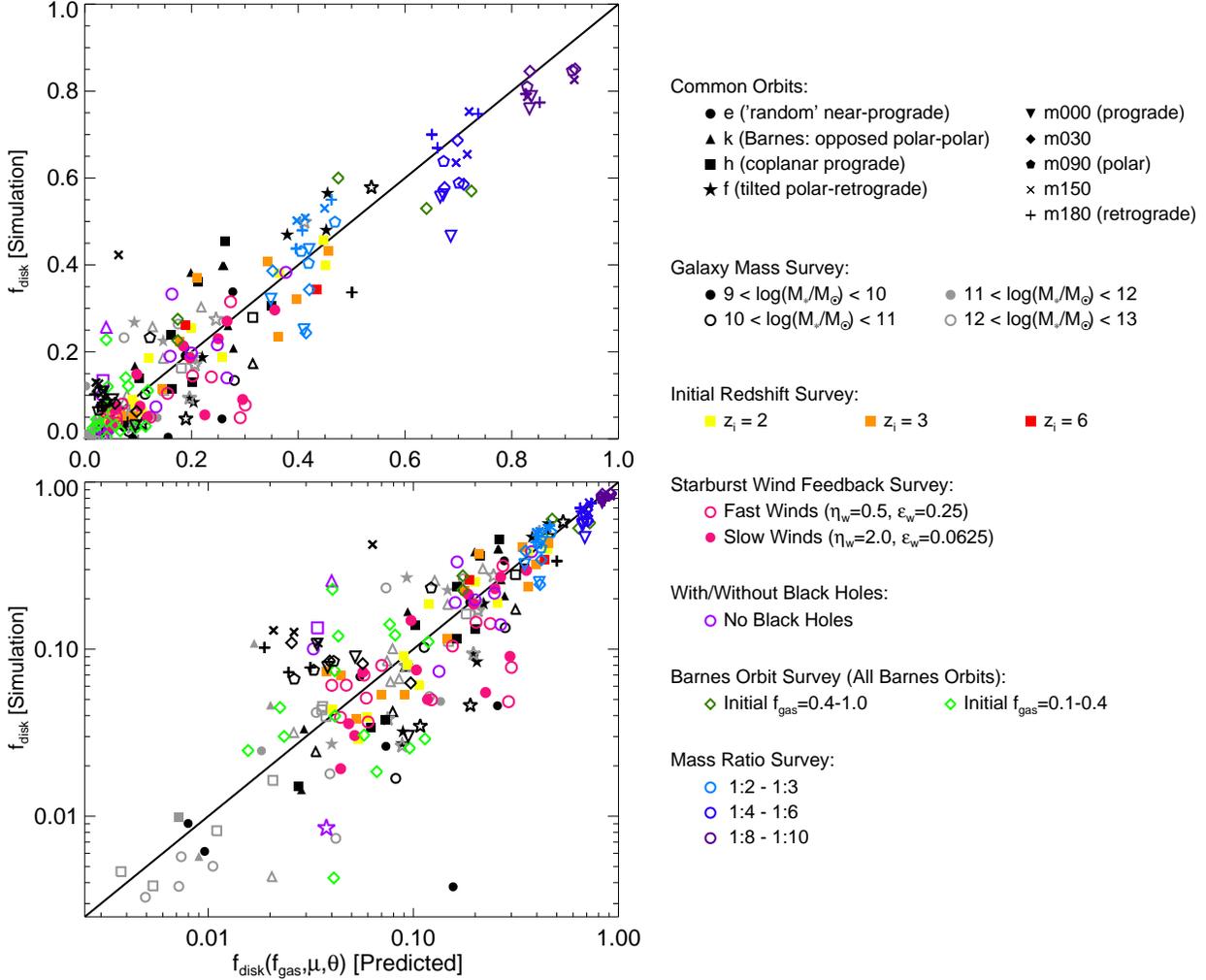}
    \caption{Summary of our comparison between simulations and 
    analytic model for the mass of disks in merger remnants as a function of 
    appropriate orbital parameters, merger mass ratio, and pre-merger 
    cold gas content. We plot our model prediction versus the simulation remnant 
    disk fraction for all $\sim400$ full hydrodynamic merger simulations considered 
    in this paper (shown in both a linear and logarithmic scale). 
    Symbols encode some of the parameter studies we consider: 
    orbital parameters, galaxy masses, initial merger redshift, 
    choice of feedback prescription, merger mass ratio, and presence or 
    absence of black holes, as labeled. For each subset of simulations, 
    we sample a wide range in initial and pre-merger gas fractions $\fgas=0-1$. 
    Solid line is a one-to-one relation. 
    In all cases, our predictions agree well with the simulations, with no systematic 
    offsets owing to any of the parameters we have varied. 
    At high $f_{\rm disk}$, our predictions are accurate to an absolute uncertainty 
    $\sim0.05-0.10$ in $f_{\rm disk}$. At low $f_{\rm disk}\lesssim0.1$, our predictions 
    are accurate to a factor $\sim2-3$ (down to $f_{\rm disk}\lesssim1\%$, where 
    it is difficult to reliably identify disks in the remnant). 
    \label{fig:summary}}
\end{figure*}

Figure~\ref{fig:summary} summarizes our results for the ensemble of our 
simulations. We compare the fraction of the baryonic galaxy mass in the 
merger remnant 
that is in a surviving post-merger disk to that predicted by our simple model scalings, 
and find good agreement over the entire range in disk and bulge mass 
fractions sampled, with surprisingly small scatter given the complexity of 
behavior in mergers. 
We highlight several of the parameter studies, 
showing that -- for fixed mass ratio, orbital parameters, and gas content 
{\em at the time of the final merger}, none of these choices systematically 
affect our predictions (note that these are not the only parameters varied -- the 
complete list is discussed in \S~\ref{sec:sims}, but it is representative). 
That is not to say they cannot affect them indirectly, 
by e.g.\ altering how much gas is available at the time of merger -- but it 
emphasizes that the processes we model and use to form our 
predictions, the processes that dominate violent relaxation and the loss 
of angular momentum in gas in mergers, are fundamentally dynamical. 

This allows us to make robust, accurate physical predictions independent of 
the (considerable) uncertainty in feedback physics and sub-resolution 
physics of the ISM. Regardless of how those physics alter the ``initial'' 
conditions, they do not change basic dynamical processes, 
and so do not introduce significant uncertainties in our model. 

In turn, this means that we can use our model to understand just why and how 
feedback is important for the cosmological survival of disks. Why, in short, 
have various works \citep[see e.g.][]{springel:spiral.in.merger,robertson:disk.formation,
governato:disk.formation} concluded that strong feedback is essential for 
enabling disk survival in mergers? 
Our results show that it is not that feedback somehow makes the disk 
more robust to the dynamical torques within the merger, in any 
instantaneous sense. 
These torques, at least within the critical radii where 
the gravitational perturbation from the merger is large and 
in resonance, are sufficiently strong that any reasonable feedback 
prescription is a dynamically negligible restoring force. 
Rather, feedback has two important effects that 
fundamentally alter the conditions in the merger: first, 
it allows the galaxy to retain much higher gas content going into the 
merger. Without feedback from e.g.\ star formation and supernovae 
contributing to heating and pressurizing the ISM and 
redistributing gas spatially, isolated gas-rich disks may be unstable to 
fragmentation. Even if fragmentation is avoided, 
it is well-known that star formation 
in simulations proceeds efficiently under these conditions. This would leave the 
disks essentially pure stars \citep[even for idealized simulations 
beginning with $\sim100\%$ gas disks; see e.g.][]{springel:models} 
by the time of the merger, which guarantees that a major merger 
will inevitably violently relax the stars (this is a simple collisionless 
mixing process, and under such circumstances is 
inescapable). With large gas fractions, however, 
the system relies on stripping angular momentum from the gas to 
form new bulge stars, which in turn relies on internal torques from 
induced asymmetries in the stellar disk. If the gas fractions are sufficiently 
large, there is little stellar disk to do any such torquing, and the 
gas survives largely intact. 

Second, feedback from supernovae and stellar winds moves the 
gas to large radii, where it does not feel significant torques from the 
merger. Again, recall that the most efficient torquing is driven 
by the internal stellar disk of the galaxy, and as such is most efficient 
at torquing gas within small radii (this can be thought of as 
analogous to the well-known co-rotation condition for isolated 
disk bars). If star formation-driven feedback has blown much of the 
gas to large radii, then there is little gas inside the radius 
where torques can efficiently strip angular momentum, yielding little 
induced starburst and largely preserving the gas disk at large radii. 

Not only can we qualitatively identify these requirements for feedback 
processes, but we can more precisely use our model to set quantitative 
limits on how much gas must be retained and/or the radii it must be 
redistributed to in order to enable disk survival under various 
conditions. This also clearly implies that disks must be able to avoid 
fragmentation and strong local gravitational instabilities when they 
achieve these gas fractions. 
This provides a valuable constraint for feedback models -- 
how those models affect star formation efficiencies, the ``blowout'' of 
gas, and the local hydrodynamic state (effective equations of 
state and phase structure) of ISM gas -- 
and should be useful for calibrating their (still largely 
phenomenological) implementations in both numerical 
and semi-analytic models of galaxy formation. 

Our predictions are also of interest in any cosmological model for the 
emergence of the Hubble sequence, since they apply not just to 
disk-dominated galaxies but to small disks in bulge-dominated 
systems. 
We give a number of simple prescriptions for 
application of our conclusions to analytic and semi-analytic models 
of galaxy formation, which can be used to predict 
the distribution of bulge to disk ratios in cosmological ensembles. 
But even without reference to a full such model, a number of 
interesting consequences are immediately apparent. 

First, it is a well-known problem that theoretical models systematically 
overpredict the abundance and mass fractions of bulges in 
(especially) low-mass galaxies. This is true even in e.g.\ semi-analytic 
models, which are not bound by resolution requirements and can adopt a 
variety of prescriptions for behavior in mergers. 
However, it is also well-established observationally that disk gas fractions tend 
to be very high in this regime, with large populations of gas-dominated 
disks at $M_{\ast}\ll 10^{10}\,M_{\sun}$ \citep{belldejong:tf,kannappan:gfs,mcgaugh:tf}. 
Our models predict that bulge formation 
should, therefore, be strongly suppressed in precisely the regime 
required by observations. For e.g.\ disks with $M_{\ast}<10^{9}\,M_{\sun}$ where 
observations suggest typical gas fractions $\sim60-80\%$, our results 
show that even a 1:1 major merger would typically yield a remnant with 
only $\sim30\%$ bulge by mass -- let alone a more typical 
1:3-1:4 mass-ratio merger, which should yield a remnant with $<20\%$ bulge. 
That is not to say that it is impossible to form a bulge-dominated system 
at these masses, but it should be much more difficult than at high masses, 
requiring either unusually gas-poor systems, violent merger histories, or 
rarer merging orbits that are more efficient at destroying disks. 
Our conclusions therefore have dramatic implications for the abundance of 
bulges and typical morphologies and bulge-to-disk ratios 
at low galaxy masses and in gas-rich systems. Low-mass systems, 
when a proper dynamical model of bulge formation in mergers is considered, 
should have lower bulge-to-disk ratios -- by factors of several, at least -- 
than have been assumed and modeled in previous 
theoretical models. 
Whether this alone is sufficient to resolve the discrepancies with the observations 
remains to be seen, but it is clearly of fundamental importance that future 
generations of models incorporate this scaling. 

Second, the importance of this suppression owing to gas content in disks 
will be even more significant at high redshifts.
Observations suggest \citep[see e.g.][]{erb:lbg.gasmasses} 
that by $z\sim2$, even systems with masses near $\sim L_{\ast}$ 
($M_{\ast}\sim 10^{10}-10^{11}\,M_{\sun}$) may have gas fractions as 
high as $\fgas\sim0.6$. In this regime, the same argument as above should apply, 
dramatically suppressing the ability of mergers to destroy disks. 
Moreover, 
since most of the mass density is near $L_{\ast}$, this can change not just 
the behavior in a specific mass regime but significantly suppress the global 
mass density of spheroids, modifying the predicted redshift history of bulge formation. 
(Note that this will not change when {\em stars} form by very much, so it has little or 
no effect on e.g.\ the ages of $z=0$ spheroids). 

This redshift evolution may also explain the 
solution to a fundamental problem in reconciling observed disk populations 
with CDM cosmologies. Integrated far enough back in time, every galaxy 
is expected to have experienced a significant amount of major merging. 
In extreme cases, the mass of the system when it had its last such merger 
may be so small that it would not be noticed today, but in general, 
it does not require going far back in redshift (to perhaps 
$z\sim2-4$ before almost every $z=0$ galaxy should have had such a 
merger). How, then, can the abundance of systems with relatively 
small (or even no) visible bulges be explained? Our conclusions here 
highlight at least part of the answer: as you go back in time, 
the gas fractions of systems are also higher, nearing unity. So even though, integrating 
sufficiently far in time, every system has experienced major mergers, 
it is also true that the systems were increasingly gas-rich, and therefore that 
the impact of those mergers was more and more suppressed. Only mergers 
at later times, below certain gas fraction thresholds, will typically destroy disks. 

Third, to the extent that bulge formation is suppressed at 
increasing redshifts, the existence of 
an $M_{\rm BH}-M_{\rm bulge}$ relation \citep[e.g.][]{magorrian} implies 
that black hole growth should also be suppressed. Indeed, 
bulge formation is suppressed specifically because gas cannot efficiently 
lose angular momentum in mergers if the systems are gas-dominated -- 
if the gas cannot lose angular momentum efficiently, then it certainly 
cannot efficiently be accreted by the nuclear black hole. 
Since this pertains to gas on the scales of galactic disks, 
it is probably not relevant for the formation of ``seed'' black holes at 
very high redshift, but it will in general inhibit the growth of black holes 
owing to early merging activity. At the same time, of course, 
higher gas fractions in general imply increasing fuel supplies for black 
hole growth, so the effects are not entirely clear, and more detailed 
models are needed to see how this impacts the history of black hole 
growth and quasar luminosity functions. Nevertheless, this may in part 
explain why, above $z\sim2$ (where, for the argument above, these 
effects become important for the global mass density of spheroids), 
the global rate of black hole growth (i.e.\ total quasar luminosity 
density) appears to decline much more rapidly with increasing 
redshift than the star formation rate density \citep[compare e.g.][]{hopkinsbeacom:sfh,
hopkins:groups.qso,hopkins:bol.qlf}. 

Fourth, our models imply that a large fraction of bulges and disks 
survive mergers together, rather than being formed entirely separately. 
It is often assumed that classical bulges -- being similar to 
small ellipticals in most of their properties -- were formed initially in 
major mergers, as entirely bulge-dominated systems, and then accreted 
new gaseous and stellar disks at later times. Although nothing in our modeling 
would prevent this from happening, our analytic and simulation results 
generically lead to the expectation that a large (perhaps even dominant) fraction 
of the bulge population did {\em not} form in this manner, but rather 
formed {\em in situ} from minor mergers or less efficient major mergers (in e.g.\ 
very gas-rich systems). 
Observations tracing the evolution of disk components, 
kinematics, and morphology in the last $\sim10\,$Gyr 
increasingly suggest that such co-formation or disk regeneration scenario 
is common \citep[see e.g.][and references therein]{hammer:obs.disks.w.mergers,
conselice:tf.evolution,flores:tf.evolution,puech:tf.evol}.
In short, a system with a mass fraction $\sim0.1-0.2$ in a 
bulge could be the remnant of an early, violent major merger (when the system 
was $\sim0.1$ times its present mass) with a re-accreted disk, or could be 
the remnant of a typical (low to intermediate gas fraction) 1:10-1:5 mass ratio 
minor merger, or could even be the remnant of a gas-rich major merger 
(mass ratio $\lesssim1:3$, if $f_{\rm gas}$ is sufficiently large). 

Based on 
a simple comparison of typical merger histories, we would actually expect that 
the minor merger mechanism should be most common, but all may be 
non-negligible. Fundamentally, the physics forming the bulge (torquing the gas within 
some radius owing to internal asymmetries and violently relaxing stars within 
a corresponding radius) are the same in all three cases, and moreover other indicators 
such as their stellar populations will be quite similar \citep[in all cases, the bulge will 
appear old: this is both because the central stars in even present-day disks 
are much older than those at more typical radii, and because in any case star formation 
will cease within the bulge itself, as opposed to the ongoing star formation in the disk, 
and stellar population age estimates are primarily sensitive to the amount of 
recent or ongoing star formation; see e.g.][]{trager:ages}. 
This is also not to say that mergers are the only means of producing 
bulges. Secular evolution of e.g.\ barred disks probably represents 
an increasingly important channel for bulge evolution in later-type 
and more gas-rich systems \citep[see e.g.][]{christodoulou:bar.crit.1,sheth:bar.frac.evol,
mayer:lsb.disk.bars,debattista:pseudobulges.a,jogee:bar.frac.evol,
kormendy.kennicutt:pseudobulge.review,marinova:bar.frac.vs.freq}, and 
may even be related (albeit through longer timescales of ``isolated,'' 
post-merger evolution and different physics) to initial bar formation 
or ``triggering'' in mergers. 
More detailed theoretical 
work and analysis of cosmological simulations is needed to develop observational 
probes that can distinguish between these histories. 

Further work is specifically needed to investigate the processes at work in 
minor mergers with mass ratios $\sim$1:10 ($\mu\sim0.05-0.1$), 
which cosmological simulations suggest 
are an important contributor to the growth of disks, especially 
in later-type systems 
\citep{maller:sph.merger.rates,
fakhouri:halo.merger.rates,stewart:mw.minor.accretion}. 
In more minor mergers $\mu \ll 0.1$, the secondaries are sufficiently small and dynamical 
friction times sufficiently 
long that the disk is unlikely to feel significant external perturbations. 
More major mergers $\mu \gtrsim 0.1$, the cases of interest here, 
induce sufficiently large responses in the disk and evolve sufficiently rapidly 
that they can be considered ``merger-dominated'' 
for the reasons in \S~\ref{sec:model.orbit} \&\ \ref{sec:model.secular}. 
But in the intermediate regime, internal amplification of instabilities in a 
traditional secular fashion may occur on a timescale comparable to 
or shorter than the evolution of the secondary orbit, potentially 
leading to a more complex interplay between the two. It is not entirely 
clear whether such a system would remain ``locked'' to the driven 
perturbation, or function as a purely secular system (merely initially 
driven by the presence of the secondary), or some nonlinear combination 
of both. A more detailed comparison of the relevant timescales 
for these processes and their relation to e.g.\ cosmological triggering of 
bars and large-scale non-axisymmetric modes in disks will be the 
subject of future study (in preparation).

Our results are also of direct interest to models of spheroid formation in 
ellipticals and S0 galaxies. As discussed in \S~\ref{sec:intro}, it is increasingly 
clear that embedded sub-components -- constituting surviving gaseous 
and stellar disks -- are both ubiquitously observed and critical 
for theoretical models to match the detailed kinematics and isophotal 
shapes of observed systems \citep{naab:gas,
cox:xray.gas,cox:kinematics,robertson:fp,jesseit:kinematics,
hopkins:cusps.ell,hopkins:cusps.fp}. We have developed a model 
that allows us to make specific predictions for how disks survive mergers, 
including both the survival of some amount of the pre-merger stellar disks 
and the post-merger re-formation of disks and rotationally supported 
components from gas that survives the merger without losing most of its 
angular momentum.

Figure~\ref{fig:summary} shows that we can extend 
these predictions with reasonable accuracy to surviving rotational systems 
containing as little as $\sim 1\%$ of the remnant stellar mass, comparable to 
small central subcomponents and subtle features giving rise to e.g.\ 
slightly disky isophotal shapes \citep[see e.g.][]{ferrarese:type12,lauer:centers,
mcdermid:sauron.profiles}. Owing to 
the combination of resolution requirements and desire to understand the 
fundamental physics involved, most theoretical studies of these detailed 
properties of ellipticals have been limited to idealized studies of individual 
mergers. Our results allow these to be placed in a more global context 
of cosmological models and merger histories. Moreover, our 
models allow the existence of such features (or lack thereof) to be translated 
into robust constraints on the possible merger histories and gas-richness 
of spheroid-forming mergers. Further, \citet{hopkins:cusps.ell,hopkins:cusps.fp}, 
studied how the dissipational starburst components arising in gas-rich 
mergers are critical to explaining the observed properties and scaling relations 
of ellipticals, and how these components can both be extracted from 
and related to observed elliptical surface brightness profiles. Because both 
the starburst and surviving disks arise from gas in mergers, the combination of 
constraints from the central stellar populations, studied therein, with 
constraints on the survival and/or loss of gas angular momentum in mergers 
studied here, should be able to break some of the degeneracies in e.g.\ 
pre-merger gas fractions and merger histories in order to enable new 
constraints and understanding of spheroid merger histories, and 
new tests of models for spheroid formation in gas-rich mergers. 

These points relate to a number of potentially testable predictions of 
our models. These include the in situ formation of bulges from various types of 
mergers, and possible associated stellar population signatures, the 
presence of embedded disks in ellipticals, and how their sizes and mass 
fractions scale with e.g.\ the masses and formation times of ellipticals 
(and how this relates to gas fractions and stellar 
populations in observed disks). In general, 
for similar merger histories, the increasing prevalence of later type 
galaxies (S0's and S0a's) at lower masses where disks are characteristically 
more gas rich is a natural consequence of our predictions here, and 
it is straightforward to convert our predicted scalings into detailed predictions 
for the abundance and mass fractions of disks given some simplified merger histories. 
To the extent that these processes also give rise to disk heating 
and/or increasing velocity dispersions in disks, or changing kinematics in 
both disks and bulges, then there should be corresponding relationships 
between galaxy shapes, kinematics, and bulge-to-disk ratios along the Hubble 
sequence. We investigate these possible correlations and tests in 
subsequent papers (in preparation). 

Altogether, our results here elucidate 
the relevant physics important for both dissipational and dissipationless 
bulge formation in mergers. They 
support a new paradigm in 
which to view bulge and disk formation: gas-richness is not simply 
a ``tweak'' to existing models of bulge formation and disk 
destruction in mergers. Rather, if disks are sufficiently gas rich, 
the qualitative character of mergers is different, with inefficient 
angular momentum loss giving rise to disk-dominated 
remnants. This process is not inherently governed by poorly-understood 
feedback physics (although such feedback may be critical for 
establishing the conditions necessary in the first place), 
but rather by well-understood gravitational physics, and as such is 
robust and fundamentally inescapable. Aspects of 
galaxy populations such as the continuum of relative bulge 
and disk mass ratios are not simply consequences of e.g.\ different 
amounts of accretion, but can arise owing to the continuum in 
efficiencies of disk destruction as a function of merger 
mass ratios, orbital parameters, and gas content. 
The relative (lack of) abundance of bulges at low galaxy masses and high 
redshift is a basic consequence of the dynamics of 
how gas loses angular momentum in mergers, even for similar 
merger histories. In short, the baryonic physics of mergers 
ensures that, despite the near self-similarity of the physics and merger histories 
of their host halos, disk and bulge formation are not a self-similar 
process, influenced dramatically (well out of proportion to the absolute 
cold gas mass fractions) by the gas-richness of the baryonic systems.

\acknowledgments 
We thank Shardha Jogee and Rachel Somerville 
for helpful discussions, and thank the anonymous referee for helpful 
suggestions and clarification. This work
was supported in part by NSF grants ACI 96-19019, AST 00-71019, AST
02-06299, and AST 03-07690, and NASA ATP grants NAG5-12140,
NAG5-13292, and NAG5-13381. Support for 
TJC was provided by the W.~M.\ Keck 
Foundation. 

\bibliography{/Users/phopkins/Documents/lars_galaxies/papers/ms}

\end{document}